\documentclass{amsart}
\usepackage{fullpage,graphicx,subfigure,mathpazo,color}
\usepackage{amsmath,amscd,tikz}
\usepackage[normalem]{ulem}

\newtheorem{theorem}{Theorem}

\newtheorem{remark}{Remark}

\newcommand{\ii}{\mathrm{i}}
\newcommand{\ee}{\mathrm{e}}
\newcommand{\dd}{\mathrm{d}}
\newcommand{\T}{\mathrm{T}}
\newcommand{\sn}{\mathrm{sn}}
\newcommand{\cn}{\mathrm{cn}}
\newcommand{\dn}{\mathrm{dn}}

\makeatletter
\renewcommand{\theequation}{\arabic{section}.\arabic{equation}}
\@addtoreset{equation}{section}
\makeatother

\begin{document}

\title{Multi-breathers and high order rogue waves for the nonlinear Schr\"odinger equation on the elliptic function background}

\author{Bao-Feng Feng}
\address{School of Mathematical and Statistical Sciences, The University of Texas Rio Grande Valley Edinburg TX, 78541-2999, USA}
\email{baofeng.feng@utrgv.edu}
\author{Liming Ling}
\address{School of Mathematics, South China University of Technology, Guangzhou, 510641, China}
\email{linglm@scut.edu.cn; lingliming@qq.com}
\author{Daisuke A. Takahashi}
\address{Research and Education Center for Natural Sciences, Keio University, Hiyoshi 4-1-1, Yokohama, Kanagawa 223-8521, Japan}
\email{daisuke.takahashi@keio.jp}
\date{\today}

\begin{abstract}
We construct the multi-breather solutions of the focusing nonlinear Schr\"odinger equation (NLSE) on the background of elliptic functions by the Darboux transformation, and express them in terms of the determinant of theta functions. The dynamics of the breathers in the presence of various kinds of backgrounds such as dn, cn, and non-trivial phase-modulating elliptic solutions are presented, and their behaviors dependent on the effect of backgrounds are elucidated. We also determine the asymptotic behaviors for the multi-breather solutions with different velocities in the limit $t\to\pm\infty$, where the solution in the neighborhood of each breather tends to the simple one-breather solution. Furthermore, we exactly solve the linearized NLSE using the squared eigenfunction and determine the unstable spectra for elliptic function background. By using them, the Akhmediev breathers arising from these modulational instabilities are plotted and their dynamics are revealed. Finally, we provide the rogue-wave and higher-order rogue-wave solutions by taking the special limit of the breather solutions at branch points and the generalized Darboux transformation. The resulting dynamics of the rogue waves involves rich phenomena: depending on the choice of the background and possessing different velocities relative to the background. We also provide an example of the multi- and higher-order rogue wave solution.
\end{abstract}

\maketitle
\section{Introduction}
As a universal model in the description of the propagation of a quasi-monochromatic wave in a weakly nonlinear medium, the nonlinear Schr\"odinger equation (NLSE) has played a central role in mathematical physics and integrable partial differential equations \cite{FaddeevTakhtajan}. It is relevant in numerous applications of natural sciences such as water
waves, nonlinear optics, plasmas, Bose-Einstein condensates. For the zero background, a variety of methods either for the analysis of the Cauchy problem or the systematic construction of the multi-soliton solutions
has been well established for a long time. A nonzero with plane wave background can be unstable under the perturbation, and this modulation instability (MI) is considered as the main cause for the formation of rogue
waves (RWs) in nature. The integrable nature of the NLSE allows one to construct a family of exact solutions corresponding to this nonzero plane wave background
by different methods such as the generalized Darboux transformation method,  the dressing method, the inverse scattering method and Hirota bilinear method. Since the 70s, there are lots of works on the integrable NLSE model; giving a complete account of the reference on the relative subject is out of the scope of this introduction.

Recent study by Grinevich and Santini \cite{GrinevichS18} (equations (48)-(49) ) shows finite-gap soliton solutions of genius-1 or genius-2 may be developed from a Cauchy problem of the NLSE with a very small perturbation. For the genius-1 case, a nonzero background expressed by elliptic function is formed before the rogue wave solution or the Akhmediev breather solution is formed. In this sense, the rogue wave or Akhmediev breather solutions with complicated backgrounds such as elliptic functions are more natural and are expected to occur in experiments.  Actually, Akhmediev et al. have studied this kind of solution in a simple setting \cite{KedzioraAA14}. Due to the mathematical difficulty, the study of constructing soliton solutions with elliptic function background is much less and nontrivial. In this paper, one aim is to construct general breathers and rogue waves of the NLSE with the elliptic function background systematically.

From the theoretical point of view, the systematic construction of the multi-soliton solutions on the elliptic function background can be broadly classified into two categories.
The first method is to take the limit in the higher-genus solutions constructed by the algebro-geometric method \cite{BelokolosBEIM94} or the Riemann-Hilbert method. The other way is using the inverse scattering method or the Darboux transformation \cite{BilmanB18,BilmanLM18,Cieslieski09,GuHZ05,GuoLL12,MatveevS91} to construct the breathers by the linear algebra. However, both methods have different difficulties. For the first way, it is hard to analyze the different limitation and deal with lots of elliptic integration, while for the second way, the obtained formula by the linear algebra is easy, but it is difficult to analyze the properties for the determinant.

Firstly, we review several known results on this aspect. In the section 4.5 of monograph \cite{BelokolosBEIM94}, the authors considered the degenerate finite-gap solution---multi-phase modulations of the cnoidal wave for the defocusing NLSE by taking the limit and used the complex coordinate transformation $x\to\ii x$ to derive multi Akhmediev breathers on the ${\rm dn}$ function background \cite{BelokolosBEIM94}. Recently, Shin considered the single soliton dynamics in phase-modulated lattices by Darboux transformation by modified squared wavefunction approach \cite{Shin12}. A compact formula for the single breather solution was not, however, given explicitly, and hence the shift of the crest and phase was not analyzed. The exact solitons on the elliptic function background are constructed by the eigenfunction symmetry method \cite{ChengLC14,LouCT14}. Breathers, rogue waves and higher order rogue waves on the dn and cn background are obtained by numerically solving the Lax pair equation and Darboux transformation \cite{AshourCNB,KedzioraAA14}. The solitons for the defocusing NLSE on the elliptic function background was obtained by the inverse scattering method \cite{Takahashi16}, where the seed eigenfunction of the Lax pair was obtained using the result of \cite{Takahashi12}, and the solution was expressed by the determinant of theta function. The rogue waves for the focusing NLSE on the cn and dn background are obtained by combining the Darboux transformation with nonlinearization of the Lax pair \cite{ChenP18}.
Meanwhile, there are some relevant works on the rogue waves on the higher genus solutions \cite{BertolaET16}, in which the analytical criterion to generalized rogue waves are obtained by the Riemann-Hilbert method. The bounded ultra-elliptic solutions for the defocusing NLSE are considered by the effective integration method \cite{Wright17}. The various initial data for the semiclassical NLSE are related with the rogue-wave phenomena \cite{LyngM07,BertolaT13,ElKT16}.

Even though there are some contributions on this subject, there is not systematic theory to construct the breathers and rogue waves on the elliptic function background. What is more important is that there is no method to analyze the solutions generated by the Darboux transformation. In this work, we would like to address these issues. The key step to generate the solution by the Darboux transformation is solving the Lax pair equation with the seed solution. Once it can be obtained, plugging the vector solutions of the Lax pair into the B\"acklund transformation induced by the Darboux transformation, we get the new solutions. Since the formula of the B\"acklund transformation is algebraic, the new solutions are obtained exactly, and furthermore, this operation can be iterated as many times as desired. Therefore, the family of the multi-fold and higher-order solutions can be obtained recursively. The closed-form multi-breather solutions  can be easily plotted by just setting the parameters, and no numerical computation is necessary.

However, since the solutions generated by the Darboux transformation involve lots of free non-trivial parameters, if we want to understand their effect precisely, the formulas found from the B\"acklund transformation are not always useful for observation of the general laws of the dynamics. Thus we need to reduce the formula in a proper form, so that the dynamics of solutions can be analyzed. In this work, we use the addition formulas of theta functions to rewrite the expressions of the multi-soliton solutions. In this way, the general dynamics can be analyzed directly through the asymptotic analysis of the determinant formula.

It is well known that the breather solutions in the focusing NLSE involve the special and important limiting cases, which have a close relation to the modulational instability.
The elliptic-function solutions for the focusing NLSE are also known to be unstable \cite{DeconinckS17}. It is therefore natural to construct the exact solutions to describe the unstable process. Thus, the second aim in this work is to construct the Akhmediev breathers on the elliptic function background, which is helpful to understand the mechanism of modulational instability.

Finally, for the rogue waves, it is also related with the modulational instability with infinite period. The comprehensive construction for higher order rogue waves with the closed expressions has been open so far. In this work, we prove that the formulas of higher order rogue waves can be represented as a rational functions of $x$, $t$, the Jacobi elliptic functions ${\rm sn}$, ${\rm cn}$, ${\rm dn}$ and the second type of elliptic integral.

The rest of this paper is organized as the follows. In section \ref{sec2} we solve the Lax pair with the elliptic function by combining the algebro-geometric method and the modified squared wavefunction approach. As a byproduct of the modified squared wavefunction approach, we also solve the linearized NLSE directly, which is useful for the analysis of the modulational instability. In section \ref{sec4}, we construct the breathers and multi-breathers on the elliptic function background. Furthermore, with the aid of the addition formula for the theta function, we rewrite them in a determinant form. Their dynamics are also investigated for various backgrounds. In section \ref{sec5}, we determine the unstable spectra for the linearized NLSE, and analyze the corresponding Akhmediev breathers on the elliptic-function background, which is related with the modulational instability. In section \ref{sec6}, we provide the rogue waves and high order rogue waves. We further discuss the effect of the choice of backgrounds to the dynamics of rogue waves. Section \ref{sec7} is devoted to the conclusion and discussion.

\section{The elliptic function solution of NLSE and its Lax pair}\label{sec2}

In this section, we derive the elliptic function solution for NLSE firstly and then the corresponding solutions for the Lax pair are solved with the aid of algebraic geometry method. The scheme is given in details in the monograph \cite{BelokolosBEIM94}. Here we just focus on the genus one case, in which we can use the elliptic functions and integrals to represent these solutions instead of the Abel map and Riemann theta function. In the following, we give the details to construct the solutions.

The NLSE reads
\begin{equation}\label{eq:nlse}
{\rm i}\psi_t+\frac{1}{2}\psi_{xx}+|\psi|^2\psi=0,
\end{equation}
which admits the following Lax pair:
\begin{equation}\label{eq:lax-nlse}
\begin{split}
{\Phi}_x&=\mathbf{U}(\mathbf{Q};\lambda) {\Phi}, \,\, \mathbf{U}(\mathbf{Q};\lambda)=-{\rm i}\lambda \sigma_3+{\rm i}\mathbf{Q}, \\
{\Phi}_t&=\mathbf{V}(\mathbf{Q};\lambda) {\Phi}, \,\, \mathbf{V}(\mathbf{Q};\lambda)=-{\rm i}\lambda^2\sigma_3+\lambda {\rm i}\mathbf{Q}+\mathbf{V}_0,\,\,\,\, \mathbf{V}_0=\frac{1}{2}\sigma_3({\rm i}\mathbf{Q}^2-\mathbf{Q}_{x}),
\end{split}
\end{equation}
where
\begin{equation*}
\sigma_3=\begin{bmatrix}
1&0\\
0&-1
\end{bmatrix},\,\,\,\,\,\, \mathbf{Q}=\begin{bmatrix}
0&\psi\\
\psi^*&0
\end{bmatrix}.
\end{equation*}
In other words, the compatibility condition $\Phi_{xt}=\Phi_{tx}$ or the zero curvature condition $\mathbf{U}_t-\mathbf{V}_x+[\mathbf{U},\mathbf{V}]=0$ yields the NLSE \eqref{eq:nlse}. It is readily to see the symmetry for the potential function $\mathbf{U}$ and $\mathbf{V}$: $\mathbf{U}^{\dag}(\mathbf{Q};\lambda^*)=-\mathbf{U}(\mathbf{Q};\lambda)$ and $\mathbf{V}^{\dag}(\mathbf{Q};\lambda^*)=-\mathbf{V}(\mathbf{Q};\lambda)$, where $^\dag$ denotes the hermitian conjugate. Based on the above symmetry, if $\Phi(x,t;\lambda)$ is a matrix solution for Lax pair \eqref{eq:lax-nlse}, then $\Phi^{\dag}(x,t;\lambda^*)$ is a matrix solution for the adjoint Lax pair of \eqref{eq:lax-nlse}. As a general symmetry, if $\Phi(x,t;\lambda)$ satisfies the Lax pair \eqref{eq:lax-nlse}, then $({\rm i}\sigma_2)\Phi(x,t;\lambda)^\T(-{\rm i}\sigma_2)$ satisfies the adjoint Lax pair of \eqref{eq:lax-nlse}, where $\sigma_2$ is a Pauli matrix:
\begin{equation*}
\sigma_2=\begin{bmatrix}
0&-\ii\\
\ii&0
\end{bmatrix}.
\end{equation*}
We define a new matrix function $\mathbf{L}(x,t;\lambda)=-\frac{1}{2}\Phi(x,t;\lambda)\sigma_3 ({\rm i}\sigma_2)\Phi(x,t;\lambda)^{\T}(-{\rm i}\sigma_2)$ which satisfies the following stationary zero curvature equations:
\begin{equation}\label{eq:station-lax-nlse}
\begin{split}
\mathbf{L}_x&=[\mathbf{U},\mathbf{L}], \\
\mathbf{L}_t&=[\mathbf{V},\mathbf{L}].
\end{split}
\end{equation}
It should be stressed that the compatibility condition for the above stationary zero curvature conditions also give the NLSE \eqref{eq:nlse} by the compatibility condition of the Lax pair and Jacobi identity.

\subsection{Linearized NLSE and its solution}\label{subsec:linearizednlse}
The above stationary zero curvature equation \eqref{eq:station-lax-nlse} can be used to construct the solutions of linearized nonlinear Schr\"odinger equation:
\begin{equation}\label{eq:linear-nlse}
{\rm i}p_t+\frac{1}{2}p_{xx}+2|\psi|^2p+\psi^2p^*=0,
\end{equation}
where $p(x,t)$ is a perturbed function.
To give the exact solution from the stationary zero curvature equations, we rewrite the matrix function $\mathbf{L}(x,t;\lambda)$ with in the form:
\begin{equation}\label{eq:l}
\mathbf{L}(x,t;\lambda)=\begin{bmatrix}
-{\rm i}{f}(x,t;\lambda) &{g}(x,t;\lambda) \\
{h}(x,t;\lambda) & {\rm i}{f}(x,t;\lambda)
\end{bmatrix},
\end{equation}
where
\begin{equation}
\begin{split}
{f}(x,t;\lambda)&=-\frac{{\rm i}}{2}\left[\phi_1(x,t;\lambda)\varphi_2(x,t;\lambda)+\varphi_1(x,t;\lambda)\phi_2(x,t;\lambda)\right],\\
{g}(x,t;\lambda)&=\phi_1(x,t;\lambda)\phi_2(x,t;\lambda),\,\,\,\,\,
{h}(x,t;\lambda)=-\varphi_1(x,t;\lambda)\varphi_2(x,t;\lambda),
\end{split}
\end{equation}
and
\[
\Phi(x,t;\lambda)=\begin{bmatrix}
\phi_1(x,t;\lambda) & \phi_2(x,t;\lambda)  \\
\varphi_1(x,t;\lambda) & \varphi_2(x,t;\lambda)
\end{bmatrix},
\]
$[\phi_i,\varphi_i]^{\rm T}$ are two vector solutions for Lax pair \eqref{eq:lax-nlse}.
Then we write the stationary zero curvature equation in more details as
\begin{equation}\label{eq:station-lax-nlse-x}
\begin{split}
f_x=&\psi^* g-\psi h,\,\,\,\, g_x=-2(\psi f+{\rm i}\lambda g),\,\,\,\, h_x=2(\psi^*f+{\rm i}\lambda h),\\
f_t=&\left(\lambda \psi^*-\frac{{\rm i}}{2}\psi^*_x\right)g-\left(\psi\lambda+\frac{{\rm i}}{2}\psi_x\right) h,\,\,\,\,\,
g_t=-2\left[\left(\lambda\psi +\frac{\rm i}{2}\psi_x\right)f+\left({\rm i}\lambda^2-\frac{\rm i}{2}|\psi|^2 \right)g\right],  \\
h_t =&2\left[\left(\lambda\psi^* -\frac{\rm i}{2}\psi_x^*\right)f+\left({\rm i}\lambda^2-\frac{\rm i}{2}|\psi|^2 \right)h\right].\\
\end{split}
\end{equation}
From the above equations, we obtain
\begin{equation*}
-\frac{\rm i}{2}[\frac{1}{2}g_{xx}+\psi(\psi^*g-\psi h)]=\frac{1}{2}({\rm i}\psi_x +2\lambda \psi) f+{\rm i}\lambda^2 g,
\end{equation*}
then we find that
\begin{equation}\label{eq:linear-g}
{\rm i}g_t+\frac{1}{2}g_{xx} +2|\psi|^2 g-\psi^2 h=0.
\end{equation}
Similarly, we have
\begin{equation}\label{eq:linear-h}
-{\rm i}h_t+\frac{1}{2}h_{xx} +2|\psi|^2 h-\psi^{*2} g=0.
\end{equation}
If $h=-g^*$, then the above equation turns into the linearized NLSE \eqref{eq:linear-nlse}. But in general, $h$ does not equals to $-g^*$ in the above setting.   To satisfy this requirement, we use the symmetry to construct the combination of solution. Since $[\phi_i(\lambda),\varphi_i(\lambda)]^{\rm T}$, $i=1,2$ is a solution for Lax pair \eqref{eq:lax-nlse}, then $[-\varphi_i^*(\lambda),\phi_i^*(\lambda)]^{\rm T}$ is a solution for Lax pair \eqref{eq:lax-nlse} by replacing $\lambda$ with $\lambda^*$. Thus, we find that the function $g(\lambda^*)=\varphi_1^*(x,t;\lambda)\varphi_2^*(x,t;\lambda)$ and $h(\lambda^*)=-\phi_1^*(x,t;\lambda)\phi_2^*(x,t;\lambda)$ satisfy the linearized equation \eqref{eq:linear-g} and \eqref{eq:linear-h} respectively. Then we conclude that the exact solution for the linearized NLSE \eqref{eq:linear-nlse} is $p=\phi_1(x,t;\lambda)\phi_2(x,t;\lambda)+\varphi_1^*(x,t;\lambda)\varphi_2^*(x,t;\lambda).$ This solution is useful to analyze the spectral stability of the background solution.

\subsection{Elliptic function solution for NLSE and the corresponding solution of Lax pair}

Starting from the stationary zero curvature equations, we set the solution $\mathbf{L}$ as the polynomial of $\lambda$. For the genus one solution, we assume that the matrix function $\mathbf{L}$ is a quadratic polynomials of $\lambda$:
\begin{equation}\label{eq:L-poly}
\mathbf{L}=\mathbf{L}_0(x,t)\lambda^2+\mathbf{L}_1(x,t)\lambda+\mathbf{L}_2(x,t).
\end{equation}
Plugging the above equation \eqref{eq:L-poly} into equation \eqref{eq:station-lax-nlse} and comparing the coefficient of $\lambda$, we obtain that: for the $x$-equation
\begin{equation}\label{eq:l-x-part}
\begin{split}
&\sigma_3\mathbf{L}_0^{\rm off}=0, \\
&\mathbf{L}_{0,x}+2{\rm i}\sigma_3\mathbf{L}_1^{\rm off}-{\rm i}[\mathbf{Q},\mathbf{L}_0]=0, \\
&\mathbf{L}_{1,x}+2{\rm i}\sigma_3\mathbf{L}_2^{\rm off}-{\rm i}[\mathbf{Q},\mathbf{L}_1]=0, \\
&\mathbf{L}_{2,x}-{\rm i}[\mathbf{Q},\mathbf{L}_2]=0, \\
\end{split}
\end{equation}
which infers that
$\mathbf{L}=-{\rm i}\sigma_3 \alpha_0\lambda^2+(-{\rm i}\sigma_3 \alpha_1+{\rm i}\alpha_0 \mathbf{Q} )\lambda+\alpha_0\mathbf{V}_0+{\rm i}\alpha_1 \mathbf{Q}+\alpha_2(-{\rm i}\sigma_3)$;
for the $t$-equation
\begin{equation}\label{eq:l-t-part}
\begin{split}
&\sigma_3\mathbf{L}_0^{\rm off}=0, \\
&2{\rm i}\sigma_3\mathbf{L}_1^{\rm off}-{\rm i}[\mathbf{Q},\mathbf{L}_0]=0, \\
&\mathbf{L}_{0,t}+2{\rm i}\sigma_3\mathbf{L}_2^{\rm off}-{\rm i}[\mathbf{Q},\mathbf{L}_1]-[\mathbf{V}_0,\mathbf{L}_0]=0, \\
&\mathbf{L}_{1,t}-{\rm i}[\mathbf{Q},\mathbf{L}_2]-[\mathbf{V}_0,\mathbf{L}_1]=0, \\
&\mathbf{L}_{2,t}-[\mathbf{V}_0,\mathbf{L}_2]=0, \\
\end{split}
\end{equation}
from which we obtain that $\alpha_{0,t}=\alpha_{1,t}=\alpha_{2,t}=0$ and
\begin{equation}\label{eq:Qt-Qx}
\mathbf{Q}_t-2{\rm i}\alpha_2\sigma_3 \mathbf{Q}+\alpha_1\mathbf{Q}_x=0.
\end{equation}
On the other hand, suppose that the matrix solution $\mathbf{L}(x,t;\lambda)$ can be solved formally with the form:
\begin{equation*}
\begin{split}
\mathbf{L}(x,t;\lambda)= &\exp\left[\int_{(x_0,t_0)}^{(x,t_0)}\mathbf{U}(x',t';\lambda){\rm d}x'+\int_{(x,t_0)}^{(x,t)}\mathbf{V}(x',t';\lambda){\rm d}t'\right]\mathbf{M}(\lambda) \\
&\,\,\,\,\,\,\,\,\,\,\,\,\,\,\,\,\,\,\,\,\,\,\,\,\,\,\,\,\,\,\,\,\,\,\,\,\,\,\,\,\exp\left[-\int_{(x_0,t_0)}^{(x,t_0)}\mathbf{U}(x',t';\lambda){\rm d}x'-\int_{(x,t_0)}^{(x,t)}\mathbf{V}(x',t';\lambda){\rm d}t'\right]
\end{split}
\end{equation*}
which is non-degenerate if the constant matrix $\mathbf{M}(\lambda)$ is non-singular. Thus, combining with the above ansatz \eqref{eq:L-poly}, the determinant of the matrix $\mathbf{L}(x,t;\lambda)$ can take the following form
\begin{equation}\label{eq:L-det}
\begin{split}
\det(\mathbf{L}(x,t;\lambda))=f(x,t;\lambda)^2-g(x,t;\lambda)h(x,t;\lambda)=P(\lambda), \\
P(\lambda):=\prod_{i=1}^{4}(\lambda-\lambda_i)=\lambda^4-s_1\lambda^3+s_2\lambda^2-s_3\lambda+s_4,
\end{split}
\end{equation}
where $\lambda_i$s are zeros of the polynomial which characterize the genus one solution. Here we see that $\alpha_0=1$. By the symmetry relationship and the existence and uniqueness of ordinary differential equation, we know that the solution  satisfies $\mathbf{L}(x,t;\lambda)=-\mathbf{L}^{\dag}(x,t;\lambda^*).$ Moreover, through the determinant relation \eqref{eq:L-det}, we obtain that $\prod_{i=1}^{4}(\lambda-\lambda_i)=\prod_{i=1}^{4}(\lambda-\lambda_i^*).$ In this way, we can suppose $\lambda_{j}=\lambda_{j+2}^*$, $j=1,2.$ In summary, the functions $f(x,t;\lambda)$, $g(x,t;\lambda)$ and $h(x,t;\lambda)$ are given by:
\begin{equation}\label{eq:f-g-h}
f(x,t;\lambda)=\lambda^2+\alpha_1\lambda +(\alpha_2-\frac{1}{2}|\psi|^2),\,\,\, g(x,t;\lambda)=-h^*(x,t;\lambda^*)=
{\rm i}\lambda \psi-\frac{1}{2}\psi_x+{\rm i} \alpha_1\psi=\ii\psi(\lambda-\mu)
\end{equation}
where we introduced $ \mu(x,t):=-\frac{{\rm i}}{2}(\ln\psi)_x-\alpha_1$ in the last expression. Now we give the relationship between the coefficients $s_{1\leq i\leq 4}$ and functions $\psi(x,t)$, $\mu(x,t)$, $\nu=|\psi|^2$, $\alpha_2.$ By comparing the equation \eqref{eq:L-det}, we find that
\begin{equation}\label{eq:rela-alpha-s}
\begin{split}
2\alpha_1&=-s_1,\,\,\,\,\, \alpha_1^2+2\alpha_2=s_2, \\
2\alpha_1(\alpha_2-\frac{1}{2}\nu)-\nu(\mu+\mu^*)&=-s_3,\,\,\,\,\,\,\, (\alpha_2-\frac{1}{2}\nu)^2+\nu\mu\mu^*=s_4. \\
\end{split}
\end{equation}
Moreover, from the latter two equations of \eqref{eq:rela-alpha-s}, we obtain the algebraic relation between $\mu$, $\mu^*$ and $\nu$:
\begin{equation}\label{eq:mu-nu}
\mu+\mu^*=\frac{s_1}{2}-\frac{q}{\nu},\,\,\,\,\,\, \mu\mu^*=-\frac{1}{\nu}(\nu^2-2p \nu+p^2-4s_4),
\end{equation}
where $q=\frac{s_1}{2}(s_2-\frac{s_1^2}{4})-s_3$, $p=s_2-\frac{s_1^2}{4}.$ Then $\mu$, $\mu^*$ can be solved exactly:
\begin{equation}\label{eq:mu-mu*}
\mu,\mu^*=\frac{s_1}{4}-\frac{q\pm {\rm i}\sqrt{-R(\nu)}}{2\nu},
\end{equation}
where
\begin{equation}\label{eq:discrim}
R(\nu)=\nu^3+(s_1^2/4-2p)\nu^2-(s_1q+4s_4)\nu+q^2.
\end{equation}
If there exists a root $\nu_1$ such that $R(\nu_1)=0$, then the polynomials \eqref{eq:L-det} $\det(\mathbf{L}(x,t;\lambda))$ can be decomposed into
\begin{equation}\label{eq:det-rewritten}
[\lambda^2+(\alpha_1+{\rm i}\sqrt{\nu_1})\lambda+(\alpha_2-\nu_1/2)-{\rm i}\sqrt{\nu_1}\mu][\lambda^2+(\alpha_1-{\rm i}\sqrt{\nu_1})\lambda+(\alpha_2-\nu_1/2)+{\rm i}\sqrt{\nu_1}\mu].
\end{equation}
Based on the relation between the roots and coefficients in equations \eqref{eq:L-det} and \eqref{eq:det-rewritten}, we know that
\[
\alpha_1+{\rm i}\sqrt{\nu_1}=-\lambda_1-\lambda_3,\,\,\,\,\, \alpha_1-{\rm i}\sqrt{\nu_1}=-\lambda_2-\lambda_4.
\]
Furthermore, we have the relation
\begin{equation}\label{eq:rela-nu-lambda}
\nu_1=-\frac{1}{4}(\lambda_1+\lambda_3-\lambda_2-\lambda_4)^2.
\end{equation}
In a similar procedure, for the other two roots $\nu_2$ and $\nu_3$, we have
\begin{equation}\label{eq:rela-nu-lambda-1}
\nu_2=-\frac{1}{4}(\lambda_1+\lambda_4-\lambda_2-\lambda_3)^2,\,\,\,\,\,\,
\nu_3=-\frac{1}{4}(\lambda_1+\lambda_2-\lambda_3-\lambda_4)^2.
\end{equation}
Here, we take advantage of the effective integration technique in \cite{Kamchatnov97}. In what follows, we will see that the parameters $\nu_1$, $\nu_2$ and $\nu_3$ will determine the dynamics of solution $\psi$. Because of the above relationship between $\nu_i$ $(i=1,2,3)$ and $\lambda_j$ $(j=1,2,3,4)$, the parameters $\lambda_j$ also determine the dynamics of solution.

Next we derive the Dubrovin type equation for the root $\mu(x,t)$. Since $g_x=-2(\psi f+{\rm i}\lambda g)$ and $g={\rm i}\psi (\lambda-\mu)$ [equations \eqref{eq:station-lax-nlse-x} and \eqref{eq:f-g-h}], by taking $\lambda=\mu$, it follows that
\begin{equation}\label{eq:dubrovin-x}
\mu_x=-2{\rm i}f(\mu)=-2{\rm i} \sqrt{P(\mu)},
\end{equation}
and
\begin{equation}\label{eq:dubrovin-t}
\mu_t=-2{\rm i}\left(\mu+\frac{\rm i}{2}(\ln \psi)_x\right) \sqrt{P(\mu)}=-{\rm i} s_1\sqrt{P(\mu)}.
\end{equation}
Furthermore, from equations \eqref{eq:Qt-Qx} and \eqref{eq:mu-mu*} and the relation $\mu-\mu^*=-\frac{{\rm i}}{2}\frac{\nu_x}{\nu}$, we know that
\begin{equation}\label{eq:nv-equation}
\nu_t=\frac{s_1}{2}\nu_x=s_1\sqrt{-R(\nu)}.
\end{equation}
Introduce the notations $\lambda_{1,3}=\alpha\pm{\rm i} \gamma$, $\lambda_{2,4}=\beta\pm {\rm i}\delta$, then it follows from formulas \eqref{eq:rela-nu-lambda} and \eqref{eq:rela-nu-lambda-1} that
\begin{equation}\label{eq:nv-3root}
\nu_1=-(\alpha-\beta)^2,\,\,\, \nu_2=(\gamma-\delta)^2,\,\,\, \nu_3=(\gamma+\delta)^2.
\end{equation}
We can see that $\nu_1\leq0\leq\nu_2<\nu_3$. By simple mathematical analysis on the differential equation \eqref{eq:nv-equation}, we find that $\nu$ is a periodic function on the interval $[\nu_2,\nu_3]$, which is indeed written by an elliptic function:
\begin{equation}\label{eq:elliptic-function}
\nu(x+\frac{s_1}{2}t)=\nu_3+(\nu_2-\nu_3){\rm sn}^2(\sqrt{\nu_3-\nu_1}(x+\frac{s_1}{2}t)|m)
\end{equation}
where $m=(\nu_3-\nu_2)/(\nu_3-\nu_1).$ Moreover, from equations \eqref{eq:Qt-Qx}, \eqref{eq:mu-mu*}, and \eqref{eq:nv-equation}, we know that
\begin{equation}\label{psi-x-t}
\psi_x =2{\rm i}(\mu-s_1/2) \psi=-{\rm i}\left(\frac{s_1}{2}+\frac{2q+{\rm i}\nu_x}{2\nu}\right)\psi,\,\,\,\, \psi_t={\rm i}(s_2-s_1^2/4)\psi+\frac{s_1}{2} \psi_x.
\end{equation}
When $\nu_2>0$, the function $\psi(x,t)$ can integrate exactly
\begin{equation}\label{psi-sol}
\psi=\sqrt{\nu(x+\frac{s_1}{2}t)}\exp\left[-\frac{{\rm i}}{2}s_1(x+s_1t)-{\rm i}q\int_{0}^{x+\frac{s_1}{2}t}\frac{ {\rm d}s}{\nu(s)}+{\rm i} s_2 t\right].
\end{equation}
When $\nu_2=0$, since there is a singularity located at $\sqrt{\nu_3-\nu_1}(x+s_1t)=0\mod 2K$ in equations \eqref{psi-x-t}, ($K\equiv K(m)$ is the complete elliptic integral of the first kind), the function $\psi(x,t)$ can be represented as
\begin{equation}\label{psi-sol-c}
\psi=(\gamma+\delta){\rm cn}\left((\alpha-\beta)(x+\frac{s_1}{2}t)|m\right)\exp\left[-\frac{{\rm i}}{2}s_1(x+s_1t)+{\rm i} s_2 t\right].
\end{equation}

\begin{remark}
We consider the Galilean transformation to conceal the parameter $s_1$, using the symmetry of the NLSE $\psi(x,t)\rightarrow \psi(x-\frac{s_1}{2}t,t){\rm e}^{{\rm i}\frac{\theta}{2}},$ $\theta=s_1(x-\frac{s_1}{4}t)$. We also apply a similar transformation on the Lax pair to achieve the symmetry $(\lambda,\Phi)\rightarrow (\widehat{\lambda},\widehat{\Phi})$, where $\widehat{\lambda}=\lambda-\frac{s_1}{4}$, $\widehat{\Phi}={\rm e}^{\frac{\rm i}{2}\theta \sigma_3} \Phi(x-\frac{s_1}{2}t,t)$. Thus we have the correpsponce between the spectral curve for $\lambda$ and $\widehat{\lambda}$, where the new spectral curve is given by  $\widehat{P}(\widehat{\lambda}):=\widehat{\lambda}^4-\hat{s}_1\widehat{\lambda}^3+\hat{s}_2\widehat{\lambda}^2-\hat{s}_3\widehat{\lambda}+\hat{s}_4$ with $\hat{s}_1=0$, $\hat{s}_2=s_2-\frac{3}{8}s_1^2$, $\hat{s}_3=
\frac{s_1^3}{8} -\frac{s_1 s_2}{2} + s_3$, and $\hat{s}_4=-\frac{3 s_1^4}{256} + \frac{s_1^2 s_2}{16}-\frac{s_1 s_3}{4} + s_4$. The variables $\nu_{1\leq i \leq 3}$ are invariant under the transformation. So is the parameter $q=\sqrt{-\nu_1\nu_2\nu_3}$. Thus, under the Galilean transformation, the solution \eqref{psi-sol} can convert into a stationary solution which is better to analyze the dynamics without loss the generality:
\begin{equation}\label{psi-sol-station}
\psi=\sqrt{\nu(x)}\exp\left[-{\rm i}q\int_{0}^{x}\frac{ {\rm d}s}{\nu(s)}+{\rm i} \hat{s}_2 t\right]= (\gamma+\delta){\rm cn}\left((\alpha-\beta)x|m\right)\exp\left[{\rm i} \hat{s}_2 t\right].
\end{equation}
So in the following, we set the parameter $s_1=0$ based on the above analysis.
\end{remark}

\begin{remark}\label{rem:2}
We give the Weierstrass elliptic function representation. Set $\nu(x)=4(\frac{s_2}{6}-\wp(2x+\omega_3))$, then $\wp(2x+\omega_3)$ satisfies $[\frac{{\rm d}\wp(2x+\omega_3)}{{\rm d}x}]^2=16\times \left[4\wp(2x+\omega_3)-g_2\wp(2x+\omega_3)-g_3\right]=64(\wp(2x+\omega_3)-e_1)(\wp(2x+\omega_3)-e_2)(\wp(2x+\omega_3)-e_3)$, where $g_2=\frac{s_2^2}{12}+s_4$ and $g_3=\frac{s_2^3}{216}+\frac{s_3^2}{16}-\frac{s_2s_4}{6}$, $e_1=\frac{1}{12}(\nu_2+\nu_3-2\nu_1)$, $e_2=\frac{1}{12}(\nu_1+\nu_3-2\nu_2)$, $e_3=\frac{1}{12}(\nu_1+\nu_2-2\nu_3)$. And $\wp(\omega_i)=e_i$, $i=1,2,3,$ $\omega_1$ and $\omega_3$ are the half-period of the $\wp$ function. This is consistent with the Jacobi elliptic function representation since it takes the maximum at $x=0$.
\end{remark}

We now proceed to find the solutions of the corresponding Lax pair. Let us introduce the notation $y^2=\det(\mathbf{L}(x,t;\lambda))$, i.e. $\det({\rm i}y\mathbb{I}-\mathbf{L}(x,t;\lambda))=0$. So the kernel of matrix ${\rm i}y\mathbb{I}-\mathbf{L}(x,t;\lambda)$ are $K_1(x,t;\lambda)=(1,r_1(x,t;\lambda))^{\rm T}$ and $K_2(x,t;\lambda)=(1,r_2(x,t;\lambda))^{\rm T},$ where
$$r_1(x,t;\lambda)={\rm i}\frac{f(x,t;\lambda)+y}{g(x,t;\lambda)}={\rm i}\frac{h(x,t;\lambda)}{f(x,t;\lambda)-y} ,\,\,\,\, r_2(x,t;\lambda)={\rm i}\frac{f(x,t;\lambda)-y}{g(x,t;\lambda)}={\rm i}\frac{h(x,t;\lambda)}{f(x,t;\lambda)+y}.$$
Suppose there exists a fundamental matrix solution for the Lax pair \eqref{eq:lax-nlse}: $\widehat{\Phi}(x,t;\lambda)$ with the initial data $\widehat{\Phi}(0,0;\lambda)=\mathbb{I}$, and the solution matrix $\widehat{\Phi}(x,t;\lambda)\mathbf{L}(0,0;\lambda)$ is also a matrix solution for the Lax pair \eqref{eq:lax-nlse}.  On the other hand, we know that $\mathbf{L}(x,t;\lambda) \widehat{\Phi}(x,t;\lambda)$ is also a matrix solution for the Lax pair \eqref{eq:lax-nlse}. By the uniqueness and existence of ordinary differential equation, we arrive at $\widehat{\Phi}(x,t;\lambda)\mathbf{L}(0,0;\lambda)=\mathbf{L}(x,t;\lambda) \widehat{\Phi}(x,t;\lambda).$ Now we again consider the solution $\Phi(x,t;\lambda)\equiv(\Psi_1(x,t;\lambda),\Psi_2(x,t;\lambda))$. Since the Lax pair is linear, we can normalized the first component of the vector solutions $\Psi_1(0,0;\lambda),$ $\Psi_2(0,0;\lambda)$ to be unity, and set their second components as $\psi_1(0,0;\lambda)$ and $\psi_2(0,0;\lambda)$, respectively. In other words, \[\Phi(x,t;\lambda)=\widehat{\Phi}(x,t;\lambda)(K_1(0,0;\lambda),K_2(0,0;\lambda)).\]  Based on this, we have $[{\rm i}y\mathbb{I}-\mathbf{L}(x,t;\lambda)]\Phi(x,t;\lambda)=0,$ which yields $r_i(x,t;\lambda)=\frac{\varphi_i(x,t;\lambda)}{\phi_i(x,t;\lambda)}$, $i=1,2.$

Since $f(x,t;\lambda)$ and $g(x,t;\lambda)$ are known, then the functions $\psi_i(x,t;\lambda)$, $i=1,2$
can be exactly obtained. The left unknown variables for Lax pair are $\phi_i(x,t;\lambda)$, $i=1,2.$ Indeed, based on the previous known variables, $\phi_i(x,t;\lambda)$, $i=1,2$ can be derived from the integration:
\begin{equation}\label{eq:integration}
\phi_{1,x}=(-{\rm i}\lambda+{\rm i}\psi r_1) \phi_1,\,\,\,\,\, \phi_{1,t}=\left[-{\rm i}\lambda^2+\frac{\rm i}{2}\nu+({\rm i}\lambda \psi-\frac{1}{2}\psi_x) r_1\right] \phi_1.
\end{equation}
Inserting equation \eqref{eq:f-g-h} into above equation \eqref{eq:integration}, we have
\begin{equation}\label{eq:phi1-1}
\phi_{1}(x,t;\lambda)=\sqrt{\frac{\nu(x)-\beta_1}{\nu(0)-\beta_1}} \exp(\theta_1),\,\,\, \theta_1={\rm i}\int_{0}^{x}\frac{C_1{\rm d}s}{\nu(s)-\beta_1}+{\rm i}\lambda x+{\rm i}(\frac{s_2}{2}+y)t,
\end{equation}
where $\beta_1=2(\lambda^2+\frac{s_2}{2}-y)$, $C_1=2\lambda \beta_1-s_3.$ By using a similar integration, we obtain
\begin{equation}\label{eq:phi2}
\phi_{2}(x,t;\lambda)=\sqrt{\frac{\nu(x)-\beta_2}{\nu(0)-\beta_2}} \exp(\theta_2 ),\,\,\,\,\, \theta_2={\rm i}\int_{0}^{x}\frac{C_2{\rm d}s}{\nu(s)-\beta_2}+{\rm i}\lambda x+{\rm i}(\frac{s_2}{2}-y)t,
\end{equation}
where $\beta_2=2(\lambda^2+\frac{s_2}{2}+y)$ and $C_2=2\lambda \beta_2-s_3.$ We also have  $\varphi_i=\phi_ir_i,$ $i=1,2.$ Thus, the solution for the Lax pair with the elliptic potential function are determined uniquely.

\begin{remark}
Though the expressions \eqref{eq:phi1-1} and \eqref{eq:phi2} seem to have singularities on the points $\nu(x)-\beta_i=0$, $i=1,2$, they in fact have no singularity. Here we check it for, e.g., $\phi_1$.
  Let $\wp(\omega_{\pm})=-\lambda^2/2-s_2/12\pm \sqrt{P}/2$, then
$C_1=\mp4{\rm i} \wp'(\omega_+)$ and $C_2=\mp4{\rm i}\wp'(\omega_-)$ since $[\wp'(\omega_{\pm})]^2=4\wp(\omega_{\pm})-g_2\wp(\omega_{\pm})-g_3.$ Now we can expand at the neighborhood of $(\omega_+-\omega_3)/2$: $\nu(x)-\beta_1=4\left[\wp(\omega_+)-\wp(2x+\omega_3)\right]=\mp8\wp'(\omega_+)(x-(\omega_+-\omega_3)/2)+O((x-(\omega_+-\omega_3)/2)^2)$. It follows that $\exp(\theta_1)=O([(x-(\omega_+-\omega_3)/2]^{\mp1/2})$. This proves that $\phi_1$ has no singularity.
\end{remark}

To look for a fundamental matrix solution with better symmetry, we give the integral representation for $\varphi_i$, $i=1,2$: \begin{equation}\label{eq:varphi12}
\begin{split}
\varphi_{1}(x,t;\lambda)=D_1\sqrt{\frac{\nu(x)-\beta_2}{\nu(0)-\beta_2}} \exp(-\theta_2),\\
\varphi_{2}(x,t;\lambda)=D_2\sqrt{\frac{\nu(x)-\beta_1}{\nu(0)-\beta_1}} \exp(-\theta_1),
\end{split}
\end{equation}
where $$D_1={\rm i}r_1(0,0;\lambda)={\rm i}\sqrt{\frac{f(0,0;\lambda)+y}{f(0,0;\lambda)-y}\frac{h(0,0;\lambda)}{g(0,0;\lambda)}}={\rm i}\sqrt{\frac{\nu(0)-\beta_2}{\nu(0)-\beta_1}}$$
and $D_2={\rm i}\sqrt{\frac{\nu(0)-\beta_1}{\nu(0)-\beta_2}}.$ By ignoring the constant factors of vector solutions, we obtain the following theorem:
\begin{theorem}\label{thm1}
One fundamental solution for Lax pair \eqref{eq:lax-nlse} with the elliptic functions \eqref{psi-sol} or \eqref{psi-sol-c} can be represented as
\begin{equation}\label{eq:fund-sol}
\Phi(x,t;\lambda)=\begin{bmatrix}
\sqrt{\nu(x)-\beta_1}{\rm e}^{\theta_1}&\sqrt{\nu(x)-\beta_2}{\rm e}^{\theta_2}\\[5pt]
{\rm i}\sqrt{\nu(x)-\beta_2}{\rm e}^{-\theta_2} & {\rm i}\sqrt{\nu(x)-\beta_1}{\rm e}^{-\theta_1}\\
\end{bmatrix},
\end{equation}
where $\theta_1$, $\theta_2$ are given in equations \eqref{eq:phi1-1}, \eqref{eq:phi2}.
\end{theorem}
The above formulas involve the incomplete elliptic integral of the third kind.  A direct way is using the integral representation. In what follows, we use the theta function to represent the integration.

The fundamental solutions \eqref{eq:fund-sol} can be normalized at $(x,t)=(0,0)$:
\begin{equation}\label{eq:fund-sol-norm}
\Phi^{\rm N}(x,t;\lambda)=\Phi(x,t;\lambda)\Phi^{-1}(0,0;\lambda)
\end{equation}
with $\Phi^{\rm N}(0,0;\lambda)=\mathbb{I},$ which can be proved by the fact that it is an entire function of $\lambda$ by the Proposition 2.1 in \cite{BilmanM17}.

\subsection{Theta function representation for solutions}

Let $z_0=K+2{\rm i}l$ such that $l\in [-K'/2,K'/2]$ and $\alpha=\sqrt{\nu_3-\nu_1}$, $m=\frac{\nu_2-\nu_1}{\nu_3-\nu_1}$, and we introduce the parameterization:
\begin{equation}\label{eq:para-nu123}
\nu_1=-\alpha^2 {\rm dn}^2(K+2{\rm i}l|m),\,\,\,\, \nu_2=-m\alpha^2  {\rm cn}^2(K+2{\rm i}l|m),\,\,\,\,\, \nu_3=m\alpha^2 {\rm sn}^2(K+2{\rm i}l|m).
\end{equation}
Thus $q={\rm i}m\alpha^3 {\rm sn}(K+2{\rm i}l|m){\rm cn}(K+2{\rm i}l|m){\rm dn}(K+2{\rm i}l|m)$, $\nu(x)=m\alpha^2({\rm sn}^2(K+2{\rm i}l|m)-{\rm sn}^2(\alpha x|m))$. Hereafter, for simplicity we omit the modulus parameter $m$.  Rewriting the formula \eqref{eq:appBPi},
\begin{equation}\label{eq:third-type}
\int_0^{x} \frac{{\rm sn}(v){\rm cn}(v){\rm dn}(v)}{{\rm sn}^2(u)-{\rm sn}^2(v)} {\rm d}u=\frac{1}{2} \ln \frac{\vartheta_1(\frac{v-x}{2K})}{\vartheta_1(\frac{v+x}{2K})}+xZ(v),
\end{equation}
and using it, we obtain that
\begin{equation}\label{eq:phase}
\int_0^{x} \frac{-{\rm i}q}{\nu(u)} {\rm d}u=-\int_0^{\alpha x} \frac{{\rm sn}(K+2{\rm i}l){\rm cn}(K+2{\rm i}l){\rm dn}(K+2{\rm i}l)}{{\rm sn}^2(u)-{\rm sn}^2(K+2{\rm i}l)} {\rm d}u=-\frac{1}{2} \ln \frac{\vartheta_1(\frac{K+2{\rm i}l-\alpha x}{2K})}{\vartheta_1(\frac{K+2{\rm i}l+\alpha x}{2K})}-\alpha Z(K+2{\rm i}l) x.
\end{equation}
On the other hand, on account of addition formula for theta function (see  Appendix~\ref{sec:appB} or \cite{KharchevZ15}), we obtain
\begin{equation}\label{eq:add-nu}
\nu(x)=\alpha^2 \frac{\vartheta_2^2}{\vartheta_3^2}\left(\frac{\vartheta_1^2(\frac{K+2{\rm i}l}{2K})}{\vartheta_4^2(\frac{K+2{\rm i}l}{2K})}-\frac{\vartheta_1^2(\frac{\alpha x}{2K})}{\vartheta_4^2(\frac{\alpha x}{2K})}\right)=\alpha^2 \frac{\vartheta_2^2}{\vartheta_3^2} \frac{\vartheta_4^2\vartheta_1(\frac{K+2{\rm i}l+\alpha x}{2K})\vartheta_1(\frac{K+2{\rm i}l-\alpha x}{2K})}{\vartheta_4^2(\frac{K+2{\rm i}l}{2K})\vartheta_4^2(\frac{\alpha x}{2K})}.
\end{equation}
Then we obtain the theta function representation:
\begin{equation}\label{eq:psi-theta}
\psi=\alpha \frac{\vartheta_2\vartheta_4}{\vartheta_3} \frac{\vartheta_2(\frac{\alpha x+2{\rm i}l}{2K})}{\vartheta_3(\frac{2{\rm i}l}{2K})\vartheta_4(\frac{\alpha x}{2K})}{\rm e}^{-\alpha Z(K+2{\rm i}l)x +{\rm i}s_2 t}.
\end{equation}
From the formula $Z(u)=Z(u+2K)$ and $Z(-u)=-Z(u)$, we know that $Z(K+2{\rm i}l)\in {\rm i}\mathbb{R}.$

Furthermore, we derive the theta-function representation for the solutions of the Lax pair. Firstly, we need to parameterize the spectral curve $y^2=\prod_{i=1}^4(\lambda-\lambda_i)$ with the aid of equation \eqref{eq:dubrovin-x}: $\lambda(z)=\mu({\rm i}(z-l)/\alpha)$ and $y(z)=\frac{\alpha}{2}\frac{{\rm d}\mu({\rm i}(z-l)/\alpha)}{{\rm d} z}$, where
\begin{equation*}
\begin{split}
\lambda_{1,3}= &\frac{1}{2}\sqrt{-\nu_1}\pm \frac{{\rm i}}{2}(\sqrt{\nu_2}+\sqrt{\nu_3})=\frac{\alpha}{2}k'\,{\rm cd}(l|m')\pm \frac{{\rm i}\alpha }{2}k(k'\,{\rm sd}(l|m')+{\rm nd}(l|m')), \\
\lambda_{2,4}= &-\frac{1}{2}\sqrt{-\nu_1}\pm \frac{{\rm i}}{2}(\sqrt{\nu_2}-\sqrt{\nu_3})=-\frac{\alpha}{2}k'\,{\rm cd}(l|m')\pm \frac{{\rm i}\alpha }{2}k(k'\,{\rm sd}(l|m')-{\rm nd}(l|m')),  \\
\end{split}
\end{equation*}
and $k=\sqrt{m},$  $k'=\sqrt{1-m}.$ So $s_2=\lambda_{1I}^2+\lambda_{2I}^2-2\lambda_{R}^2=\frac{1}{2}(\nu_1+\nu_2+\nu_3).$
 We use the addition formula for the Jacobi elliptic function to reduce $\mu({\rm i}(z-l)/\alpha)$ with a better representation:
\begin{equation}\label{eq:mu(iz)}
\begin{split}
\lambda(z)=&\frac{\alpha {\rm i}}{2}\left(\frac{-{\rm sn}(K+2{\rm i}l){\rm cn}(K+2{\rm i}l){\rm dn}(K+2{\rm i}l)+{\rm sn}({\rm i}(z-l)){\rm cn}({\rm i}(z-l)){\rm dn}({\rm i}(z-l))}{{\rm sn}^2(K+2{\rm i}l)-{\rm sn}^2({\rm i}(z-l))}\right) \\
=&\frac{\alpha {\rm i}}{2}\left[\frac{{\rm dn}({\rm i}(z-l))[{\rm sn}({\rm i}(z-l)){\rm cn}({\rm i}(z-l)){\rm dn}(K+2{\rm i}l)-{\rm sn}(K+2{\rm i}l){\rm cn}(K+2{\rm i}l){\rm dn}({\rm i}(z-l))] }{({\rm sn}^2(K+2{\rm i}l)-{\rm sn}^2({\rm i}(z-l))){\rm dn}(K+2{\rm i}l)}\right.  \\
&+\left.\frac{-{\rm sn}(K+2{\rm i}l){\rm cn}(K+2{\rm i}l)[{\rm dn}^2(K+2{\rm i}l)-{\rm dn}^2({\rm i}(z-l)))] }{({\rm sn}^2(K+2{\rm i}l)-{\rm sn}^2({\rm i}(z-l))){\rm dn}(K+2{\rm i}l)} \right] \\
=&\frac{\alpha {\rm i}}{2}\left[\frac{{\rm dn}({\rm i}(z-l)) {\rm cs}(-K-{\rm i}(z+l))+m\, {\rm sn}(K+2{\rm i}l){\rm cn}(K+2{\rm i}l)}{{\rm dn}(K+2{\rm i}l)}  \right] \\
=&\frac{\alpha }{2}\left[\frac{-{\rm dn}({\rm i}(z-l)) {\rm dn}({\rm i}K'+K-{\rm i}(z+l))+{\rm i}m\, {\rm sn}(K+2{\rm i}l){\rm cn}(K+2{\rm i}l)}{{\rm dn}(K+2{\rm i}l)}  \right],
\end{split}
\end{equation}
where we used the addition formula for Jacobi elliptic function:
\[
{\rm cs}(u+v)=\frac{{\rm sn}\, u\, {\rm cn}\, u\, {\rm dn}\,v-{\rm sn}\, v\, {\rm cn}\, v \,{\rm dn}\,u}{{\rm sn}^2\,u-{\rm sn}^2\, v}.
\]
If $l=\pm K'/2$, the above formula can be replaced by $\lambda(z)=\frac{\alpha \ii}{2}k^2{\rm sn}({\rm i}(z-l))\, {\rm cn}({\rm i}(z-l))/{\rm dn}({\rm i}(z-l)).$
For the parameter $y(z)$, we can rewrite it as
\begin{equation}\label{eq:dmu(iz)}
\begin{split}
y(z)=&\frac{\alpha^2 m{\rm i}}{4}\frac{{\rm sn}({\rm i}(z-l)){\rm cn}({\rm i}(z-l)){\rm dn}(C-{\rm i}(z+l))-{\rm sn}(C-{\rm i}(z+l)){\rm cn}(C-{\rm i}(z+l)){\rm dn}({\rm i}(z-l))}{{\rm dn}(K+2{\rm i}l)} \\
=&\frac{\alpha^2 m{\rm i}}{4{\rm dn}(K+2{\rm i}l)} \left[{\rm sn}^2({\rm i}(z-l))-{\rm sn}^2(C-{\rm i}(z+l))\right]{\rm cs}(C-2{\rm i}l)\\
=&\frac{\alpha^2}{4}\left({\rm dn}^2(C-{\rm i}(z+l))-{\rm dn}^2({\rm i}(z-l))\right),
\end{split}
\end{equation}
where $C=K+\ii K'.$
Moreover, by addition formula of the Jacobi elliptic function, we obtain a simple formula of $\lambda^2(z)$:
\begin{equation}\label{eq:lambda2}
\lambda^2(z)=\frac{\alpha^2}{4}\left({\rm dn}^2(K+2\ii l)+m-2+{\rm dn}^2(\ii (z-l))+{\rm dn}^2(\ii K'+K-\ii (z+l))\right).
\end{equation}
Furthermore, combining the above result, we have
\begin{equation}\label{eq:beta1}
\begin{split}
\beta_1=&2(\lambda^2(z)+s_2/2-y(z))=m\alpha^2\left({\rm sn}^2(K+2\ii l)-{\rm sn}^2(\ii (z-l))\right),\\     \beta_2=&2(\lambda^2(z)+s_2/2+y(z))=m\alpha^2\left({\rm sn}^2(K+2\ii l)-{\rm sn}^2(\ii K'+K-\ii (z+l))\right).
\end{split}
\end{equation}
It follows that
\begin{equation}\label{eq:nubeta1}
\begin{split}
\nu(x)-\beta_1=&m\alpha^2\left({\rm sn}^2(\ii (z-l))-{\rm sn}^2(\alpha x)\right),     \\
\nu(x)-\beta_2=&m\alpha^2\left({\rm sn}^2(\ii K'+K-\ii (z+l))-{\rm sn}^2(\alpha x)\right).
\end{split}
\end{equation}
Moreover, we have
\begin{equation}\label{eq:C1}
\begin{split}
C_1=&2\lambda \beta_1-s_3=\alpha \frac{\rm d}{{\rm d}z}(y(z)-\lambda(z)^2) \\
=&\frac{\alpha^3}{4}\frac{\rm d}{{\rm d}z}\left(-{\rm dn}^2(-K+2\ii l)-m+2-2\,{\rm dn}^2(\ii (z-l))\right)    \\
=&\ii \alpha^3m\, {\rm sn}(\ii (z-l)) {\rm cn}(\ii (z-l)){\rm dn}(\ii (z-l))
\end{split}
\end{equation}
and
\begin{equation}\label{eq:C2}
C_2=\ii \alpha^3m\, {\rm sn}(\ii K'+K-\ii (z+l)) {\rm cn}(\ii K'+K-\ii (z+l)){\rm dn}(\ii K'+K-\ii (z+l)).
\end{equation}
On account of the above equations, through the similar calculation as \eqref{eq:phase} and \eqref{eq:add-nu}, we obtain
\begin{equation}\label{eq:phi1}
\begin{split}
\phi_1&=\sqrt{\nu(x)-\beta_1}\,\exp(\theta_1)=\alpha \frac{\vartheta_2\vartheta_4}{\vartheta_3}\frac{\vartheta_1(\frac{\ii (z-l)-\alpha x}{2K})}{
\vartheta_4(\frac{\ii (z-l)}{2K})\vartheta_4(\frac{\alpha x}{2K})}{\rm e}^{(\alpha Z(\ii (z-l))+\ii \lambda)x+\ii (\frac{s_2}{2}+y)t}, \\
\phi_2&=\sqrt{\nu(x)-\beta_2}\,\exp(\theta_2)=\alpha \frac{\vartheta_2\vartheta_4}{\vartheta_3}\frac{\vartheta_1(\frac{\ii K'-K-\ii (z+l)-\alpha x}{2K})}{
\vartheta_4(\frac{\ii K'-K-\ii (z+l)}{2K})\vartheta_4(\frac{\alpha x}{2K})}{\rm e}^{(\alpha Z(\ii K'-K-\ii (z+l))+\ii \lambda)x+\ii (\frac{s_2}{2}-y)t}.
\end{split}
\end{equation}
By the addition formula for theta functions (see Appendix~\ref{sec:appB}),
we obtain
\begin{equation}\label{eq:lambda-mu}
\begin{split}
\lambda(z)-\mu(x)&=\frac{\alpha \ii }{2\,{\rm nd}(2\ii l)}\left(\frac{\vartheta_1(\frac{\ii (z+l)}{2K})\vartheta_3(\frac{\ii (z-l)}{2K})}{\vartheta_2(\frac{\ii (z+l)}{2K})\vartheta_4(\frac{\ii (z-l)}{2K})}-\frac{\vartheta_1(\frac{\alpha\, x+2\ii \,l}{2K})\vartheta_3(\frac{\alpha\, x}{2K})}{\vartheta_2(\frac{\alpha\, x+2\ii l}{2K})\vartheta_4(\frac{\alpha\, x}{2K})}\right)   \\
&=\frac{\alpha \ii }{2\,{\rm nd}(2\ii l)}\frac{\vartheta_1(\frac{\ii (z-l)-\alpha\, x}{2K})\vartheta_2\vartheta_3(\frac{\ii (z+l)+\alpha\, x}{2K})\vartheta_4(\frac{2\ii l}{2K})}{\vartheta_2(\frac{\ii (z+l)}{2K})\vartheta_4(\frac{\ii (z-l)}{2K})\vartheta_2(\frac{\alpha\, x+2\ii l}{2K})\vartheta_4(\frac{\alpha\, x}{2K})}.
\end{split}
\end{equation}
Similarly, we have
\begin{equation}\label{eq:lambda-mu1}
\begin{split}
\lambda(z)-\mu(x)^*&=\frac{\alpha \ii }{2\,{\rm nd}(2\ii l)}\left(\frac{\vartheta_1(\frac{\ii (z+l)}{2K})\vartheta_3(\frac{\ii (z-l)}{2K})}{\vartheta_2(\frac{\ii (z+l)}{2K})\vartheta_4(\frac{\ii (z-l)}{2K})}-\frac{\vartheta_1(\frac{-\alpha\, x+2\ii \,l}{2K})\vartheta_3(\frac{-\alpha\, x}{2K})}{\vartheta_2(\frac{-\alpha\, x+2\ii l}{2K})\vartheta_4(\frac{-\alpha\, x}{2K})}\right)   \\
&=\frac{\alpha \ii }{2\,{\rm nd}(2\ii l)}\frac{\vartheta_1(\frac{\ii (z-l)+\alpha\, x}{2K})\vartheta_2\vartheta_3(\frac{\ii (z+l)-\alpha\, x}{2K})\vartheta_4(\frac{2\ii l}{2K})}{\vartheta_2(\frac{\ii (z+l)}{2K})\vartheta_4(\frac{\ii (z-l)}{2K})\vartheta_2(\frac{-\alpha\, x+2\ii l}{2K})\vartheta_4(-\frac{\alpha\, x}{2K})}.
\end{split}
\end{equation}
Thus
\begin{equation}\label{eq:muratio}
\begin{split}
\frac{\lambda(z)-\mu(x)^*}{\lambda(z)-\mu(x)}&=
\frac{\vartheta_1(\frac{\ii (z-l)+\alpha\, x}{2K})\vartheta_3(\frac{\ii (z+l)-\alpha\, x}{2K})}{\vartheta_1(\frac{\ii (z-l)-\alpha\, x}{2K})\vartheta_3(\frac{\ii (z+l)+\alpha\, x}{2K})}\frac{\vartheta_2(\frac{\alpha\, x+2\ii l}{2K})}{\vartheta_2(\frac{-\alpha\, x+2\ii l}{2K})}.
\end{split}
\end{equation}
Meanwhile, we have
\begin{equation}\label{eq:expphi0}
\begin{split}
\exp(\ii \phi_0(x))&=\sqrt{\frac{\vartheta_2(\frac{2\ii l-\alpha\, x}{2K})}{\vartheta_2(\frac{\alpha\, x+2\ii l}{2K})}}{\rm e}^{\alpha Z(2\ii l+K)x},\,\,\,\,\, \phi_0(x)=q\int_{0}^{x}\frac{ {\rm d}s}{\nu(s)},
\end{split}
\end{equation}
and
\begin{equation}\label{eq:nuratio}
\begin{split}
\frac{\nu(x)-\beta_2}{\nu(x)-\beta_1}&=\frac{\vartheta_1(\frac{\ii K'-K-\ii (z+l)+\alpha\, x}{2K})\vartheta_1(\frac{\ii K'-K-\ii (z+l)-\alpha\, x}{2K})\vartheta_4^2(\frac{\ii (z-l)}{2K})}{\vartheta_1(\frac{\ii (z-l)+\alpha\, x}{2K})\vartheta_1(\frac{\ii (z-l)-\alpha\, x}{2K})\vartheta_4^2(\frac{\ii K'-K-\ii (z+l)}{2K})}\\
&=\frac{\vartheta_1(\frac{K+\ii (K'-z-l)+\alpha\, x}{2K})\vartheta_1(\frac{K+\ii (K'-z-l)-\alpha\, x}{2K})\vartheta_4^2(\frac{\ii (z-l)}{2K})}{\vartheta_1(\frac{\ii (z-l)+\alpha\, x}{2K})\vartheta_1(\frac{\ii (z-l)-\alpha\, x}{2K})\vartheta_4^2(\frac{K+\ii (K'-z-l)}{2K})}.
\end{split}
\end{equation}
Moreover, we have
\begin{equation}\label{eq:r1-1}
\begin{split}
r_1=&\frac{\varphi_1}{\phi_1}=\ii  \sqrt{\frac{\nu(x)-\beta_2}{\nu(x)-\beta_1}}\sqrt{\frac{\lambda-\mu^*}{\lambda-\mu}}{\rm e}^{\ii  \phi_0(x)-\ii s_2 t}\\
=&\ii \frac{\vartheta_4(\frac{\ii (z-l)}{2K})}{\vartheta_4(\frac{K+\ii K'-\ii (z+l)}{2K})}
\frac{\ee^{\alpha [Z(2\ii l+K)+\frac{\ii \pi}{2K}]x-\ii s_2t}}{\vartheta_1(\frac{\ii (z-l)-\alpha\, x}{2K})}\sqrt{\frac{\vartheta_1(\frac{K+\ii (K'-z-l)+\alpha\, x}{2K})\vartheta_1(\frac{K+\ii (K'-z-l)-\alpha\, x}{2K})\vartheta_3(\frac{\ii (z+l)-\alpha\,x}{2K})}{\vartheta_3(\frac{\ii (z+l)+\alpha\, x}{2K})}}\\
=&\ii \frac{\vartheta_4(\frac{\ii (z-l)}{2K})}{\vartheta_4(\frac{K+\ii K'-\ii (z+l)}{2K})}
\frac{\vartheta_1(\frac{K+\ii K'-\ii (z+l)+\alpha\, x}{2K})}{\vartheta_1(\frac{\ii (z-l)-\alpha\, x}{2K})}\ee^{\alpha Z(2\ii l+K)x-\ii s_2t},
\end{split}
\end{equation}
where we used the following shift formula in the second equality:
\begin{equation*}
\vartheta_3(\frac{\ii (z+l)\pm \alpha\, x}{2K})=-\ii {\rm e}^{-\pi \ii (\frac{\ii (z+l)\pm\alpha\, x-K-\ii K'}{2K}+\frac{\tau}{4})}\vartheta_1(\frac{K+\ii K'-\ii (z+l)\mp\alpha\, x}{2K}),\,\,\,\,\, \tau=K'/K.
\end{equation*}
In a similar procedure as above, we have
\begin{equation}\label{eq:r2-1}
r_2=\ii \frac{\vartheta_4(\frac{K+\ii K'-\ii (z+l)}{2K})}{\vartheta_4(\frac{\ii (z-l)}{2K})}
\frac{\vartheta_1(\frac{\ii (z-l)+\alpha\, x}{2K})}{\vartheta_1(\frac{K+\ii K'-\ii (z+l)-\alpha\, x}{2K})}{\rm e}^{\alpha [Z(2\ii l+K)+\frac{\ii \pi}{2K}]x-\ii s_2t}.
\end{equation}
Together with the equations \eqref{eq:phi1}, \eqref{eq:r1-1} and \eqref{eq:r2-1}, we conclude that
\begin{theorem}\label{thm2}
The vector solutions can be represented with the theta function form:
\begin{equation}\label{eq:vector-1}
\begin{bmatrix}
\phi_1(x,t;z) \\[8pt]
\varphi_1(x,t;z) \\
\end{bmatrix}=\alpha \frac{\vartheta_2\vartheta_4}{\vartheta_3 \vartheta_4(\frac{\alpha\,x}{2K})}{\rm e}^{[\alpha Z(\ii (z-l))+\ii \lambda]x+\ii (\frac{s_2}{2}+y)t}\begin{bmatrix}
\frac{\vartheta_1(\frac{-\alpha\, x+\ii (z-l)}{2K})}{\vartheta_4(\frac{\ii (z-l)}{2K})} \\[8pt]
\ii \frac{\vartheta_1(\frac{K+\ii K'-\ii (z+l)+\alpha\,x}{2K})}{\vartheta_4(\frac{K+\ii K'-\ii (z+l)}{2K})}{\rm e}^{\alpha [Z(2\ii l+K)+\frac{\ii \pi}{2K}]x-\ii s_2t} \\
\end{bmatrix}
\end{equation}
and
\begin{equation}\label{eq:vector-2}
\begin{bmatrix}
\phi_2(x,t;z) \\[8pt]
\varphi_2(x,t;z) \\
\end{bmatrix}=\alpha \frac{\vartheta_2\vartheta_4}{\vartheta_3 \vartheta_4(\frac{\alpha\,x}{2K})}{\rm e}^{[\alpha Z(K+\ii K'-\ii (z+l))+\ii \lambda]x+\ii (\frac{s_2}{2}-y)t}\begin{bmatrix}
 \frac{\vartheta_1(\frac{-\alpha\, x+K+\ii K'-\ii (z+l)}{2K})}{\vartheta_4(\frac{K+\ii K'-\ii (z+l)}{2K})} \\[8pt]
\ii\frac{\vartheta_1(\frac{\ii (z-l)+\alpha\,x}{2K})}{\vartheta_4(\frac{\ii (z-l)}{2K})}{\rm e}^{\alpha [Z(2\ii l+K)+\frac{\ii \pi}{2K}]x-\ii s_2t} \\
\end{bmatrix}.
\end{equation}
\end{theorem}
In what follows, we would like use the different versions of vector solutions to construct the breathers and rogue waves on the elliptic function background by the Darboux transformation. For the breather solution, we use the theta function, which is helpful to give asymptotic analysis on that. For the rogue wave and high order rogue waves, it is convenient to construct the exact solutions by the Jacobi cn, sn and dn function to represent them.

\section{Multi-breather solutions}\label{sec4}
In this section, we consider the multi-breather solution by the multi-fold Darboux matrix (see  Appendix~\ref{sec:appA}). This step is just by inserting the solution into the formula of B\"acklund transformation. However, if we would like to analyze the dynamics of the solution, the expressions should be simplified into a proper form.

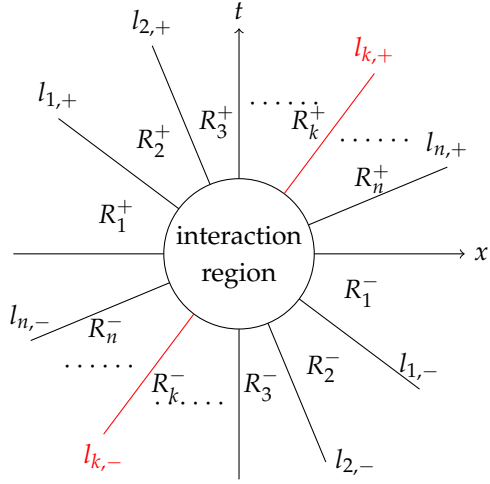
\begin{figure}
\begin{tikzpicture}
\draw[] (0,-3)--(0,-1);
\draw[->] (0,1)--(0,3) node[above]{$t$};
\draw[] (-3,0)--(-1,0);
\draw[->] (1,0)--(3,0) node[right]{$x$};
\draw[](0,0) circle (1);
\node[above] at (0,0) {interaction };

\node[below] at (0,0) {region};

\draw[red] (3/5,4/5)--(9/5,12/5) node[above]{$l_{k,+}$};
\draw[] (-4/5,3/5)--(-12/5,9/5) node[above]{$l_{1,+}$};
\draw[] (12/13,5/13)--(3*12/13,3*5/13)node[above]{$l_{n,+}$};
\draw[] (-5/13,12/13)--(-3*5/13,3*12/13)node[above]{$l_{2,+}$};

\node[right] at (0,2) {$\cdots\cdots$};
\node[left] at (0,-2) {$\cdots\cdots$};

\node[right] at (1.2,3/2) {$\cdots\cdots$};
\node[left] at (-1.2,-3/2) {$\cdots\cdots$};

\node[right] at (-2,1/2) {$R_1^+$};
\node[right] at (-3/2,3/2) {$R_2^+$};
\node[right] at (-2/3,7/4) {$R_3^+$};
\node[left] at (1.3,7/4) {$R_k^+$};
\node[right] at (1.4,1) {$R_n^+$};

\node[left] at (2,-1/2) {$R_1^-$};
\node[left] at (3/2,-3/2) {$R_2^-$};
\node[left] at (2/3,-7/4) {$R_3^-$};
\node[right] at (-1.3,-7/4) {$R_k^-$};
\node[left] at (-1.4,-1) {$R_n^-$};

\draw[red] (-3/5,-4/5)--(-9/5,-12/5)node[below]{$l_{k,-}$};
\draw[] (4/5,-3/5)--(12/5,-9/5)node[above]{$l_{1,-}$};
\draw[] (-12/13,-5/13)--(-3*12/13,-3*5/13)node[above]{$l_{n,-}$};
\draw[] (5/13,-12/13)--(3*5/13,-3*12/13)node[right]{$l_{2,-}$};

\coordinate (A) at (2,2);
\end{tikzpicture}
\caption{\label{fig:interaction}The sketch map for the interaction for $N$-breathers}
\end{figure}

We consider the general $N$- breather solution on the elliptic function background. Let us choose a special solution using \eqref{eq:vector-1} and \eqref{eq:vector-2}
\begin{align}
\Phi(x,t;z_j)&=
\begin{bmatrix}
\phi_1(x,t;z) \\[8pt]
\varphi_1(x,t;z) \\
\end{bmatrix}+\alpha_j\begin{bmatrix}
\phi_2(x,t;z) \\[8pt]
\varphi_2(x,t;z) \\
\end{bmatrix}
\nonumber\\
&=\alpha \frac{\vartheta_2\vartheta_4}{\vartheta_3\vartheta_4(\frac{\alpha\,x}{2K})}\left(E(x,t;z_j)\begin{bmatrix}
d(-x;z_j)\\
\ii \,d(x;z_j'){\rm e}^{\alpha \frac{\ii \pi}{2K}x+\omega(x,t)}\\
\end{bmatrix}+\alpha_jE(x,t;z_j')\begin{bmatrix}
d(-x;z_j')\\
\ii d(x;z_j){\rm e}^{\alpha \frac{\ii \pi}{2K}x+\omega(x,t)}\\
\end{bmatrix}\right),
\end{align}
where $\alpha_j$ is a complex parameter, $z_j'=-\ii (K+\ii K')-z_j,\ $   $\omega(x,t)=\alpha\,Z(K+2\ii l)\,x-\ii s_2\,t$, and
\[
d(x;z_j):=\frac{\vartheta_1(\frac{\alpha\,x+\ii (z_j-l)}{2K})}{\vartheta_4(\frac{\ii (z_j-l)}{2K})},\,\,\,\,\,\,\, E(x,t;z_j):=\exp[(\alpha\, Z(\ii (z_j-l))+\ii \lambda(z_j))x+\ii (s_2/2+y(\lambda(z_j)))t].
\]
By the symmetry relationship, we arrive at
\[
\Phi^{*}(x,t;z_i)=\alpha \frac{\vartheta_2\vartheta_4}{\vartheta_3\vartheta_4(\frac{\alpha\,x}{2K})}\left(-E^*(x,t;z_i)\begin{bmatrix}
d(x;z_i^*)\\
\ii \,d(-x;z_i^{*}\,')  {\rm e}^{-\alpha \frac{\ii \pi}{2K}x-\omega(x,t)}\\
\end{bmatrix}+\alpha_i^*E^*(x,t;z_i')\begin{bmatrix}
d(x;z_i^*\,') \\
\ii \,d(-x;z_i^{*}){\rm e}^{-\alpha \frac{\ii \pi}{2K}x-\omega(x,t)} \\
\end{bmatrix} \right),
\]
where $z_i^{*}\,'=-\ii (K+\ii K')-z_i^*$.
Consequently the general breather solution can be expressed with the Darboux transformation:
\begin{equation}\label{eq:gene-breather}
\psi_{n}=\psi-[\Phi_1(z_1),\Phi_1(z_2),\cdots,\Phi_1(z_n)]\mathbf{M}^{-1}\begin{bmatrix}
\Phi_2^*(z_1)\\
\Phi_2^*(z_2)\\
\vdots \\
\Phi_2^*(z_n)\\
\end{bmatrix},
\end{equation}
where $\psi$ is given by \eqref{eq:psi-theta} and
\begin{equation}\label{eq:matrixM}
\mathbf{M}=\left(\frac{\Phi^{\dag}(z_i)\Phi(z_j)}{2(\lambda(z_j)-\lambda(z_i^*))}\right)_{1\leq i,j\leq n}.
\end{equation}
Let us rewrite \eqref{eq:gene-breather} in a simpler form using the Sherman-Morrison-Woodbury-type matrix identity
\begin{equation}\label{eq:SMW}
 a+\vec{v}^\dagger \mathbf{A}^{-1} \vec{u}=\frac{a^{1-n}\det(a\mathbf{A}+\vec{u}\vec{v}^\dagger)}{\det \mathbf{A}}.
\end{equation}
Firstly, the numerator of \eqref{eq:matrixM} is
\begin{equation*}
\begin{split}
\Phi^{\dag}(z_i)\Phi(z_j)=&-E(z_j)E^*(z_i)(d(-x;z_j)d(x;z_i^*)-d(x;z_j')d(-x;z_i^*\,'))\\
&+\alpha_i^*E(z_j)E^*(z_i')(d(-x;z_j)d(x;z_i^*\,')-d(x;z_j')d(-x;z_i^*))\\
&-\alpha_j E(z_j')E^*(z_i)(d(-x;z_j')d(x;z_i^*)-d(x;z_j)d(-x;z_i^*\,'))\\
&+\alpha_i^*\alpha_j E(z_j')E^*(z_i')(d(-x;z_j')d(x;z_i^*\,')-d(x;z_j)d(-x;z_i^*)).
\end{split}
\end{equation*}
By the addition formula \eqref{eq:addition} and shift formula \eqref{eq:shift}, we obtain that
\begin{align*}
d(-x;z_j)d(x;z_i^*)-d(x;z_j')d(-x;z_i^*\,')
&=-\frac{\vartheta_3(\frac{2\ii l}{2K})\vartheta_3(\frac{\ii (z_i^*+z_j)}{2K})\vartheta_4(\frac{\alpha\, x}{2K})\vartheta_4(\frac{\alpha\, x+\ii (z_i^*-z_j)}{2K})}{\vartheta_4(\frac{\ii (z_i^*-l)}{2K})\vartheta_4(\frac{\ii (z_j-l)}{2K})\vartheta_2(\frac{\ii (z_j+l)}{2K})\vartheta_2(\frac{\ii (z_i^*+l)}{2K})},\\
d(-x;z_j)d(x;z_i^*\,')-d(x;z_j')d(-x;z_i^*)&=\frac{\vartheta_3(\frac{2\ii l}{2K})\vartheta_1(\frac{\ii (z_j-z_i^*)}{2K})\vartheta_4(\frac{\alpha\, x}{2K})\vartheta_2(\frac{\alpha\, x-\ii (z_i^*+z_j)}{2K})}{\vartheta_4(\frac{\ii (z_i^*-l)}{2K})\vartheta_4(\frac{\ii (z_j-l)}{2K})\vartheta_4(\frac{\ii (z_j+l)}{2K})\vartheta_4(\frac{\ii (z_i^*+l)}{2K})}{\rm e}^{-\alpha\frac{\pi\ii }{2K} \,x},\\
d(-x;z_j')d(x;z_i^*)-d(x;z_j)d(-x;z_i^*\,')&=\frac{\vartheta_3(\frac{2\ii l}{2K})\vartheta_1(\frac{\ii (z_i^*-z_j)}{2K})\vartheta_4(\frac{\alpha\, x}{2K})\vartheta_2(\frac{\alpha\, x+\ii (z_i^*+z_j)}{2K})}{\vartheta_4(\frac{\ii (z_i^*-l)}{2K})\vartheta_4(\frac{\ii (z_j-l)}{2K})\vartheta_2(\frac{\ii (z_j+l)}{2K})\vartheta_2(\frac{\ii (z_i^*+l)}{2K})}{\rm e}^{\alpha\frac{\pi\ii }{2K} \,x},\\
d(-x;z_j')d(x;z_i^*\,')-d(x;z_j)d(-x;z_i^*)&=\frac{\vartheta_3(\frac{2\ii l}{2K})\vartheta_3(\frac{\ii (z_j+z_i^*)}{2K})\vartheta_4(\frac{\alpha\, x}{2K})\vartheta_4(\frac{\alpha\, x-\ii (z_i^*-z_j)}{2K})}{\vartheta_4(\frac{\ii (z_i^*-l)}{2K})\vartheta_4(\frac{\ii (z_j-l)}{2K})\vartheta_2(\frac{\ii (z_j+l)}{2K})\vartheta_2(\frac{\ii (z_i^*+l)}{2K})}.
\end{align*}
Using the addition formula \eqref{eq:addition}, we have
\begin{equation}\label{eq:lambdai-lambdaj}
2(\lambda(z_j)-\lambda(z_i^*))={\alpha \ii }\frac{\vartheta_2\vartheta_4}{\vartheta_3}\left(\frac{\vartheta_3(\frac{2\ii l}{2K})\vartheta_1(\frac{\ii (z_j-z_i^*)}{2K})\vartheta_3(\frac{\ii (z_j+z_i^*)}{2K})}{\vartheta_2(\frac{\ii (z_j+l)}{2K})\vartheta_4(\frac{\ii (z_j-l)}{2K})\vartheta_2(\frac{\ii (z_i^*+l)}{2K})\vartheta_4(\frac{\ii (z_i^*-l)}{2K})}\right)
\end{equation}
Thus the $(i,j)$-component of $\mathbf{M}$ is given by
\begin{equation}
\begin{split}
\frac{\Phi^{\dag}(z_i)\Phi(z_j)}{2(\lambda(z_j)-\lambda(z_i^*))}&=\frac{1}{\alpha \ii }\frac{\vartheta_3\vartheta_4(\frac{\alpha\, x}{2K})}{\vartheta_2\vartheta_4}\Delta_{i,j}, \\
\end{split}
\end{equation}
where
\begin{equation*}
\begin{split}
\Delta_{i,j}=&E(z_j)E^*(z_i)\frac{\vartheta_4(\frac{\alpha\, x+\ii (z_i^*-z_j)}{2K})}{\vartheta_1(\frac{\ii (z_j-z_i^*)}{2K})}+\alpha_i^*E(z_j)E^*(z_i')\frac{\vartheta_2(\frac{\alpha\, x-\ii (z_i^*+z_j)}{2K})}{\vartheta_3(\frac{\ii (z_j+z_i^*)}{2K})}{\rm e}^{-\alpha\frac{\pi\ii }{2K} \,x}\\
&+\alpha_j E(z_j')E^*(z_i)\frac{\vartheta_2(\frac{\alpha\, x+\ii (z_i^*+z_j)}{2K})}{\vartheta_3(\frac{\ii (z_j+z_i^*)}{2K})}{\rm e}^{\alpha\frac{\pi\ii }{2K} \,x}
+\alpha_i^*\alpha_jE(z_j')E^*(z_i')\frac{\vartheta_4(\frac{\alpha\, x-\ii (z_i^*-z_j)}{2K})}{\vartheta_1(\frac{\ii (z_j-z_i^*)}{2K})}.
\end{split}
\end{equation*}
On the other hand, we have
\begin{equation*}
\begin{split}
\Phi_1(z_j)\Phi_2^*(z_i)=&\ii \left[-E(z_j)E^*(z_i)d(-x;z_j)d(-x;z_i^*\,')+\alpha_i^*E(z_j)E^*(z_i')d(-x;z_j)d(-x;z_i^*)\right.\\ &\left.-\alpha_jE(z_j')E^*(z_i)d(-x;z_j')d(-x;z_i^*\,')+\alpha_j\alpha_i^*E(z_j')E^*(z_i')d(-x;z_j')d(-x;z_i^*)\right]{\rm e}^{-\alpha \frac{\ii \pi}{2K}x-\omega(x,t)}\\
=&\ii \left[-E(z_j)E^*(z_i)\frac{\vartheta_1(\frac{-\alpha\, x +\ii (z_j-l)}{2K})\vartheta_3(\frac{-\alpha\, x-\ii (z_i^*+l)}{2K})}{\vartheta_4(\frac{\ii (z_j-l)}{2K})\vartheta_2(\frac{-\ii (z_i^*+l)}{2K})}\right.\\
&+\alpha_i^*E(z_j)E^*(z_i')\frac{\vartheta_1(\frac{-\alpha\, x +\ii (z_j-l)}{2K})\vartheta_1(\frac{-\alpha\, x+\ii (z_i^*-l)}{2K})}{\vartheta_4(\frac{\ii (z_j-l)}{2K})\vartheta_4(\frac{\ii (z_i^*-l)}{2K})} {\rm e}^{-\alpha \frac{\ii \pi}{2K}x} \\ &-\alpha_jE(z_j')E^*(z_i)\frac{\vartheta_3(\frac{-\alpha\, x -\ii (z_j+l)}{2K})\vartheta_3(\frac{-\alpha\, x-\ii (z_i^*+l)}{2K})}{\vartheta_2(-\frac{\ii (z_j+l)}{2K})\vartheta_2(\frac{-\ii (z_i^*+l)}{2K})}{\rm e}^{\alpha \frac{\ii \pi}{2K}x}\\ &\left.+\alpha_i^*\alpha_jE(z_j')E^*(z_i')\frac{\vartheta_3(\frac{-\alpha\, x -\ii (z_j+l)}{2K})\vartheta_1(\frac{-\alpha\, x+\ii (z_i^*-l)}{2K})}{\vartheta_2(\frac{\ii (z_j+l)}{2K})\vartheta_4(\frac{\ii (z_i^*-l)}{2K})}\right]{\rm e}^{-\omega(x,t)}.
\end{split}
\end{equation*}
We can simplify $\Phi_1(z_j)\Phi_2^*(z_i)-\psi \frac{1}{\alpha \ii}\frac{\vartheta_3\vartheta_4(\frac{\alpha x}{2K})}{\vartheta_2\vartheta_4}\Delta_{j,i}$, which is necessary to calculate the numerator in the formula \eqref{eq:SMW}, using the following expressions found from the addition formula \eqref{eq:addition}
\begin{equation*}
\begin{split}
&\frac{\vartheta_4(\frac{\alpha\, x+\ii (z_i^*-z_j)}{2K})}{\vartheta_1(\frac{\ii (z_j-z_i^*)}{2K})}\frac{\vartheta_2(\frac{\alpha \,x+2\ii l}{2K})}{\vartheta_3(\frac{2\ii l}{2K})}+\frac{\vartheta_1(\frac{\alpha\, x -\ii (z_j-l)}{2K})\vartheta_3(\frac{\alpha\, x+\ii (z_i^*+l)}{2K})}{\vartheta_4(\frac{\ii (z_j-l)}{2K})\vartheta_2(\frac{-\ii (z_i^*+l)}{2K})} \\
=&\frac{\vartheta_2(\frac{\ii (z_j+l)}{2K})\vartheta_2(\frac{-\alpha x-\ii (z_i^*-z_j)-2\ii l}{2K})\vartheta_4(\frac{\alpha\,x}{2K})\vartheta_4(\frac{-\ii (z_i^*-l)}{2K})}{\vartheta_1(\frac{\ii (z_j-z_i^*)}{2K})\vartheta_3(\frac{2\ii l}{2K})\vartheta_4(\frac{\ii (z_j-l)}{2K})\vartheta_2(\frac{-\ii (z_i^*+l)}{2K})} \\
\end{split}
\end{equation*}
and
\begin{equation*}
\begin{split}
&\left(\frac{\vartheta_2(\frac{\alpha\, x-\ii (z_i^*+z_j)}{2K})}{\vartheta_3(\frac{\ii (z_j+z_i^*)}{2K})}\frac{\vartheta_2(\frac{\alpha \,x+2\ii l}{2K})}{\vartheta_3(\frac{2\ii l}{2K})}+\frac{\vartheta_1(\frac{-\alpha\, x +\ii (z_j-l)}{2K})\vartheta_1(\frac{-\alpha\, x+\ii (z_i^*-l)}{2K})}{\vartheta_4(\frac{\ii (z_j-l)}{2K})\vartheta_4(\frac{\ii (z_i^*-l)}{2K})}\right) {\rm e}^{-\alpha \frac{\ii \pi}{2K}x} \\
=&\left(\frac{\vartheta_2(\frac{\ii (z_i^*+l)}{2K})\vartheta_2(\frac{\ii (z_j+l)}{2K})\vartheta_4(\frac{\alpha\,x}{2K})\vartheta_4(\frac{\alpha\,x-\ii (z_j+z_i^*)+2\ii l}{2K})}{\vartheta_3(\frac{\ii (z_j+z_i^*)}{2K})\vartheta_3(\frac{2\ii l}{2K})\vartheta_4(\frac{\ii (z_j-l)}{2K})\vartheta_4(\frac{\ii (z_i^*-l)}{2K})}\right) {\rm e}^{-\alpha \frac{\ii \pi}{2K}x} \\
\end{split}
\end{equation*}
and
\begin{equation*}
\begin{split}
&\left(\frac{\vartheta_2(\frac{\alpha\, x+\ii (z_i^*+z_j)}{2K})}{\vartheta_3(\frac{\ii (z_j+z_i^*)}{2K})}
\frac{\vartheta_2(\frac{\alpha \,x+2\ii l}{2K})}{\vartheta_3(\frac{2\ii l}{2K})}-\frac{\vartheta_3(\frac{-\alpha\, x -\ii (z_j+l)}{2K})\vartheta_3(\frac{-\alpha\, x-\ii (z_i^*+l)}{2K})}{\vartheta_2(-\frac{\ii (z_j+l)}{2K})\vartheta_2(\frac{-\ii (z_i^*+l)}{2K})}\right){\rm e}^{\alpha \frac{\ii \pi}{2K}x} \\
=&\left(\frac{-\vartheta_4(\frac{\ii (z_i^*-l)}{2K})\vartheta_4(\frac{\ii (z_j-l)}{2K})\vartheta_4(\frac{\alpha\,x+\ii (z_i^*+z_j+2l)}{2K})\vartheta_4(\frac{\alpha\,x}{2K})}{\vartheta_3(\frac{\ii (z_j+z_i^*)}{2K})\vartheta_3(\frac{2\ii l}{2K})\vartheta_2(\frac{\ii (z_j+l)}{2K})\vartheta_2(\frac{\ii (z_i^*+l)}{2K})}\right){\rm e}^{\alpha \frac{\ii \pi}{2K}x} \\
\end{split}
\end{equation*}
and
\begin{equation*}
\begin{split}
&\frac{\vartheta_4(\frac{\alpha\, x-\ii (z_i^*-z_j)}{2K})}{\vartheta_1(\frac{\ii (z_j-z_i^*)}{2K})}
\frac{\vartheta_2(\frac{\alpha \,x+2\ii l}{2K})}{\vartheta_3(\frac{2\ii l}{2K})}+\frac{\vartheta_3(\frac{-\alpha\, x -\ii (z_j+l)}{2K})\vartheta_1(\frac{-\alpha\, x+\ii (z_i^*-l)}{2K})}{\vartheta_2(\frac{\ii (z_j+l)}{2K})\vartheta_4(\frac{\ii (z_i^*-l)}{2K})} \\
=&\frac{\vartheta_2(\frac{\ii (z_i^*+l)}{2K})\vartheta_2(\frac{\alpha\,x -\ii (z_i^*-z_j)+2\ii l)}{2K})\vartheta_4(\frac{\ii (z_j-l)}{2K})\vartheta_4(\frac{\alpha\,x}{2K})}{\vartheta_1(\frac{\ii (z_j-z_i^*)}{2K})\vartheta_2(\frac{\ii (z_j+l)}{2K})\vartheta_3(\frac{2\ii l}{2K})\vartheta_4(\frac{\ii (z_i^*-l)}{2K})}.
\end{split}
\end{equation*}
In summary, we have the following theorem:
\begin{theorem}\label{thm3}
The multi-breather solutions can be represented as the following determinant form:
\begin{equation}\label{eq:n-breather}
\psi_n=\alpha \frac{\vartheta_2\vartheta_4}{\vartheta_3\vartheta_3(\frac{2\ii l}{2K})}
\left(\frac{\vartheta_4(\frac{\alpha\,x}{2K})}{\vartheta_2(\frac{\alpha\,x+2\ii l}{2K})}\right)^{n-1}\frac{\det(\mathbf{H})}{\det(\Delta)}{\rm e}^{-\alpha Z(K+2\ii l)x+\ii s_2t}
\end{equation}
where
\[
\Delta=\left(\Delta_{ij}\right)_{1\leq i,j\leq n},\,\,\,\,\,\,\, \mathbf{H}=\left(H_{i,j}\right)_{1\leq i,j\leq n},
\]
and
\begin{equation*}
\begin{split}
H_{i,j}=&E(z_j)E^*(z_i)\frac{\vartheta_2(\frac{\alpha x+\ii (z_i^*-z_j)+2\ii l}{2K})\vartheta_4(\frac{\ii (z_i^*-l)}{2K})\vartheta_2(\frac{\ii (z_j+l)}{2K})}{\vartheta_1(\frac{\ii (z_j-z_i^*)}{2K})\vartheta_4(\frac{\ii (z_j-l)}{2K})\vartheta_2(\frac{\ii (z_i^*+l)}{2K})}\\
&+\alpha_i^*E(z_j)E^*(z_i')\left(\frac{\vartheta_4(\frac{\alpha\,x-\ii (z_j+z_i^*)+2\ii l}{2K})\vartheta_2(\frac{\ii (z_i^*+l)}{2K})\vartheta_2(\frac{\ii (z_j+l)}{2K})}{\vartheta_3(\frac{\ii (z_j+z_i^*)}{2K})\vartheta_4(\frac{\ii (z_j-l)}{2K})\vartheta_4(\frac{\ii (z_i^*-l)}{2K})}\right) {\rm e}^{-\alpha \frac{\ii \pi}{2K}\,x}\\
&+\alpha_jE(z_j')E^*(z_i)\left(\frac{-\vartheta_4(\frac{\alpha\,x+\ii (z_i^*+z_j+2l)}{2K})\vartheta_4(\frac{\ii (z_i^*-l)}{2K})\vartheta_4(\frac{\ii (z_j-l)}{2K})}{\vartheta_3(\frac{\ii (z_j+z_i^*)}{2K})\vartheta_2(\frac{\ii (z_j+l)}{2K})\vartheta_2(\frac{\ii (z_i^*+l)}{2K})}\right){\rm e}^{\alpha \frac{\ii \pi}{2K}x}\\
&+\alpha_j\alpha_i^*E(z_j')E^*(z_i')\frac{\vartheta_2(\frac{\alpha\,x -\ii (z_i^*-z_j)+2\ii l)}{2K})\vartheta_2(\frac{\ii (z_i^*+l)}{2K})\vartheta_4(\frac{\ii (z_j-l)}{2K})}{\vartheta_1(\frac{\ii (z_j-z_i^*)}{2K})\vartheta_2(\frac{\ii (z_j+l)}{2K})\vartheta_4(\frac{\ii (z_i^*-l)}{2K})}.
\end{split}
\end{equation*}
\end{theorem}
In what follows, we analyze the interaction between breathers with different velocity. Introduce the notation $\beta_i=\alpha {\rm Re}(Z(\ii(z_i'-l))-Z(\ii(z_i-l)))$ and $v_i=-\frac{2{\rm Im}(y(z_i))}{\beta_i}.$ Assume that $\beta_i>0$ and $v_1<v_2<\cdots<v_k<\cdots<v_n.$ Firstly, we split the $(x,t)$ plane into $2n$ different pieces of regions $R_i^{\pm}$ exterior to the region of interaction (we approximate region as $\{(x,t)| x^2+t^2>r \}$, $r$ is big enough) by the $n$ lines $l_i:=\{x=v_i\,t+\gamma_{i,\pm}\}$, $i=1,2,\cdots, n$ (Fig.~\ref{fig:interaction}).

Rewrite the expression of $\Delta_{i,j}$:
\begin{equation*}
\begin{split}
\Delta_{i,j}&=\begin{bmatrix}
E^*(z_i) ,& \alpha_i^* E^*(z_i')
\end{bmatrix}\begin{bmatrix}
\frac{\vartheta_4(\frac{\alpha\, x+\ii (z_i^*-z_j)}{2K})}{\vartheta_1(\frac{\ii (z_j-z_i^*)}{2K})}&\frac{\vartheta_2(\frac{\alpha\, x+\ii (z_i^*+z_j)}{2K})}{\vartheta_3(\frac{\ii (z_j+z_i^*)}{2K})}{\rm e}^{\alpha\frac{\pi\ii }{2K} \,x}
\\ \frac{\vartheta_2(\frac{\alpha\, x-\ii (z_i^*+z_j)}{2K})}{\vartheta_3(\frac{\ii (z_j+z_i^*)}{2K})}{\rm e}^{-\alpha\frac{\pi\ii }{2K} \,x}
&\frac{\vartheta_4(\frac{\alpha\, x-\ii (z_i^*-z_j)}{2K})}{\vartheta_1(\frac{\ii (z_j-z_i^*)}{2K})}\\
\end{bmatrix}\begin{bmatrix}
E(z_j) \\[8pt] \alpha_j E(z_j')
\end{bmatrix}\\
&=\begin{bmatrix}
E^*(z_i) ,& \alpha_i^* E^*(z_i')
\end{bmatrix}\begin{bmatrix}
-\frac{\vartheta_4(\frac{\alpha\, x+\ii (z_i^*-z_j)}{2K})}{\vartheta_1(\frac{\ii (z_i^*-z_j)}{2K})}&-\frac{\vartheta_4(\frac{\alpha\, x+\ii (z_i^*+z_j)-K-\ii K'}{2K})}{\vartheta_1(\frac{\ii (z_j+z_i^*)-K-\ii K'}{2K})}
\\ \frac{\vartheta_4(\frac{\alpha\, x-\ii (z_i^*+z_j)+K+\ii K'}{2K})}{\vartheta_1(\frac{-\ii (z_j+z_i^*)+K+\ii K'}{2K})}
&\frac{\vartheta_4(\frac{\alpha\, x+\ii (z_j-z_i^*)}{2K})}{\vartheta_1(\frac{\ii (z_j-z_i^*)}{2K})}\\
\end{bmatrix}\begin{bmatrix}
E(z_j) \\[8pt] \alpha_j E(z_j')
\end{bmatrix}.
\end{split}
\end{equation*}
and
\begin{equation*}
\begin{split}
H_{i,j}&=\begin{bmatrix}
E^*(z_i)\frac{\vartheta_4(\frac{\ii (z_i^*-l)}{2K})}{\vartheta_2(\frac{\ii (z_i^*+l)}{2K})}{\rm e}^{\frac{\pi l}{2K}} ,& \alpha_i^* E^*(z_i')\frac{\vartheta_2(\frac{\ii (z_i^*+l)}{2K})}{\vartheta_4(\frac{\ii (z_i^*-l)}{2K})}{\rm e}^{-\frac{\pi l}{2K}}
\end{bmatrix}\begin{bmatrix}
-\frac{\vartheta_2(\frac{\alpha\, x+\ii (z_i^*-z_j)+2\ii l}{2K})}{\vartheta_1(\frac{\ii (z_i^*-z_j)}{2K})}&\ii \frac{\vartheta_2(\frac{\alpha\, x+\ii (z_i^*+z_j)+2\ii l-K-\ii K'}{2K})}{\vartheta_1(\frac{\ii (z_j+z_i^*)-K-\ii K'}{2K})}
\\ \ii \frac{\vartheta_2(\frac{\alpha\, x-\ii (z_i^*+z_j)+2\ii l+K+\ii K'}{2K})}{\vartheta_1(\frac{-\ii (z_j+z_i^*)+K+\ii K'}{2K})}
&\frac{\vartheta_2(\frac{\alpha\, x+\ii (z_j-z_i^*)+2\ii l}{2K})}{\vartheta_1(\frac{\ii (z_j-z_i^*)}{2K})}\\
\end{bmatrix}\\
&\,\,\,\,\,\,\,\,\,\,\,\,\,\,\,\,\,\begin{bmatrix}
E(z_j)\frac{\vartheta_2(\frac{\ii (z_j+l)}{2K})}{\vartheta_4(\frac{\ii (z_j-l)}{2K})}{\rm e}^{-\frac{\pi l}{2K}}  \\[8pt] \alpha_j E(z_j') \frac{\vartheta_4(\frac{\ii (z_j-l)}{2K})}{\vartheta_2(\frac{\ii (z_j+l)}{2K})}{\rm e}^{\frac{\pi l}{2K}}
\end{bmatrix}.
\end{split}
\end{equation*}
Then the determinants $\det(\Delta)$ and $\det(\mathbf{H})$ in breather solution can be rewritten as the following matrix form:
\begin{equation}
\begin{split}
\Delta&=-\mathbf{X}_1^{\dag}\mathbf{M}_{1,1}\mathbf{X}_1-\mathbf{X}_1^{\dag}\mathbf{M}_{1,2}\mathbf{X}_2
+\mathbf{X}_2^{\dag}\mathbf{M}_{2,1}\mathbf{X}_1+\mathbf{X}_2^{\dag}\mathbf{M}_{2,2}\mathbf{X}_2,\\
\mathbf{H}&=-\mathbf{Y}_1^{\dag}\mathbf{X}_1^{\dag}\mathbf{H}_{1,1}\mathbf{X}_1\mathbf{Z}_1+\ii\mathbf{Y}_1^{\dag}\mathbf{X}_1^{\dag}\mathbf{H}_{1,2}\mathbf{X}_2\mathbf{Z}_2
+\ii\mathbf{Y}_2^{\dag}\mathbf{X}_2^{\dag}\mathbf{H}_{2,1}\mathbf{X}_1\mathbf{Z}_1+\mathbf{Y}_2^{\dag}\mathbf{X}_2^{\dag}\mathbf{H}_{2,2}\mathbf{X}_2\mathbf{Z}_2,
\end{split}
\end{equation}
where
\begin{equation}
\begin{split}
\mathbf{M}_{1,1}&=\left(\frac{\vartheta_4(\frac{\alpha\,x+\ii(z_i^*-z_j)}{2K})}{\vartheta_1(\frac{\ii(z_i^*-z_j)}{2K})}\right)_{1\leq i,j\leq n},\,\,\,\,\,\,\,\,\mathbf{M}_{1,2}=\left(\frac{\vartheta_4(\frac{\alpha\,x+\ii(z_i^*+z_j)-K-\ii K'}{2K})}{\vartheta_1(\frac{\ii(z_i^*+z_j)-K-\ii K'}{2K})}\right)_{1\leq i,j\leq n}, \\
\mathbf{M}_{2,1}&=\left(\frac{\vartheta_4(\frac{\alpha\,x-\ii(z_i^*+z_j)+K+\ii K'}{2K})}{\vartheta_1(\frac{-\ii(z_i^*+z_j)+K+\ii K'}{2K})}\right)_{1\leq i,j\leq n},\,\,\,\,\,\,\,\,\mathbf{M}_{2,2}=\left(\frac{\vartheta_4(\frac{\alpha\,x+\ii(z_j-z_i^*)}{2K})}{\vartheta_1(\frac{\ii(z_j-z_i^*)}{2K})}\right)_{1\leq i,j\leq n}, \\
\mathbf{H}_{1,1}&=\left(\frac{\vartheta_2(\frac{\alpha\,x+2\ii l+\ii(z_i^*-z_j)}{2K})}{\vartheta_1(\frac{\ii(z_i^*-z_j)}{2K})}\right)_{1\leq i,j\leq n},\,\,\,\,\,\,\,\,\mathbf{H}_{1,2}=\left(\frac{\vartheta_2(\frac{\alpha\,x+2\ii l+\ii(z_i^*+z_j)-K-\ii K'}{2K})}{\vartheta_1(\frac{\ii(z_i^*+z_j)-K-\ii K'}{2K})}\right)_{1\leq i,j\leq n}, \\
\mathbf{H}_{2,1}&=\left(\frac{\vartheta_2(\frac{\alpha\,x+2\ii l-\ii(z_i^*+z_j)+K+\ii K'}{2K})}{\vartheta_1(\frac{-\ii(z_i^*+z_j)+K+\ii K'}{2K})}\right)_{1\leq i,j\leq n},\,\,\,\,\,\,\,\,\mathbf{H}_{2,2}=\left(\frac{\vartheta_2(\frac{\alpha\,x+2\ii l+\ii(z_j-z_i^*)}{2K})}{\vartheta_1(\frac{\ii(z_j-z_i^*)}{2K})}\right)_{1\leq i,j\leq n}, \\
\end{split}
\end{equation}
and
\begin{equation}
\begin{split}
\mathbf{X}_1=&{\rm diag}\left(\overbrace{1,1,\cdots,1}^{k-1},1,\ee^{-\xi_{k+1}},\ee^{-\xi_{k+2}},\cdots,\ee^{-\xi_n}\right),\\
\mathbf{X}_2=&{\rm diag}\left(\ee^{\xi_{1}},\ee^{\xi_{2}},\cdots,\ee^{\xi_{k-1}},\ee^{\xi_k},\overbrace{1,1,\cdots,1}^{n-k}\right),\\
\mathbf{Y}_1=&\ee^{\frac{\pi\,l}{2K}}{\rm diag}\left(\frac{\vartheta_4(\frac{\ii(z_1^*-l)}{2K})}{\vartheta_2(\frac{\ii(z_1^*+l)}{2K})},\frac{\vartheta_4(\frac{\ii(z_2^*-l)}{2K})}{\vartheta_2(\frac{\ii(z_2^*+l)}{2K})},\cdots,\frac{\vartheta_4(\frac{\ii(z_n^*-l)}{2K})}{\vartheta_2(\frac{\ii(z_n^*+l)}{2K})}\right),\\
\mathbf{Y}_2=&\ee^{-\frac{\pi\,l}{2K}}{\rm diag}\left(\frac{\vartheta_4(\frac{\ii(z_1^*+l)}{2K})}{\vartheta_2(\frac{\ii(z_1^*-l)}{2K})},\frac{\vartheta_4(\frac{\ii(z_2^*+l)}{2K})}{\vartheta_2(\frac{\ii(z_2^*-l)}{2K})},\cdots,\frac{\vartheta_4(\frac{\ii(z_n^*+l)}{2K})}{\vartheta_2(\frac{\ii(z_n^*-l)}{2K})}\right),\\
\mathbf{Z}_1=&\ee^{-\frac{\pi\,l}{2K}}{\rm diag}\left(\frac{\vartheta_2(\frac{\ii(z_1+l)}{2K})}{\vartheta_4(\frac{\ii(z_1-l)}{2K})},\frac{\vartheta_2(\frac{\ii(z_2+l)}{2K})}{\vartheta_4(\frac{\ii(z_2-l)}{2K})},\cdots,\frac{\vartheta_2(\frac{\ii(z_n+l)}{2K})}{\vartheta_4(\frac{\ii(z_n-l)}{2K})}\right),\\
\mathbf{Z}_2=&\ee^{\frac{\pi\,l}{2K}}{\rm diag}\left(\frac{\vartheta_4(\frac{\ii(z_1-l)}{2K})}{\vartheta_2(\frac{\ii(z_1+l)}{2K})},\frac{\vartheta_4(\frac{\ii(z_2-l)}{2K})}{\vartheta_2(\frac{\ii(z_2+l)}{2K})},\cdots,\frac{\vartheta_4(\frac{\ii(z_n-l)}{2K})}{\vartheta_2(\frac{\ii(z_n+l)}{2K})}\right),\\
\end{split}
\end{equation}
with $\xi_k=\beta_k(x-v_k t)+\ii (\alpha {\rm Im}(Z(\ii(z_k'-l))-Z(\ii(z_k-l)))\,x-2{\rm Re}(y(z_k)) \,t)+\ln(\alpha_k).$ Fixing the line $x-v_k t={\rm const}$, and taking the limit $t\to -\infty$, we can obtain that
\begin{align*}
   \psi_n&\to\alpha\frac{\vartheta_2\vartheta_4}{\vartheta_3\vartheta_3(\frac{2\ii l}{2K})}\left(\frac{\vartheta_4(\frac{\alpha\,x}{2K})}{\vartheta_2(\frac{\alpha\,x+2\ii l}{2K})}\right)^{n-1}\ee^{-\alpha Z(K+2\ii l)\,x +\ii s_2t} \prod_{i=1}^{k-1}\frac{\gamma_i}{\gamma_i^*}\prod_{i=k+1}^{n}\frac{\gamma_i^*}{\gamma_i}\times \\
   &\frac{-\gamma_k\gamma_k^{*-1}\det(\mathbf{K}^{[1,1]})+\ii \gamma_k^{*-1}\gamma_k^{-1}\det(\mathbf{K}^{[1,2]})\ee^{\xi_k+2\pi}+\ii\gamma_k^*\gamma_k^{-1}\det(\mathbf{K}^{[2,1]})\ee^{\xi_k^*-2\pi\, l}+\gamma_k^*\gamma_k^{-1}\det(\mathbf{K}^{[2,2]})\ee^{\xi_k+\xi_k^*}}
   {-\det(\mathbf{M}^{[1,1]})-\det(\mathbf{M}^{[1,2]})\ee^{\xi_k}+\det(\mathbf{M}^{[2,1]})\ee^{\xi_k^*}+\det(\mathbf{M}^{[2,2]})\ee^{\xi_k+\xi_k^*}}
\end{align*}
where
\begin{equation*}
    \gamma_{k}=\frac{\vartheta_2(\frac{\ii(z_k+l)}{2K})}{\vartheta_4(\frac{\ii(z_k-l)}{2K})}
\end{equation*}
and
\begin{equation*}
\begin{split}
 \mathbf{M}^{[a,b]}&=\left(\frac{\vartheta_4\left(\frac{\alpha\,x+\xi_i^{[a,b]}+\eta_j^{[a,b]}}{2K}\right)}{\vartheta_1\left(\frac{\xi_i^{[a,b]}+\eta_j^{[a,b]}}{2K}\right)} \right)_{1\leq i,j\leq n}, \,\,\,\,\,\,\,\,
  \mathbf{K}^{[a,b]}=\left(\frac{\vartheta_2\left(\frac{\alpha\,x+2\ii l+\xi_i^{[a,b]}+\eta_j^{[a,b]}}{2K}\right)}{\vartheta_1\left(\frac{\xi_i^{[a,b]}+\eta_j^{[a,b]}}{2K}\right)} \right)_{1\leq i,j\leq n},\,\,\,\,\,\,\, a,b=1,2.
\end{split}
\end{equation*}
We list the concrete sequences of $\xi_i^{[a,b]}$ and $\eta_j^{[a,b]}$:
\begin{equation*}
\begin{split}
\left(\xi_1^{[1,1]},\xi_2^{[1,1]},\cdots, \xi_{k}^{[1,1]},\xi_{k+1}^{[1,1]},\cdots, \xi_n^{[1,1]}\right)=&\left(\ii z_1^*,\ii z_2^*,\cdots, \ii z_k^*,K+\ii K'-\ii z_{k+1}^*,\cdots, K+\ii K'-\ii z_{n}^*\right) \\
\left(\eta_1^{[1,1]},\eta_2^{[1,1]},\cdots, \eta_{k}^{[1,1]},\eta_{k+1}^{[1,1]},\cdots, \eta_n^{[1,1]}\right)=&\left(-\ii z_1,-\ii z_2,\cdots, -\ii z_k,\ii z_{k+1}-(K+\ii K'),\cdots, \ii z_{n}-(K+\ii K')\right) \\
\end{split}
\end{equation*}
and
\begin{equation*}
\begin{split}
\left(\xi_1^{[1,2]},\xi_2^{[1,2]},\cdots, \xi_{k}^{[1,2]},\xi_{k+1}^{[1,2]},\cdots, \xi_n^{[1,2]}\right)=&\left(\ii z_1^*,\ii z_2^*,\cdots, \ii z_k^*,K+\ii K'-\ii z_{k+1}^*,\cdots, K+\ii K'-\ii z_{n}^*\right) \\
\left(\eta_1^{[1,2]},\eta_2^{[1,2]},\cdots,\eta_{k-1}^{[1,2]},\eta_{k}^{[1,2]},\cdots, \eta_n^{[1,2]}\right)=&\left(-\ii z_1,-\ii z_2,\cdots,-\ii z_{k-1}, \ii z_k-(K+\ii K'),\cdots, \ii z_{n}-(K+\ii K')\right) \\
\end{split}
\end{equation*}
and
\begin{equation*}
\begin{split}
\left(\xi_1^{[2,1]},\xi_2^{[2,1]},\cdots, \xi_{k-1}^{[2,1]},\xi_{k}^{[2,1]},\cdots, \xi_n^{[2,1]}\right)=&\left(\ii z_1^*,\ii z_2^*,\cdots,\ii z_{k-1}^*, K+\ii K'-\ii z_k^*,\cdots, K+\ii K'-\ii z_{n}^*\right) \\
\left(\eta_1^{[2,1]},\eta_2^{[2,1]},\cdots, \eta_{k}^{[2,1]},\eta_{k+1}^{[2,1]},\cdots, \eta_n^{[2,1]}\right)=&\left(-\ii z_1,-\ii z_2,\cdots, -\ii z_k,\ii z_{k+1}-(K+\ii K'),\cdots, \ii z_{n}-(K+\ii K')\right) \\
\end{split}
\end{equation*}
and
\begin{equation*}
\begin{split}
\left(\xi_1^{[2,2]},\xi_2^{[2,2]},\cdots, \xi_{k-1}^{[2,2]},\xi_{k}^{[2,2]},\cdots, \xi_n^{[2,2]}\right)=&\left(\ii z_1^*,\ii z_2^*,\cdots, \ii z_{k-1}^*,(K+\ii K')-\ii z_{k}^*,\cdots, (K+\ii K')-\ii z_{n}^*\right)  \\
\left(\eta_1^{[2,2]},\eta_2^{[2,2]},\cdots, \eta_{k-1}^{[2,2]},\eta_{k}^{[2,2]},\cdots, \eta_n^{[2,2]}\right)=&\left(-\ii z_1,-\ii z_2,\cdots, -\ii z_{k-1},\ii z_{k}-(K+\ii K'),\cdots, \ii z_{n}-(K+\ii K')\right).\\
\end{split}
\end{equation*}
To calculate the determinant of above matrix. We use the results from the Appendix D of Ref.~\cite{Takahashi16}, which may be regarded as a higher-order version of Fay's identity.
The matrices $\mathbf{M}^{[a,b]}$ and $\mathbf{K}^{[a,b]}$s are the theta function version Cauchy matrices, whose determinants can be represented as
\begin{equation*}
\begin{split}
\det(\mathbf{M}^{[a,b]})&=\frac{\vartheta_4(\frac{\alpha\, x}{2K})^{n-1}\vartheta_4\left[\frac{\alpha\,x+{\sum_{i=1}^{k}}(\xi_i^{[a,b]}+\eta_i^{[a,b]})}{2K}\right]{\displaystyle \prod_{1\leq i<j\leq n}}\vartheta_1\left(\frac{\xi_i^{[a,b]}-\xi_j^{[a,b]}}{2K}\right)\vartheta_1\left(\frac{\eta_i^{[a,b]}-\eta_j^{[a,b]}}{2K}\right)}{{\displaystyle \prod_{i,j=1}^{n,n}}\vartheta_1\left(\frac{\xi_i^{[a,b]}+\eta_j^{[a,b]}}{2K}\right)}\\
\det(\mathbf{K}^{[a,b]})&=\frac{\vartheta_2(\frac{\alpha\, x+2\ii l}{2K})^{n-1}\vartheta_2\left[\frac{\alpha\,x+2\ii l+{\sum_{i=1}^{k}}(\xi_i^{[a,b]}+\eta_i^{[a,b]})}{2K}\right]{\displaystyle \prod_{1\leq i<j\leq n}}\vartheta_1\left(\frac{\xi_i^{[a,b]}-\xi_j^{[a,b]}}{2K}\right)\vartheta_1\left(\frac{\eta_i^{[a,b]}-\eta_j^{[a,b]}}{2K}\right)}{{\displaystyle \prod_{i,j=1}^{n,n}}\vartheta_1\left(\frac{\xi_i^{[a,b]}+\eta_j^{[a,b]}}{2K}\right)}.
\end{split}
\end{equation*}

We observe that above four pairs of vectors have different $k$-th components. The other components are the same. Furthermore, the solution can be reduced as
\begin{align*}
   \psi_n&\to\alpha\frac{\vartheta_2\vartheta_4}{\vartheta_3\vartheta_3(\frac{2\ii l}{2K})}\ee^{-\alpha Z(K+2\ii l)\,x +\ii s_2t}\prod_{i=1}^{k-1}\frac{\gamma_i}{\gamma_i^*}\prod_{i=k+1}^{n}\frac{\gamma_i^*}{\gamma_i}\times \\
   &\frac{-\gamma_k\gamma_k^{*-1}C^{[1,1]}+\ii \gamma_k^{*-1}\gamma_k^{-1}C^{[1,2]}\ee^{\xi_k+2\pi}+\ii\gamma_k^*\gamma_k^{-1}C^{[2,1]}\ee^{\xi_k^*-2\pi\, l}+\gamma_k^*\gamma_k^{-1}C^{[2,2]}\ee^{\xi_k+\xi_k^*}}
   {-D^{[1,1]}-D^{[1,2]}\ee^{\xi_k}+D^{[2,1]}\ee^{\xi_k^*}+D^{[2,2]}\ee^{\xi_k+\xi_k^*}}
\end{align*}
where
\[
C^{[a,b]}=\vartheta_4\left[\frac{\alpha\,x+{\sum_{i=1}^{k}}(\xi_i^{[a,b]}+\eta_i^{[a,b]})}{2K}\right]L^{[a,b]}
,\,\,\,\,\,\, D^{[a,b]}=\vartheta_2\left[\frac{\alpha\,x+{\sum_{i=1}^{k}}(\xi_i^{[a,b]}+\eta_i^{[a,b]})}{2K}\right]L^{[a,b]},
\]
and
\[
L^{[a,b]}=\frac{{\displaystyle \prod_{i=1}^{k-1} \vartheta_1\left(\frac{\xi_i^{[a,b]}-\xi_k^{[a,b]}}{2K}\right) \vartheta_1\left(\frac{\eta_i^{[a,b]}-\eta_k^{[a,b]}}{2K}\right)\prod_{j=k+1}^{n} \vartheta_1\left(\frac{\xi_k^{[a,b]}-\xi_j^{[a,b]}}{2K}\right) \vartheta_1\left(\frac{\eta_k^{[a,b]}-\eta_j^{[a,b]}}{2K}\right)}}{{\displaystyle \prod_{i=1}^{n} \vartheta_1\left(\frac{\xi_i^{[a,b]}+\eta_k^{[a,b]}}{2K}\right)\prod_{j=1,j\neq k}^{n} \vartheta_1\left(\frac{\xi_k^{[a,b]}+\eta_j^{[a,b]}}{2K}\right)}}.
\]
Then the asymptotic behavior for the multi-breather solution with different velocity can be concluded as the following theorem:
\begin{theorem}\label{thm4}
When $t\to -\infty$, the breather solution can be asymptotically expressed as
\[
\psi_{k}^{[-\infty]}=\alpha \beta_k^{[-\infty]}\frac{\vartheta_2\vartheta_4}{\vartheta_3\vartheta_3(\frac{2\ii l}{2K})}
\frac{H_{k,k}^{[-\infty]}}{M_{k,k}^{[-\infty]}}{\rm e}^{-\alpha Z(K+2\ii l)x+\ii s_2t}+O\left(\exp(|\alpha_k|t)\right),\,\,\,\,\,\,\, \alpha_k=\min_{i\neq k}(\beta_i|v_i-v_k|)
\]
where $\beta_k^{[-\infty]}=\prod_{i=1}^{k-1}\frac{\gamma_i}{\gamma_i^*}\prod_{i=k+1}^{n}\frac{\gamma_i^*}{\gamma_i}$
\begin{equation*}
\begin{split}
M_{k,k}^{[-\infty]}&=|E(z_k)|^2\frac{\vartheta_4\left(\frac{\alpha\, x+s_k^{[-\infty]}+\ii (z_k^*-z_k)}{2K}\right)}{\vartheta_1(\frac{\ii (z_k-z_k^*)}{2K})}+\alpha_k^{[+\infty]*}E(z_k)E^*(z_k')\frac{\vartheta_2\left(\frac{\alpha\, x+s_k^{[-\infty]}-\ii (z_k^*+z_k)}{2K}\right)}{\vartheta_3(\frac{\ii (z_k+z_k^*)}{2K})}{\rm e}^{-\frac{\pi\ii }{2K} (\alpha\,x+s_k^{[-\infty]})}\\
&+\alpha^{[-\infty]}_k E(z_k')E^*(z_k)\frac{\vartheta_2\left(\frac{\alpha\, x+s_k^{[-\infty]}+\ii (z_k^*+z_k)}{2K}\right)}{\vartheta_3(\frac{\ii (z_k+z_k^*)}{2K})}{\rm e}^{\frac{\pi\ii }{2K} (\alpha\,x+s_k^{[-\infty]})}
+|\alpha^{[+\infty]}_k|^2|E(z_k')|^2\frac{\vartheta_4\left(\frac{\alpha\, x+s_k^{[-\infty]}-\ii (z_k^*-z_k)}{2K}\right)}{\vartheta_1(\frac{\ii (z_k-z_k^*)}{2K})},
\end{split}
\end{equation*}
and $\alpha^{[-\infty]}_k=\alpha_k\Delta_k^{[-\infty]}$, $s_k^{[-\infty]}=\sum_{i=1}^{k-1}\ii (z_i^*-z_i)-\sum_{i=k+1}^{n}\ii (z_i^*-z_i),$
\begin{equation*}
\begin{split}
H_{k,k}^{[-\infty]}=&|E(z_k)|^2\frac{\vartheta_2\left(\frac{\alpha x+s_k^{[-\infty]}+\ii (z_k^*-z_k)+2\ii l}{2K}\right)\vartheta_4(\frac{\ii (z_k^*-l)}{2K})\vartheta_2(\frac{\ii (z_k+l)}{2K})}{\vartheta_1(\frac{\ii (z_k-z_k^*)}{2K})\vartheta_4(\frac{\ii (z_k-l)}{2K})\vartheta_2(\frac{\ii (z_k^*+l)}{2K})}\\
&+\alpha_k^{[-\infty]*}E(z_k)E^*(z_k')\left(\frac{\vartheta_4\left(\frac{\alpha\,x+s_k^{[-\infty]}-\ii (z_k+z_k^*)+2\ii l}{2K}\right)\vartheta_2(\frac{\ii (z_k^*+l)}{2K})\vartheta_2(\frac{\ii (z_k+l)}{2K})}{\vartheta_3(\frac{\ii (z_k+z_k^*)}{2K})\vartheta_4(\frac{\ii (z_k-l)}{2K})\vartheta_4(\frac{\ii (z_k^*-l)}{2K})}\right) {\rm e}^{-\frac{\ii \pi}{2K}(\alpha \,x+s_k^{[-\infty]})}\\
&+\alpha_k^{[-\infty]}E(z_k')E^*(z_k)\left(\frac{-\vartheta_4\left(\frac{\alpha\,x+s_k^{[-\infty]}+\ii (z_k^*+z_k+2l)}{2K}\right)\vartheta_4(\frac{\ii (z_k^*-l)}{2K})\vartheta_4(\frac{\ii (z_k-l)}{2K})}{\vartheta_3(\frac{\ii (z_k+z_k^*)}{2K})\vartheta_2(\frac{\ii (z_k+l)}{2K})\vartheta_2(\frac{\ii (z_k^*+l)}{2K})}\right){\rm e}^{\frac{\ii \pi}{2K}(\alpha x+s_k^{[-\infty]})}\\
&+|\alpha_k^{[-\infty]}|^2|E(z_k')|^2\frac{\vartheta_2\left(\frac{\alpha\,x +s_k^{[-\infty]}-\ii (z_k^*-z_k)+2\ii l)}{2K}\right)\vartheta_2(\frac{\ii (z_k^*+l)}{2K})\vartheta_4(\frac{\ii (z_k-l)}{2K})}{\vartheta_1(\frac{\ii (z_k-z_k^*)}{2K})\vartheta_2(\frac{\ii (z_k+l)}{2K})\vartheta_4(\frac{\ii (z_k^*-l)}{2K})},
\end{split}
\end{equation*}
with
\begin{equation*}
\Delta_k^{[-\infty]}=
{\displaystyle \prod_{i=1}^{k-1} }\frac{\vartheta_1\left(\frac{K+\ii K'-\ii z_i-\ii z_k}{2K}\right) \vartheta_1\left(\frac{\ii z_i^*-\ii z_k}{2K}\right)}{\vartheta_1\left(\frac{K+\ii K'-\ii z_i^*-\ii z_k}{2K}\right) \vartheta_1\left(\frac{\ii z_i-\ii z_k}{2K}\right) }
\prod_{i=k+1}^{n}  \frac{\vartheta_1\left(\frac{\ii z_k-\ii z_i}{2K}\right) \vartheta_1\left(\frac{K+\ii K'-\ii z_i^*-\ii z_k}{2K}\right)}{\vartheta_1\left(\frac{\ii z_k-\ii z_i^*}{2K}\right)\vartheta_1\left(\frac{K+\ii K'-\ii z_k-\ii z_i}{2K}\right)}.
\end{equation*}

In a similar procedure, we can consider the asymptotic analysis when $t\to+\infty$, and a similar expression for the asymptotic breather solution $\psi_k^{[+\infty]}$ can be obtained by replacing $\alpha^{[-\infty]}_k$, $s_k^{[-\infty]}$ with $\alpha^{[+\infty]}_k$, $s_k^{[+\infty]}$, where
$\alpha^{[+\infty]}_k=\alpha_k\Delta_k^{[+\infty]}$, $s_k^{[+\infty]}=-\sum_{i=1}^{k-1}\ii (z_i^*-z_i)+\sum_{i=k+1}^{n}\ii (z_i^*-z_i)$, and $\beta_k^{[+\infty]}=\prod_{i=1}^{k-1}\frac{\gamma_i^*}{\gamma_i}\prod_{i=k+1}^{n}\frac{\gamma_i}{\gamma_i^*}$
\begin{equation*}
\Delta_k^{[+\infty]}=
{\displaystyle \prod_{i=1}^{k-1} }\frac{\vartheta_1\left(\frac{\ii (z_i-z_k)}{2K}\right) \vartheta_1\left(\frac{K+\ii K'-\ii (z_i^*+z_k)}{2K}\right)}{\vartheta_1\left(\frac{\ii (z_i^*- z_k)}{2K}\right) \vartheta_1\left(\frac{K+\ii K'-\ii (z_i+z_k)}{2K}\right) }
\prod_{i=k+1}^{n}  \frac{\vartheta_1\left(\frac{\ii (z_i+z_k)-K-\ii K'}{2K}\right) \vartheta_1\left(\frac{\ii (z_i^*-z_k)}{2K}\right)}{\vartheta_1\left(\frac{\ii (z_i^*+z_k)-K-\ii K'}{2K}\right)\vartheta_1\left(\frac{\ii (z_i-z_k)}{2K}\right)}.
\end{equation*}
\end{theorem}

Actually, this proves the interaction between breathers is hard to say elastic or inelastic. Through the above asymptotic analysis, we find the expression of breathers before and after the interaction just change some parameters. Actually, these parameters will affect the dynamics of breathers, since the background solution is not uniform as the zero or plane wave background.

In the following, we give some examples to illustrate the dynamics for the single breather and their interaction. Firstly we show the single breather solution, i.e. the formula \eqref{eq:n-breather} when $n=1$. From this formula, we show the asymptotic behavior for the single breather solution. We have the asymptotic analysis:
\begin{equation}\label{eq:breather-asy-1}
\psi_1\to \alpha \frac{\vartheta_2\vartheta_4}{\vartheta_3\vartheta_3(\frac{2\ii l}{2K})}\frac{\vartheta_2(\frac{\alpha x+\ii(z_1^*-z_1)+2\ii l}{2K})}{\vartheta_4(\frac{\alpha x+\ii(z_1^*-z_1)}{2K})}\frac{\vartheta_4(\frac{\ii (z_1^*-l)}{2K})\vartheta_2(\frac{\ii (z_1+l)}{2K})}{\vartheta_4(\frac{\ii (z_1-l)}{2K})\vartheta_2(\frac{\ii (z_1^*+l)}{2K})}\ee^{-\alpha Z(K+2\ii l)x+\ii s_2 t},
\end{equation}
as $\beta_1(x-v_1t+\gamma_1)\to-\infty$, and
\begin{equation}\label{eq:breather-asy-2}
\psi_1\to \alpha \frac{\vartheta_2\vartheta_4}{\vartheta_3\vartheta_3(\frac{2\ii l}{2K})}\frac{\vartheta_2(\frac{\alpha x+\ii(z_1-z_1^*)+2\ii l}{2K})}{\vartheta_4(\frac{\alpha x+\ii(z_1-z_1^*)}{2K})}\frac{\vartheta_4(\frac{\ii (z_1-l)}{2K})\vartheta_2(\frac{\ii (z_1^*+l)}{2K})}{\vartheta_4(\frac{\ii (z_1^*-l)}{2K})\vartheta_2(\frac{\ii (z_1+l)}{2K})}\ee^{-\alpha Z(K+2\ii l)x+\ii s_2 t},
\end{equation}
as $\beta_1(x-v_1t+\gamma_1)\to+\infty.$
From the above asymptotic expression, we know the phase differences is $2\ln\left(\frac{\vartheta_4(\frac{\ii (z_1^*-l)}{2K})\vartheta_2(\frac{\ii (z_1+l)}{2K})}{\vartheta_4(\frac{\ii (z_1-l)}{2K})\vartheta_2(\frac{\ii (z_1^*+l)}{2K})}\right)$ and the shift of crest is $4\mathrm{Im}(z_1)/\alpha$. The parameter $\alpha_1$ will affect the maximum of amplitude. If ${\rm Im}(\lambda(z_1))>0$, the maximum value $\nu_3+2{\rm Im}(\lambda(z_1))$ ($\nu_3$ refers to equation \eqref{eq:para-nu123}) appears at $(x,t)=(0,0)$ by choosing $\alpha_1=-1$ and the locally minimum value $\nu_3-2{\rm Im}(\lambda(z_1))$ appears at $(x,t)=(0,0)$ by choosing $\alpha_1=1$. Instead, if ${\rm Im}(\lambda(z_1))<0$, the maximum value emerges when $\alpha_1=1$ and the locally minimum value emerges for $\alpha_1=-1.$

To show the dynamics for the single breather solution, we choose the exact parameters and plot the figure.
We discuss them in the following four cases.

\subsubsection{\textbf{Example 1: The breather solutions on the dn solution background.}}
\begin{figure}[tb]
\centering
\begin{tabular}{ccc}
\includegraphics[width=.32\textwidth]{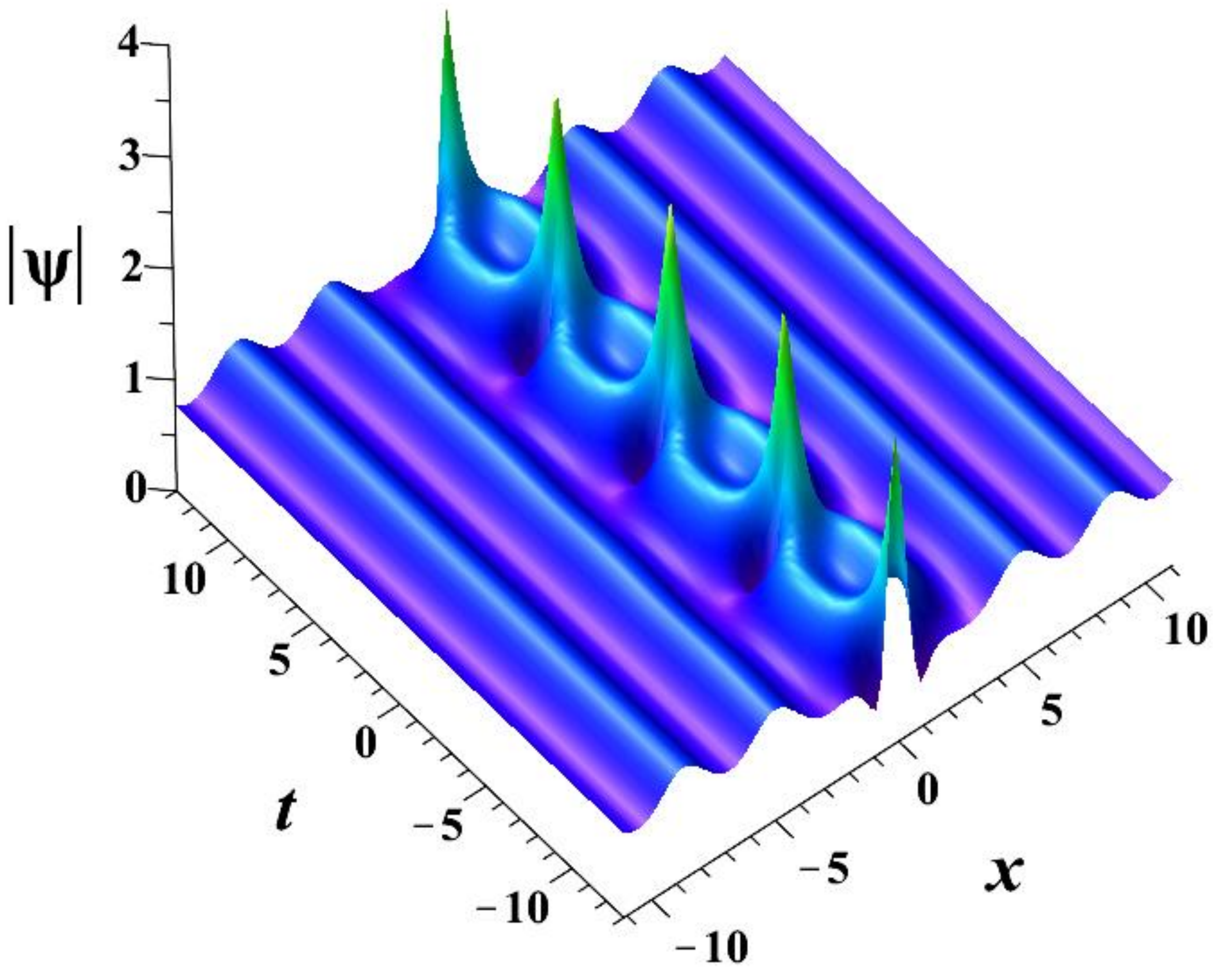} \!\!\!& \includegraphics[width=.32\textwidth]{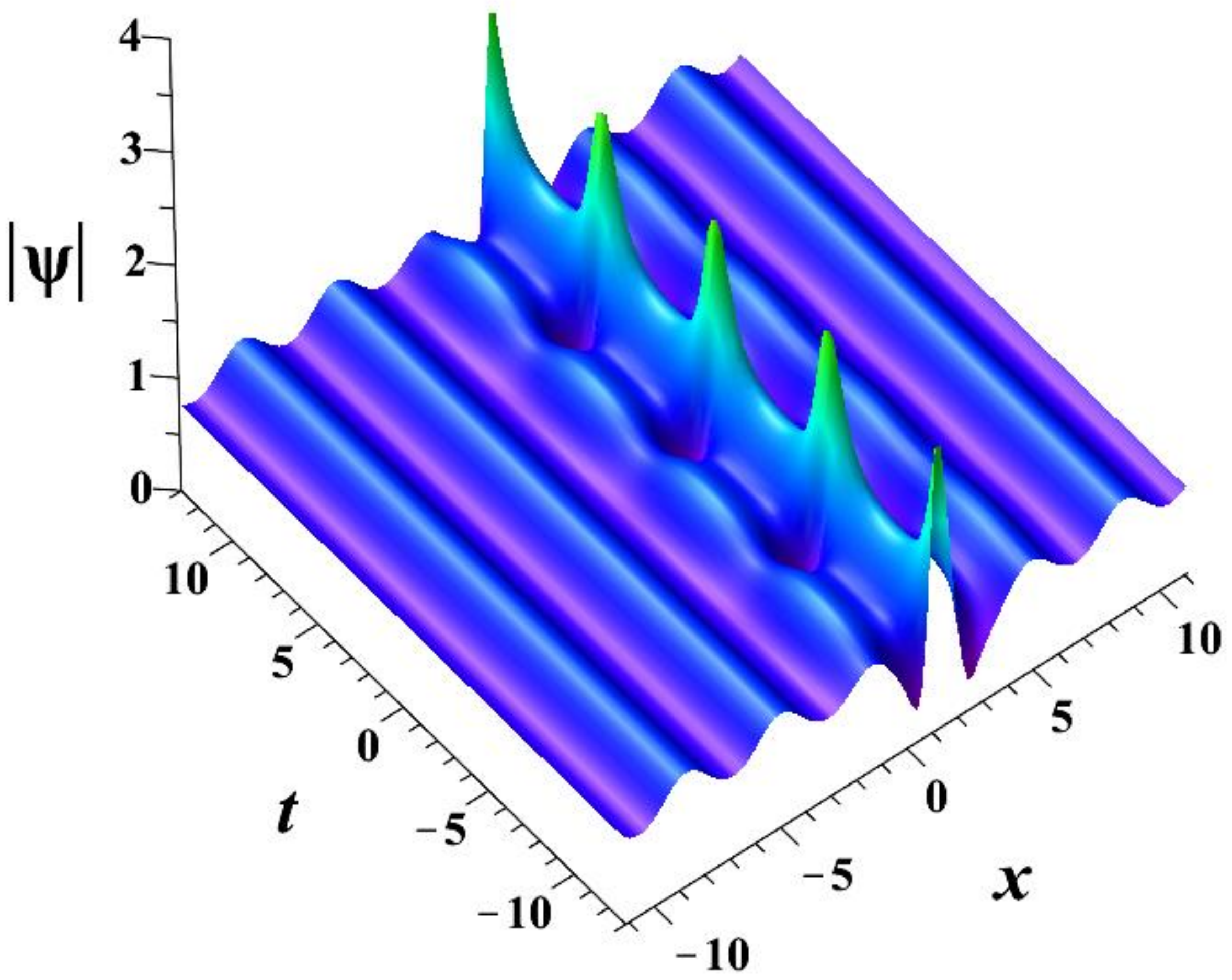}\!\!\!& \includegraphics[width=.32\textwidth]{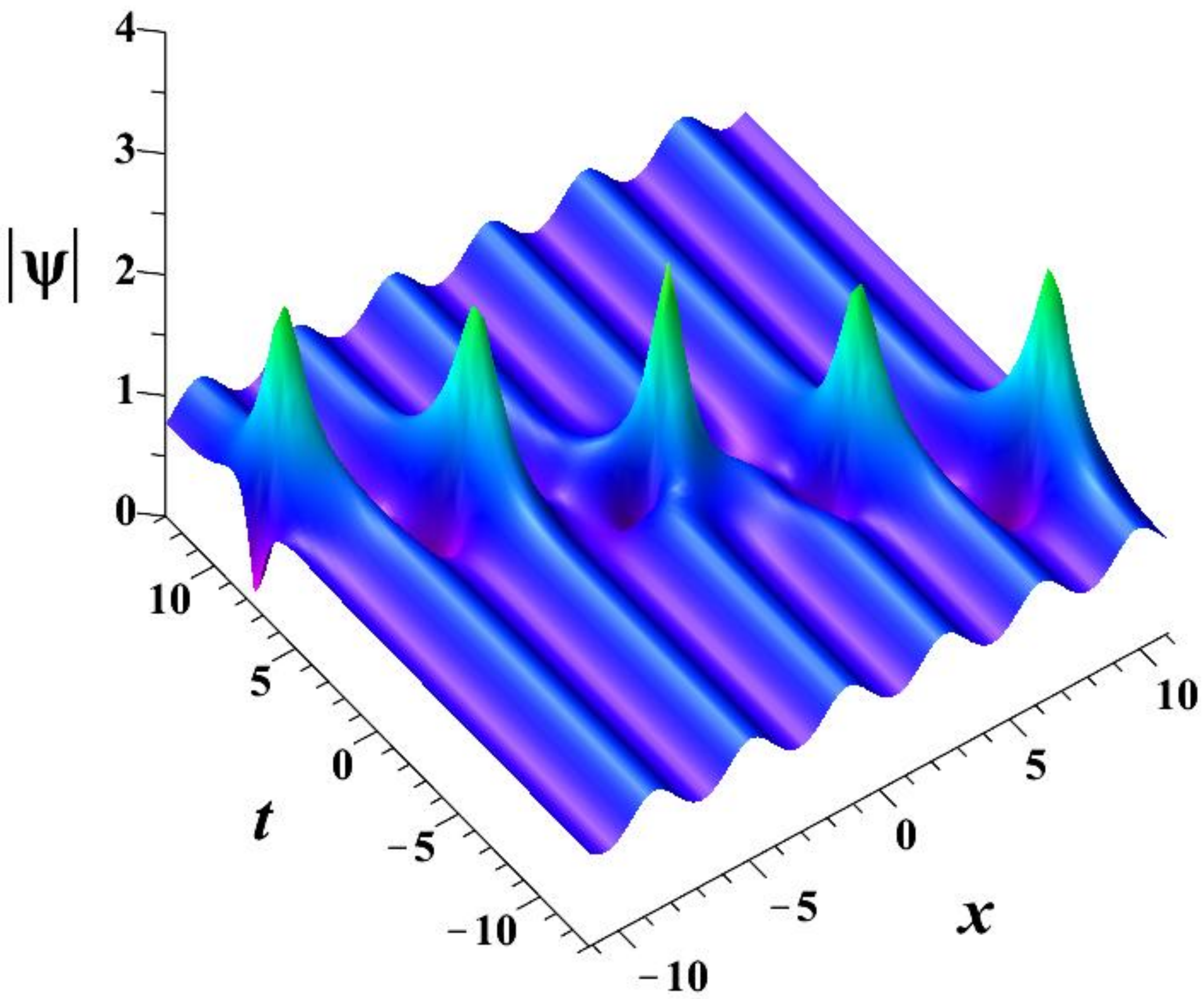}\!\!\! \\
{\footnotesize (a) single static dn breather}\!\!\!& {\footnotesize (b) another single static dn breather}\!\!\!& {\footnotesize (c) single non-static dn breather}\!\!\!
\end{tabular}
\caption{(color online): The single breather solutions with dn background, given by \eqref{eq:dnsingle} with setting $l=\frac{K'}{2},\ \alpha=1$, and $k=1/\sqrt{2}$. (a) The static breather solution with $z_1=-\frac{K'(1/2)}{2}-\frac{K(1/2)\ii}{3}$ and $\alpha_1=-1$. (b) If we adjust the parameter $\alpha_1=-4$ in (a), then the breather will change its shape due to the effect of background. (c) The example of non-static breather. Here we set $\alpha_1=-1$ and $z_1=-\frac{K+\ii K'}{3}$. The solution reaches the maximum value $2.605$ at origin.
}
\label{fig:dn-b}
\end{figure}

\begin{figure}[tb]
\centering
\begin{tabular}{cc}
\includegraphics[width=.41\textwidth]{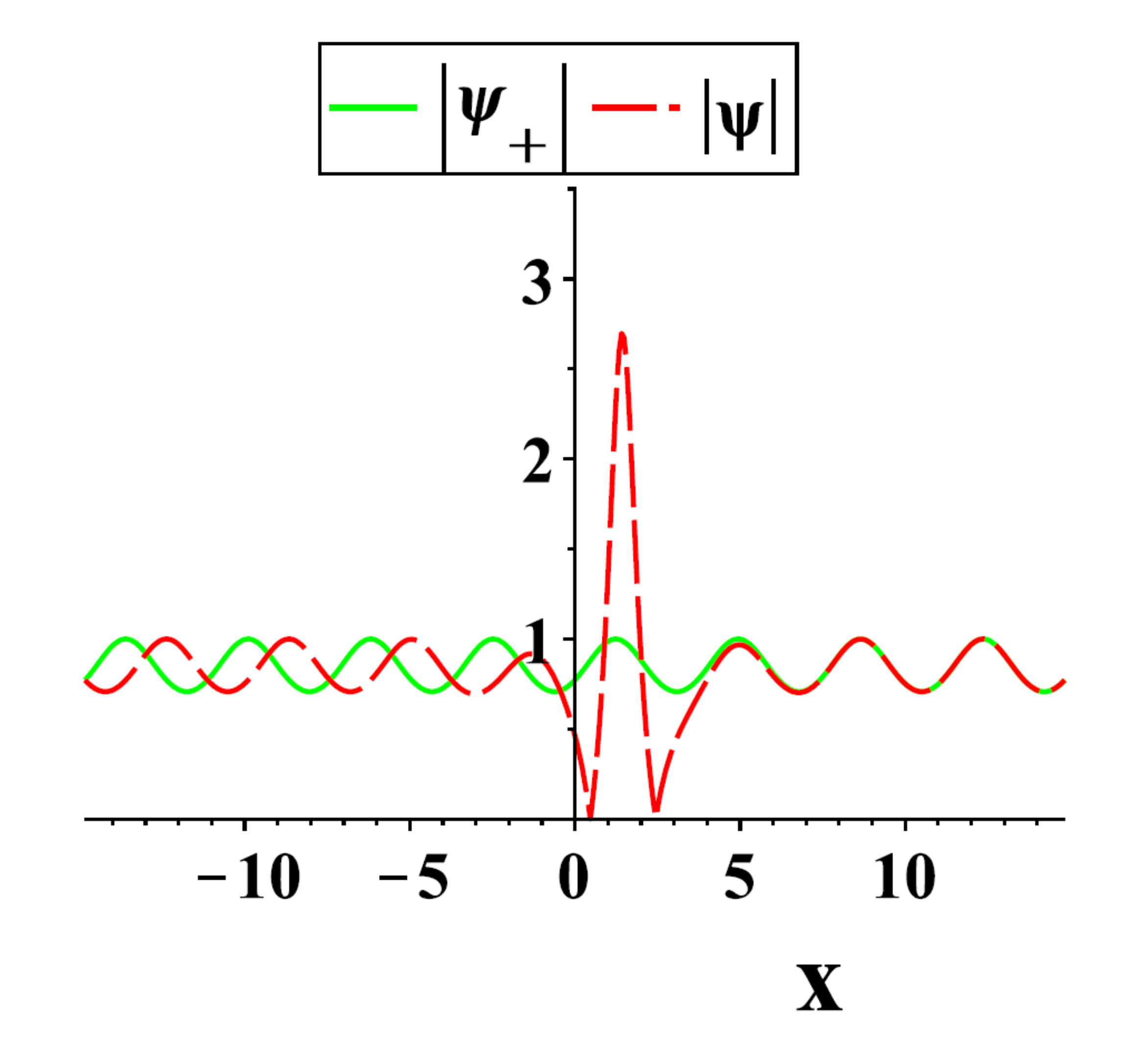} &  \includegraphics[width=.41\textwidth]{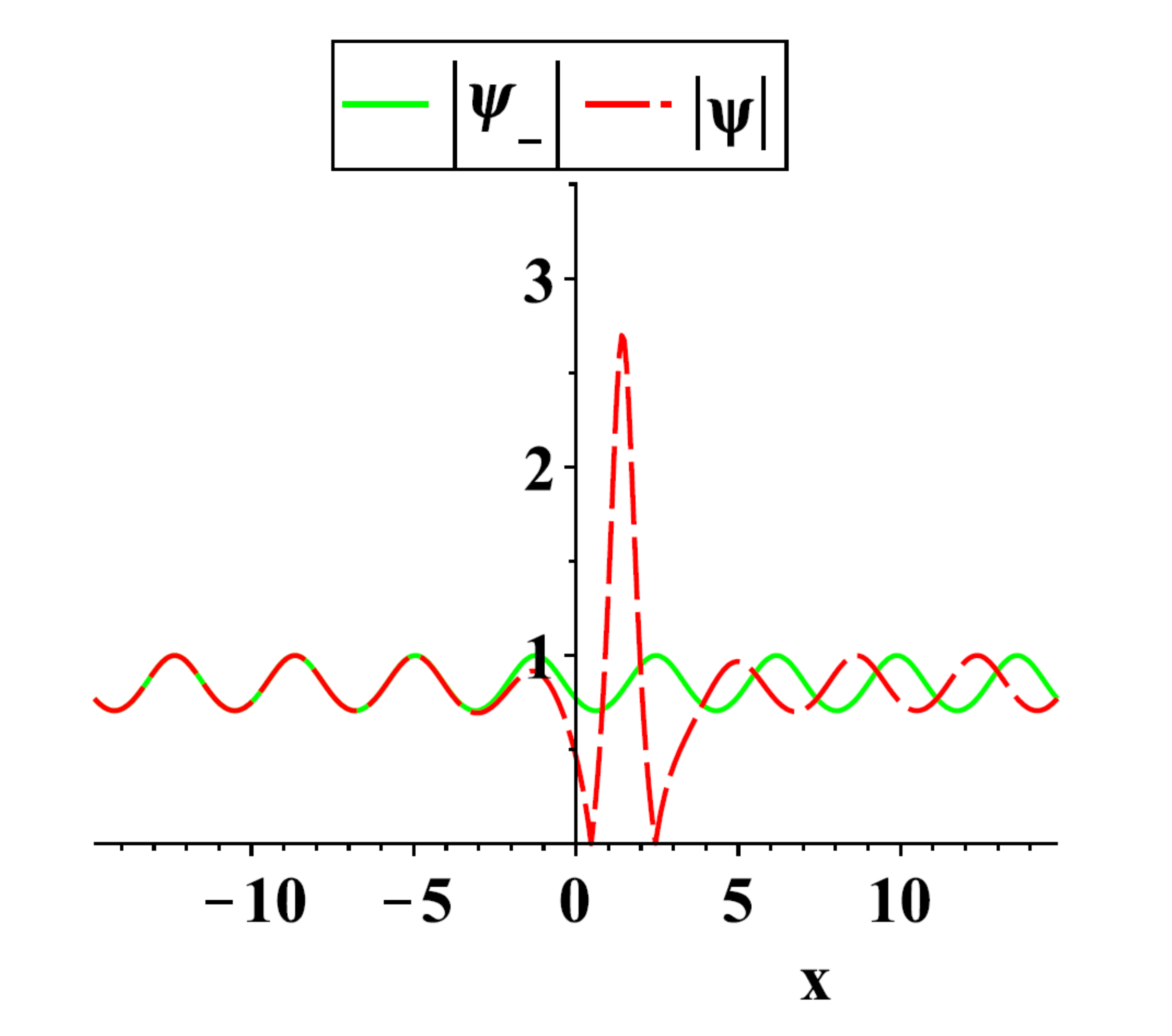} \\
{\footnotesize (a) right asymptotic form for single static breather} & {\footnotesize (b) left asymptotic form single static breather}
\end{tabular}
\caption{(color online): The asymptotic behaviors of the single breather solutions with dn background. (a) The asymptotic behavior for $x\to+\infty$. The red dashed line represents $\psi_1$ \eqref{eq:dnsingle} and the green solid one $\psi_{1+}$ \eqref{eq:asymplus}. It is seen that  $\psi_1$ tends to $\psi_{1+}$ in the limit $x\to+\infty$. (b) The asymptotic behavior for $x\to -\infty$. The green solid line represents $\psi_{1-}$ \eqref{eq:asymminus}. It is observed that $\psi_1$ tends to $\psi_{1-}$ for $x\to-\infty$.}
\label{fig:dn-b+-}
\end{figure}

In this case, we choose the parameters $l=\frac{K'}{2}$. By the shift formula (see  Appendix~\ref{sec:appB}), the single breather solution can be rewritten as the following form:
\begin{equation}\label{eq:dnsingle}
  \psi_1=\alpha\frac{\vartheta_4}{\vartheta_3}\frac{H_1}{N_1}\ee^{\ii s_2 t}
\end{equation}
where
\begin{equation*}
  \begin{split}
     H_1=&-\frac{\vartheta_4(\frac{\alpha x+\ii (z_1^*-z_1)}{2K})}{\vartheta_1(\frac{\ii (z_1^*-z_1)}{2K})}-\frac{\vartheta_4(\frac{\alpha\,x+\ii (z_1+z_1^*)-C}{2K})}{\vartheta_1(\frac{\ii(z_1+z_1^*)-C}{2K})}E_1+\frac{\vartheta_4(\frac{\alpha\,x+C-\ii (z_1+z_1^*)}{2K})}{\vartheta_1(\frac{C-\ii(z_1+z_1^*)}{2K})}E_1^*+\frac{\vartheta_4(\frac{\alpha x+\ii (z_1-z_1^*)}{2K})}{\vartheta_1(\frac{\ii (z_1-z_1^*)}{2K})}|E_1|^2,  \\
     N_1=&-\frac{\vartheta_3(\frac{\alpha x+\ii (z_1^*-z_1)}{2K})}{\vartheta_1(\frac{\ii (z_1^*-z_1)}{2K})}\frac{A_1}{A_1^*}\ee^{\frac{\pi(z_1^*-z_1)}{2K}}+\ii \frac{\vartheta_3(\frac{\alpha\,x+\ii (z_1+z_1^*)-C}{2K})}{\vartheta_1(\frac{\ii(z_1+z_1^*)-C}{2K})}\frac{E_1}{|A_1|^2}\ee^{\frac{\pi[(z_1^*+z_1)+\ii C]}{2K}}\\
     &+\ii \frac{\vartheta_3(\frac{\alpha\,x+C-\ii (z_1+z_1^*)}{2K})}{\vartheta_1(\frac{C-\ii(z_1+z_1^*)}{2K})}E_1^*|A_1|^2\ee^{-\frac{\pi[(z_1^*+z_1)+\ii C]}{2K}}+\frac{\vartheta_3(\frac{\alpha x+\ii (z_1-z_1^*)}{2K})}{\vartheta_1(\frac{\ii (z_1-z_1^*)}{2K})}\frac{A_1^*}{A_1}|E_1|^2\ee^{\frac{\pi(z_1-z_1^*)}{2K}}
  \end{split}
\end{equation*}
and $C=K+\ii K'$,
\begin{equation*}
  \begin{split}
      E_1&=\alpha_1 \exp(\alpha(Z(C-\ii(z_1+l))-Z(\ii(z_1-l)))x-2\ii y_1 t),\,\,\,\,\,\,\, \alpha_1\in \mathbb{C}, \\
      A_1&=\ee^{-\frac{\pi l}{2K}}\frac{\vartheta_2(\frac{\ii(z_1+l)}{2K})}{\vartheta_4(\frac{\ii(z_1-l)}{2K})},  \,\,\,\,\,\,
      y_1=\frac{\alpha^2}{4}({\rm dn}^2(C-\ii(z_1+l))-{\rm dn}(\ii(z_1-l))).
  \end{split}
\end{equation*}
Even though the breather solution above presents a compact form, it is not readily to obtain the properties from the above expression directly. The maximum of the modulus $|\psi_1|$ attains to $\max(|\psi_0|)+2|\textrm{Im}(\lambda(z_1))|$, where \[\lambda(z_1)=\frac{\alpha k^2 \ii}{2}\frac{{\rm sn}(\ii (z_1-l)){\rm cn}(\ii (z_1-l))}{{\rm dn}(\ii(z_1-l))}.\]
Through the expression, we find that the value of $|\psi_0|$ attains to the maximum at $x=2mK$, $m\in \mathbb{Z}.$ Thus the maximum of the breather solution can merely attain to the maximum value at these lines $x=2mK$, $m\in \mathbb{Z}.$ In what follows, we give two exact examples on the breather solutions. Firstly, we consider the stationary breather by choosing the parameters of background solution $\alpha=1$, $k=\sqrt{m}=\frac{\sqrt{2}}{2}$. It follows that $\nu_1=0$, $\nu_2=\frac{1}{2}$, $\nu_3=1$, $s_2=\frac{1}{2}(\nu_1+\nu_2+\nu_3)=\frac{3}{4}$. By taking the parameters of breathers $z_1=-\frac{K'(1/2)}{2}-\frac{K(1/2)\ii}{3}$ and $\alpha_1=-1$, we plot Fig.~\ref{fig:dn-b}. For these values, we find the parameters $\lambda_{1,3}\approx\pm \ii 0.854$ and $\lambda_{2,4}\approx \pm \ii 0.146$. Furthermore, the parameter $y_1\approx0.476$ infers that the period of breather in the direction of $t$ is $T=\frac{\pi}{|{\rm Im}(y_1)|}\approx 6.602.$ The parameters $\lambda(z_1)\approx 0.984\ii$ and $\alpha_1=-1$ can deduce that the maximum of the breather is $1+2{\rm Im}(\lambda(z_1))=2.967$ at $x=0$ and $t=mT$, $m\in \mathbb{Z}.$ The parameter $\alpha (Z(C-\ii(z_1+l))-Z(\ii(z_1-l)))\approx -0.978$ shows that $|E_1|\to \infty$ as $x\to -\infty$ and $|E_1|\to 0$ as $x\to +\infty.$ It follows that
\begin{equation}\label{eq:asymplus}
\psi_1\to \psi_{1+}\equiv \alpha\frac{\vartheta_4}{\vartheta_3}\frac{\vartheta_3(\frac{\alpha x-\frac{2}{3}K}{2K})}{\vartheta_4(\frac{\alpha x-\frac{2}{3}K}{2K})}\frac{A_1}{A_1^*} \ee^{\frac{\ii \pi}{3}+\ii\frac{3}{4} t},\,\,\,\,\,\,\, \text{as  } x\to +\infty,
\end{equation}
and
\begin{equation}\label{eq:asymminus}
\psi_1\to \psi_{1-}\equiv\alpha\frac{\vartheta_4}{\vartheta_3}\frac{\vartheta_3(\frac{\alpha x+\frac{2}{3}K}{2K})}{\vartheta_4(\frac{\alpha x+\frac{2}{3}K}{2K})}\frac{A_1^*}{A_1} \ee^{-\frac{\ii \pi}{3}+\ii\frac{3}{4} t},\,\,\,\,\,\,\, \text{as  } x\to -\infty,
\end{equation}
where
\[
A_1=\frac{\vartheta_2(\frac{1}{6})}{\vartheta_4(\frac{1}{6}-\frac{3\tau}{4})},\,\,\,\,\,\,\,\, \tau=\ii\frac{K'}{K}.
\]
The above asymptotic behaviors can be seen and verified in Fig.~\ref{fig:dn-b+-}.

Now we consider the non-stationary breather solution. In this case, we just illustrate the solutions by graphs. The properties are similar as the stationary breathers (Fig.~\ref{fig:dn-b} (c)). So we skip the detailed analysis of their dynamics.

\subsubsection{\textbf{Example 2: The breather solutions on the cn solution background.}}
\begin{figure}[tb]
\centering
\begin{tabular}{ccc}
\includegraphics[width=.32\textwidth]{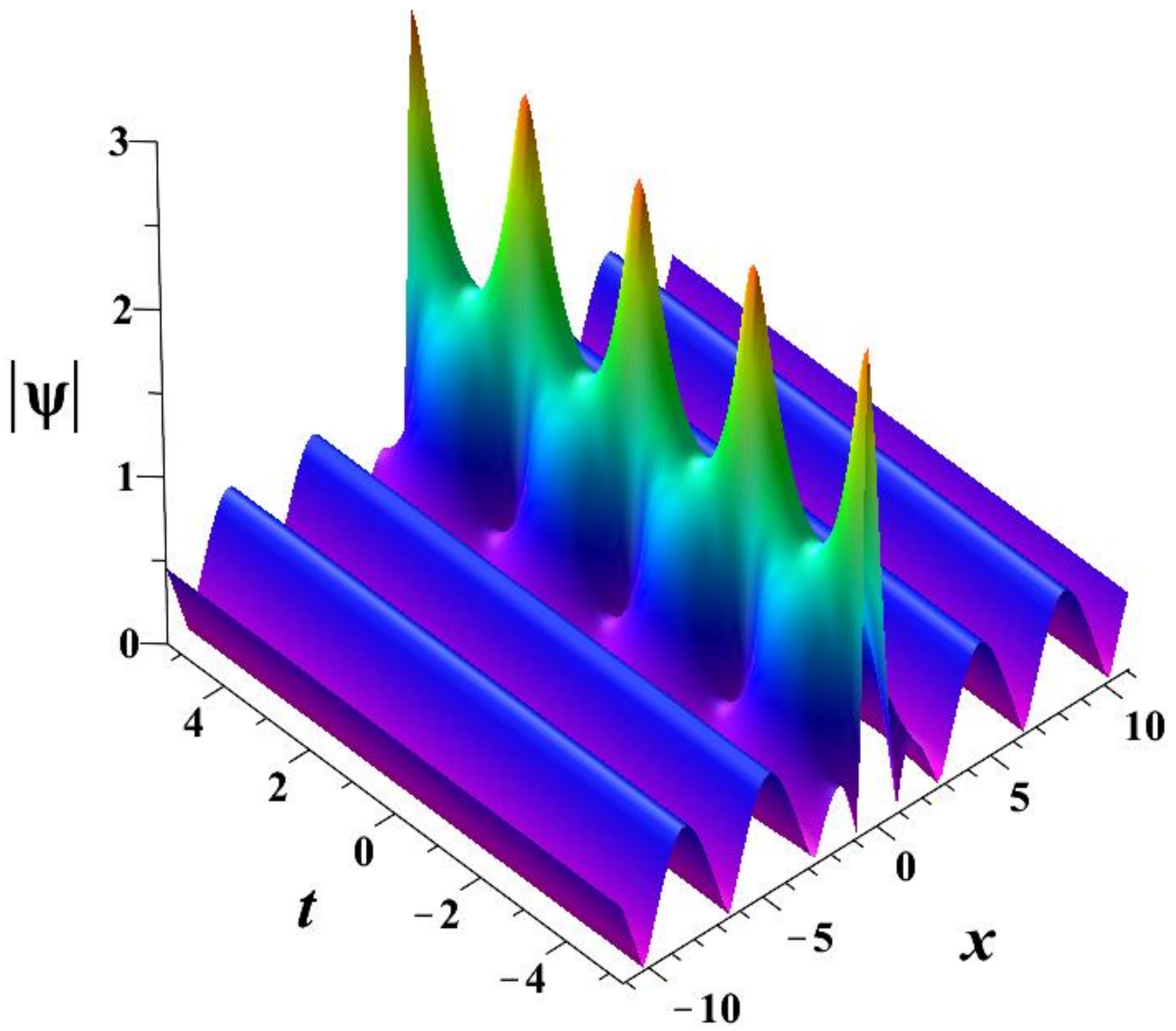}\!\!\! & \includegraphics[width=.32\textwidth]{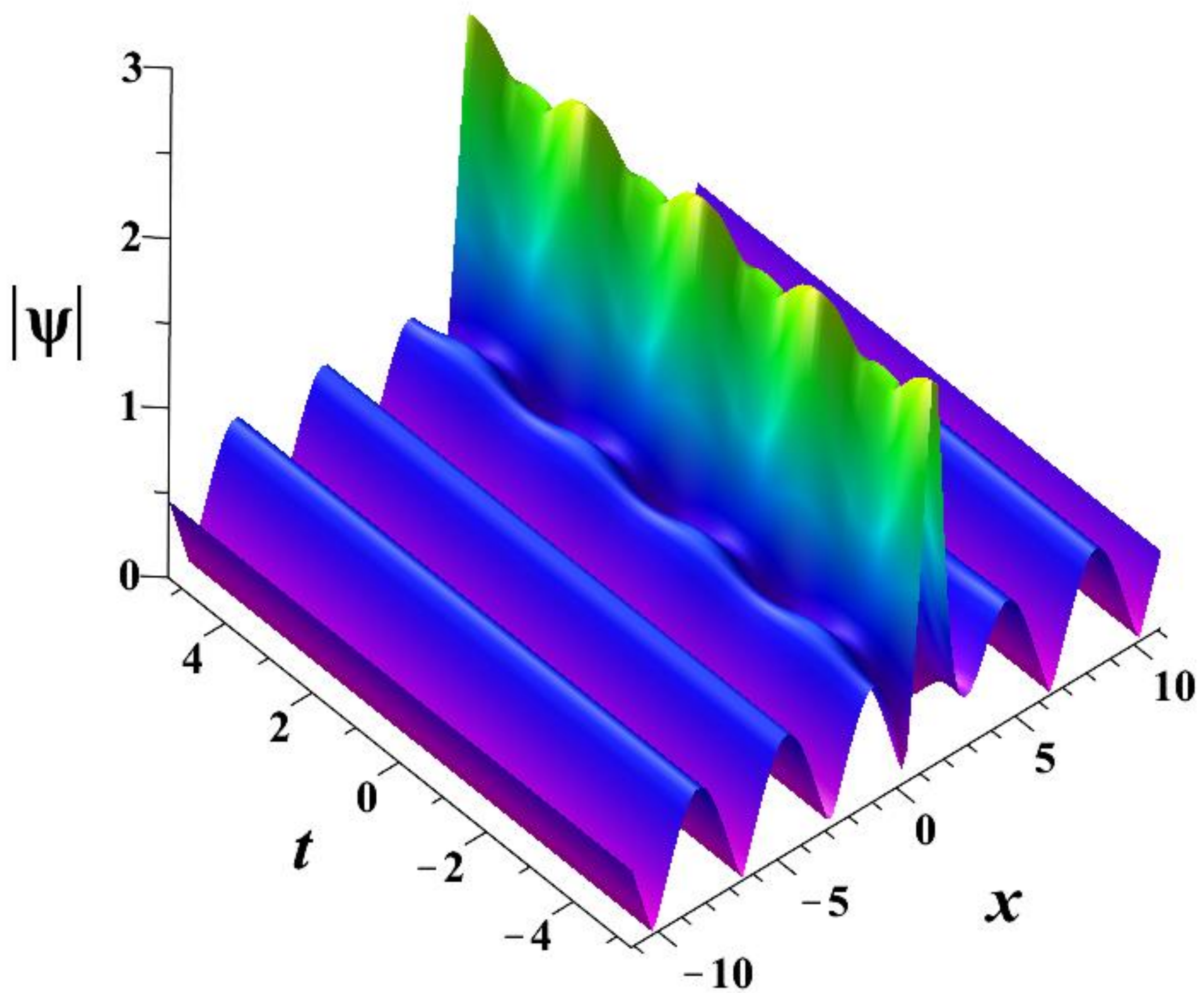}\!\!\! & \includegraphics[width=.32\textwidth]{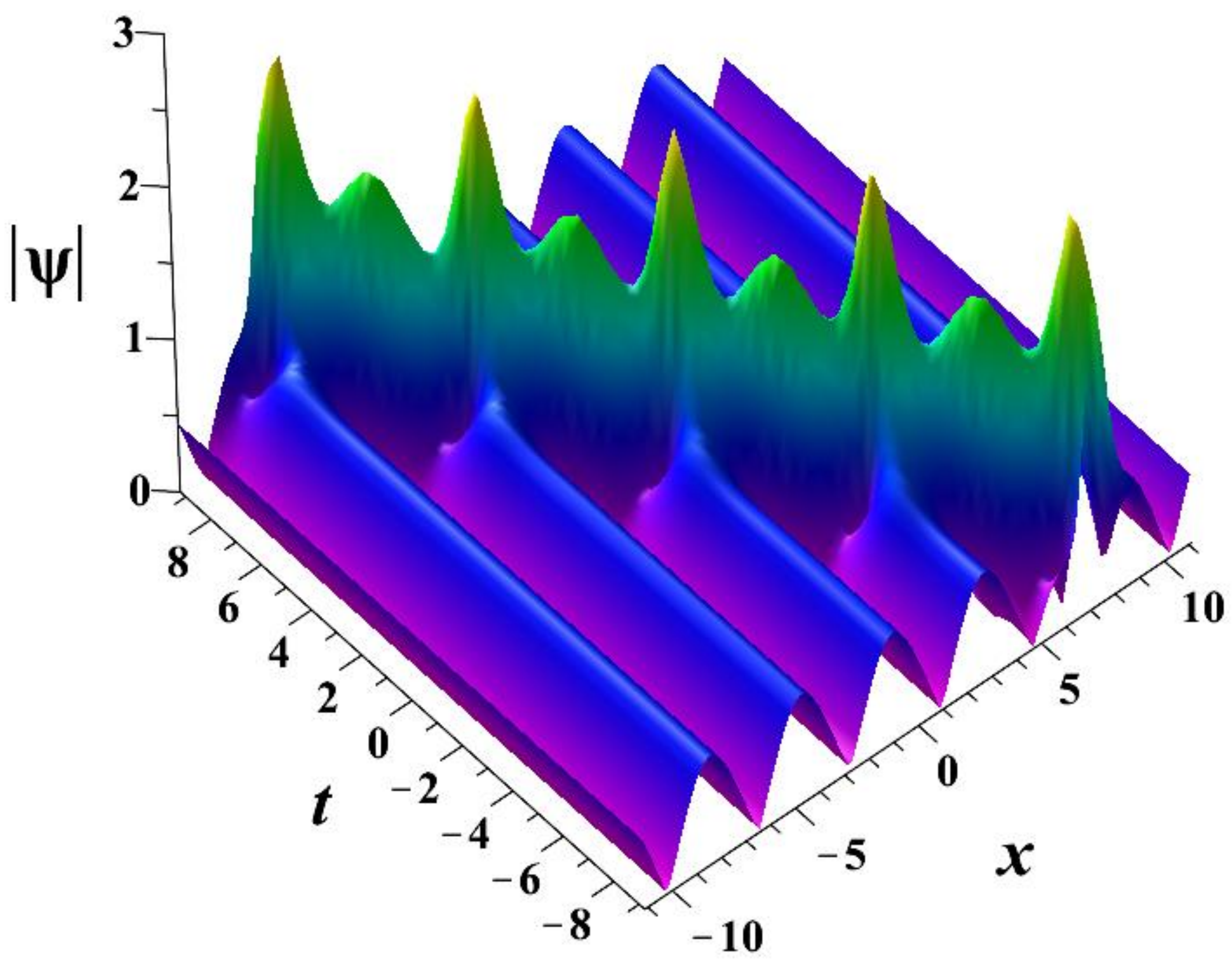}\!\!\! \\
\!\!\!\!\!\!\!{\footnotesize (a) single static cn breather}\!\!\! & {\footnotesize (b) another single static cn breather}\!\!\! & {\footnotesize (c) single non-static cn breather}\!\!\!
\end{tabular}
\caption{(color online): The single breather solutions with cn background, obtained by \eqref{eq:n-breather} with $n=1,\ l=0,\ \alpha=1$, and $k=1/\sqrt{2}$. (a) The single static breather solution with $z_1=-K(\frac{1}{2})+\frac{\ii }{4}K'(\frac{1}{2})$ and $\alpha_1=1$. (b) The same as (a), but we set $\alpha_1=\frac{1}{20}$. (c) The single non-static breather. The parameters are $z_1=-\frac{9}{10}K(1/2)+\frac{\ii}{4}K'(1/2)$ and $\alpha_1=1$.}
\label{fig:cn-non-static}
\end{figure}

Choosing the parameter $l=0$ and $n=1$ in \eqref{eq:n-breather}, we obtain the one-breather solution with  ${\rm cn}$ background. Fixing the parameters of background solution $\alpha=1$, $k=\sqrt{m}=\frac{1}{\sqrt{2}}$, we have $\nu_1=-\frac{1}{2}$, $\nu_2=0$, $\nu_3=\frac{1}{2}$, $s_2=0$. By taking the parameters $\alpha_1=1$, $z_1=-K(1/2)+\ii \frac{K'(1/2)}{4}$, we obtain the plot Fig.~\ref{fig:cn-non-static}(a). For these parameters, we obtain $\lambda(z_1)=-1.07\ii$, $y(z_1)\approx 1.19$, and the maximum value is about $2.85$ at the origin. Changing the parameters $\alpha_1=1/20$, we obtain the breather with different shape (Fig.~\ref{fig:cn-non-static}(c)). If we choose the parameters $z_1=-\frac{9}{10}K(1/2)+\frac{K'(1/2)\ii}{4}$ and $\alpha_1=1$, the non stationary breather is obtained (figure \ref{fig:cn-non-static}(b)). The maximum value is about $2.56$ at the origin. Comparing Figs. \ref{fig:dn-b} and \ref{fig:cn-non-static}, we see that the two kinds of breathers with different backgrounds possess evidently different dynamics.

\subsubsection{\textbf{Example 3: The breather solutions on the nontrivial-phase solution background.}} 
\begin{figure}[tb]
\centering
\subfigure[single static nontrivial-phase breather]{%
\includegraphics[width=.47\textwidth]{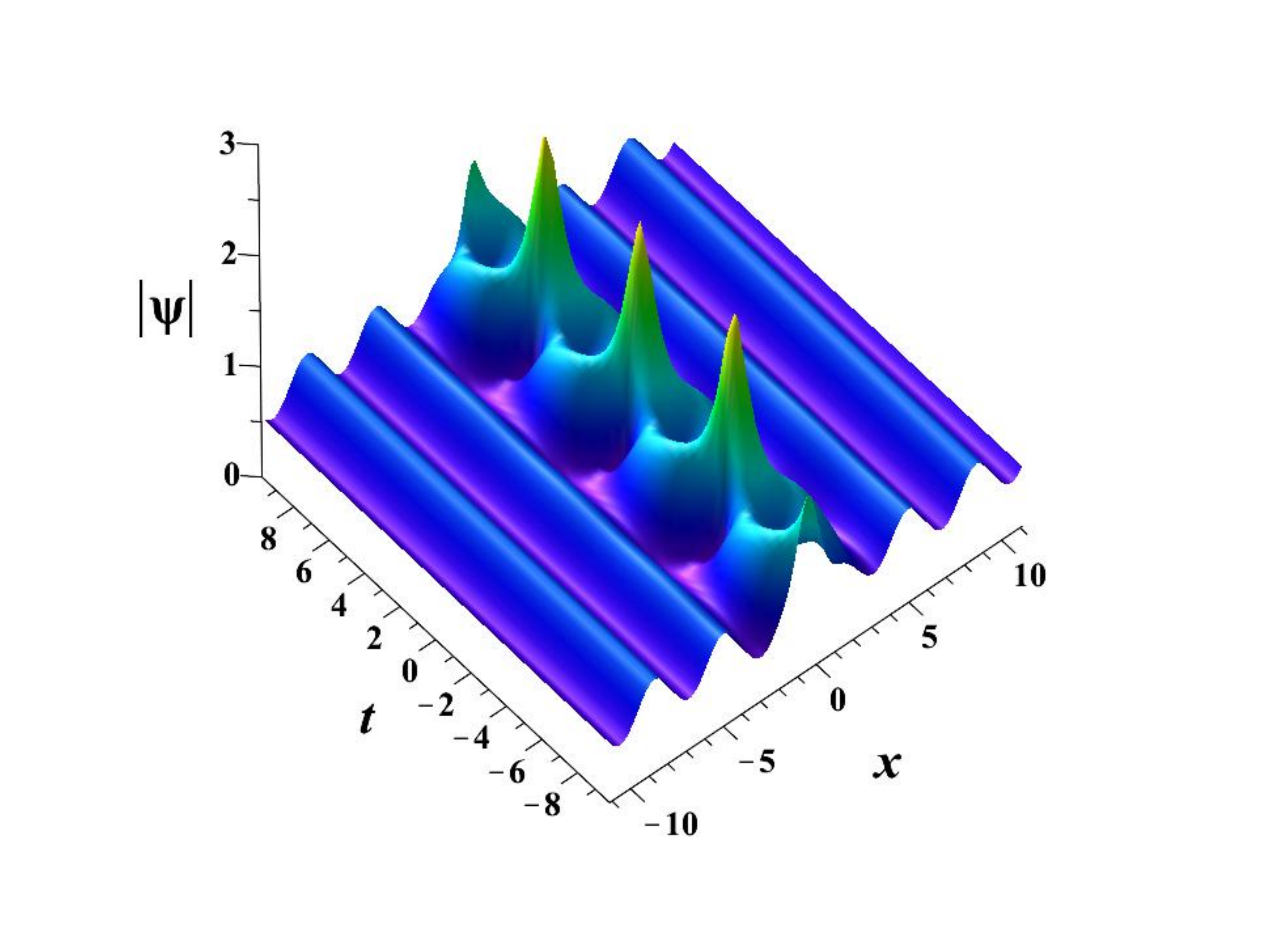}} \hfil
\subfigure[single non-static nontrivial-phase breather]{%
\includegraphics[width=.43\textwidth]{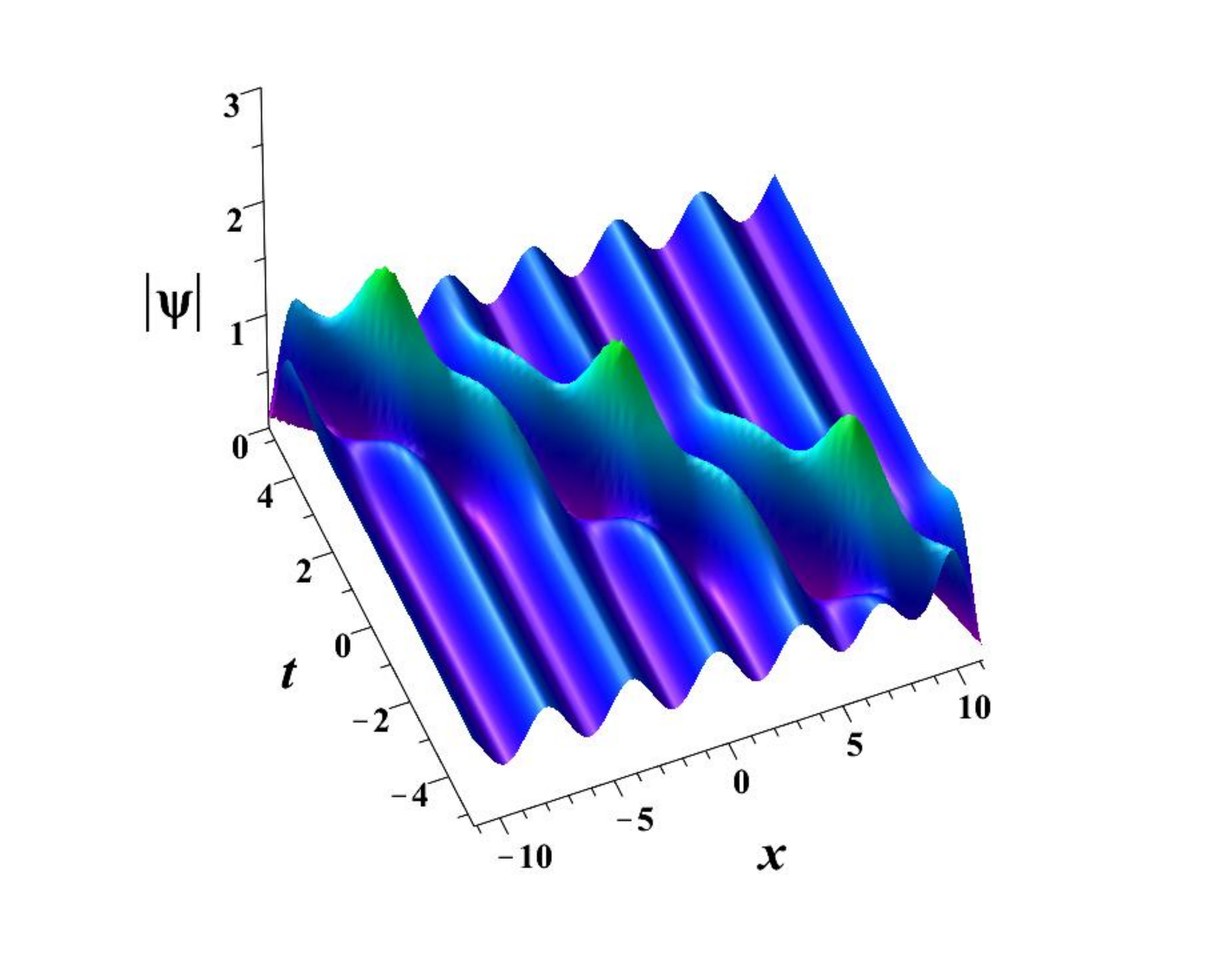}}
\caption{(color online): The single breather solutions with non-trivial phase oscillating background. (a) The single static nontrivial-phase breather. (b) The single non-static nontrivial-phase breather. The details of parameters are described  in the paragraph of Example 3.}
\label{fig:phase-non-static}
\end{figure}

Here we consider the case where the elliptic-function background has phase modulation.
In particular, we choose the parameters $k=\sqrt{m}=\frac{1}{\sqrt{2}}$, $l=\frac{K(1/2)}{4}$, $\alpha=1$, which can deduce that $\nu_1=-\alpha^2 {\rm dn}^2(K(1/2)+2\ii l)$, $\nu_2=-k^2\alpha^2 {\rm cn}^2(K(1/2)+2\ii l)$, $\nu_3=k^2\alpha^2 {\rm sn}^2(K(1/2)+2\ii l)$, $s_2=\frac{1}{2}(\nu_1+\nu_2+\nu_3)$. By choosing the parameter $z_1=-0.5+1.223\ii$, $\alpha_1=1$, we can plot the figure of breather(Fig.~ \ref{fig:phase-non-static} (a)). Moreover, $\lambda(z_1)=0.087-0.852\ii$, $y_1\approx-0.586$. The period of solution is $T\approx 5.35$. The maximum value is about $2.546$ at the origin.

For the second case, by choosing the parameters of breather $z_1=-\frac{K(1/2)}{2}+\frac{K'(1/2)\ii}{8}$, $\alpha_1=1$, which infers that $\lambda(z_1)\approx 0.797-0.480\ii$, $y_1=-0.665+0.719\ii$, we can plot the breathers(Fig.~\ref{fig:phase-non-static} (b)). The maximum value of breather is about $1.801$.

\subsubsection{\textbf{Example 4: Two-breather solutions.}}
Now we consider the two-breather solutions and its analysis. Meanwhile, we use the numeric way to verify the above asymptotic analysis.

\begin{figure}[tb]
\centering
\subfigure[two-cn-breather]{%
\includegraphics[width=.47\textwidth]{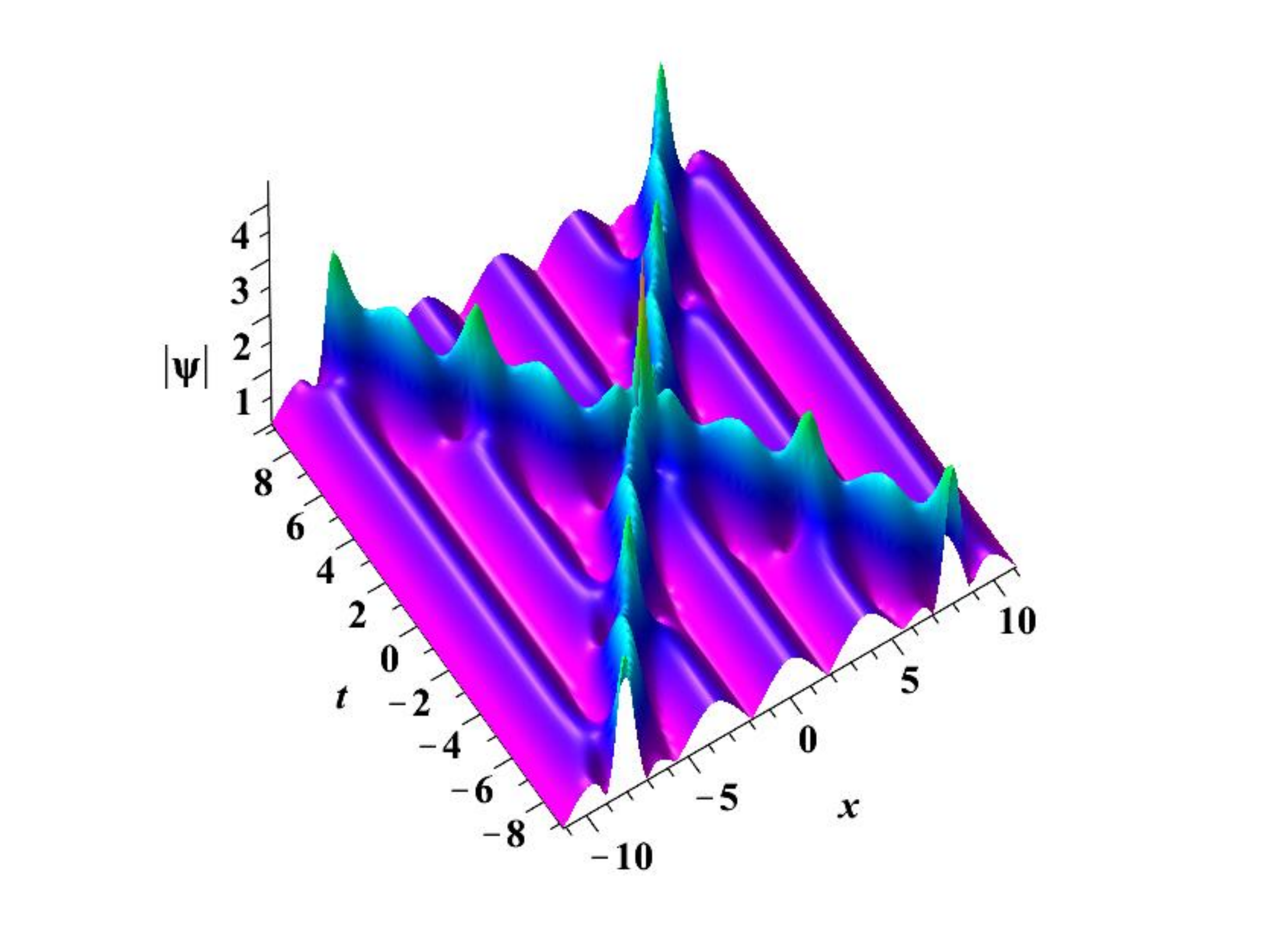}} \hfil
\subfigure[two non-trivial-phase breather]{%
\includegraphics[width=.48\textwidth]{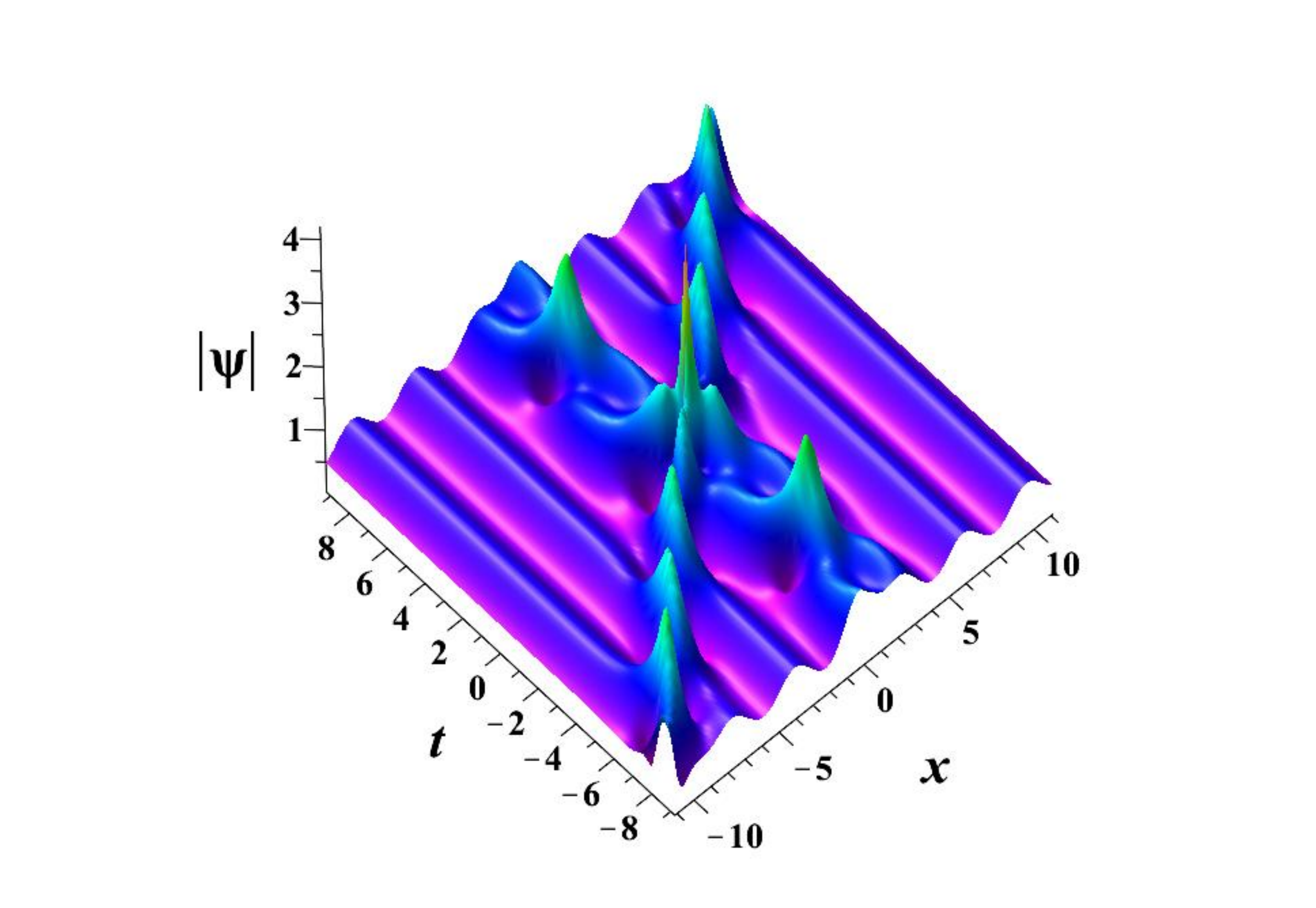}}
\caption{(color online): The two breather solutions with various kinds of backgrounds. (a) The two-cn-breather. (b) The two non-trivial-phase breather. The parameters for these solutions are given in the paragraph of Example 4.}
\label{fig:two-breathers}
\end{figure}

We consider the two-breather solution with ${\rm cn}$ background in detail. We choose the parameters of the background solution $l=0$, $\alpha=1$, $k=\frac{1}{\sqrt{2}}$, which determine the parameter $s_2=0$ and the background solution. Taking the parameters of breathers as $z_1=-\frac{9K(1/2)}{10}+\frac{K'(1/2)\ii}{4}$, $z_2=\frac{9K(1/2)}{10}+\frac{K'(1/2)\ii}{4}$, $\alpha_1=\alpha_2=1$, we can obtain the two breathers (figure \ref{fig:two-breathers} (a)). The maximum value is about $4.416$ at the origin.

\begin{figure}[tb]
\centering
\subfigure[Asymptotic form for the right breather at $t=-10$]{
\includegraphics[width=.4\textwidth]{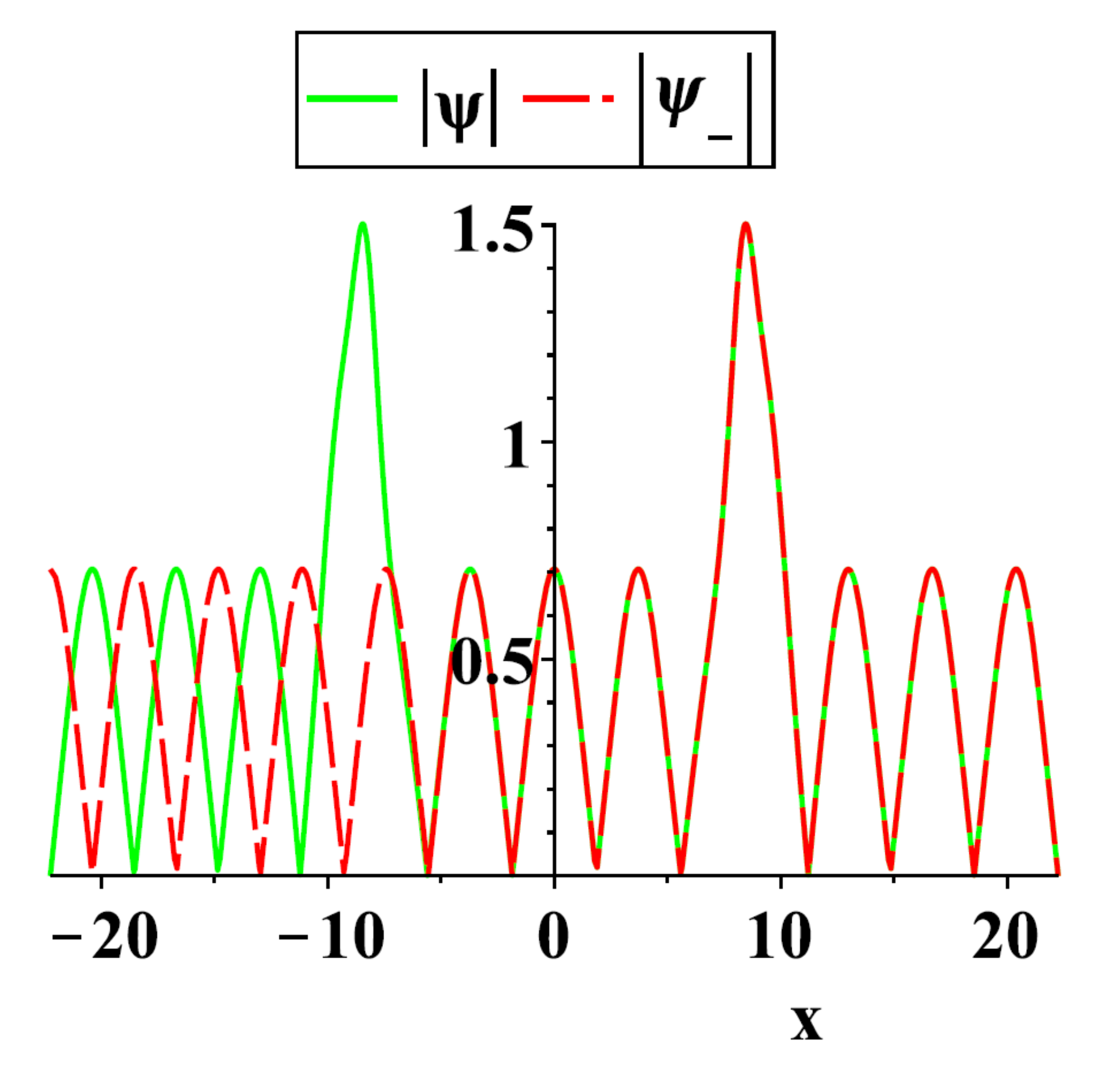}} \hfil
\subfigure[Asymptotic form for the left breather at $t=-10$]{
\includegraphics[width=.4\textwidth]{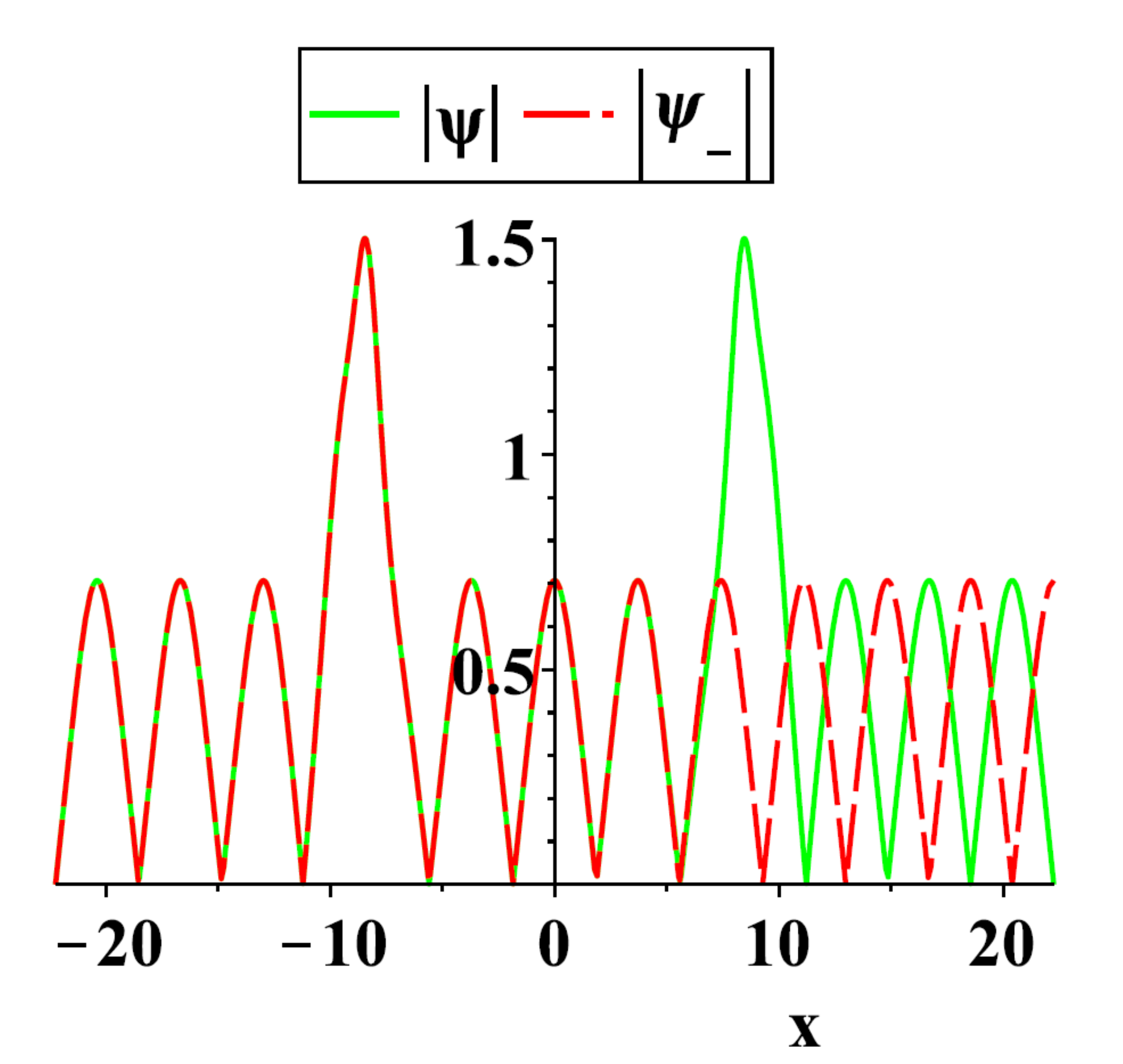}} \\
\subfigure[Asymptotic form for the left breather at $t=+10$]{
\includegraphics[width=.4\textwidth]{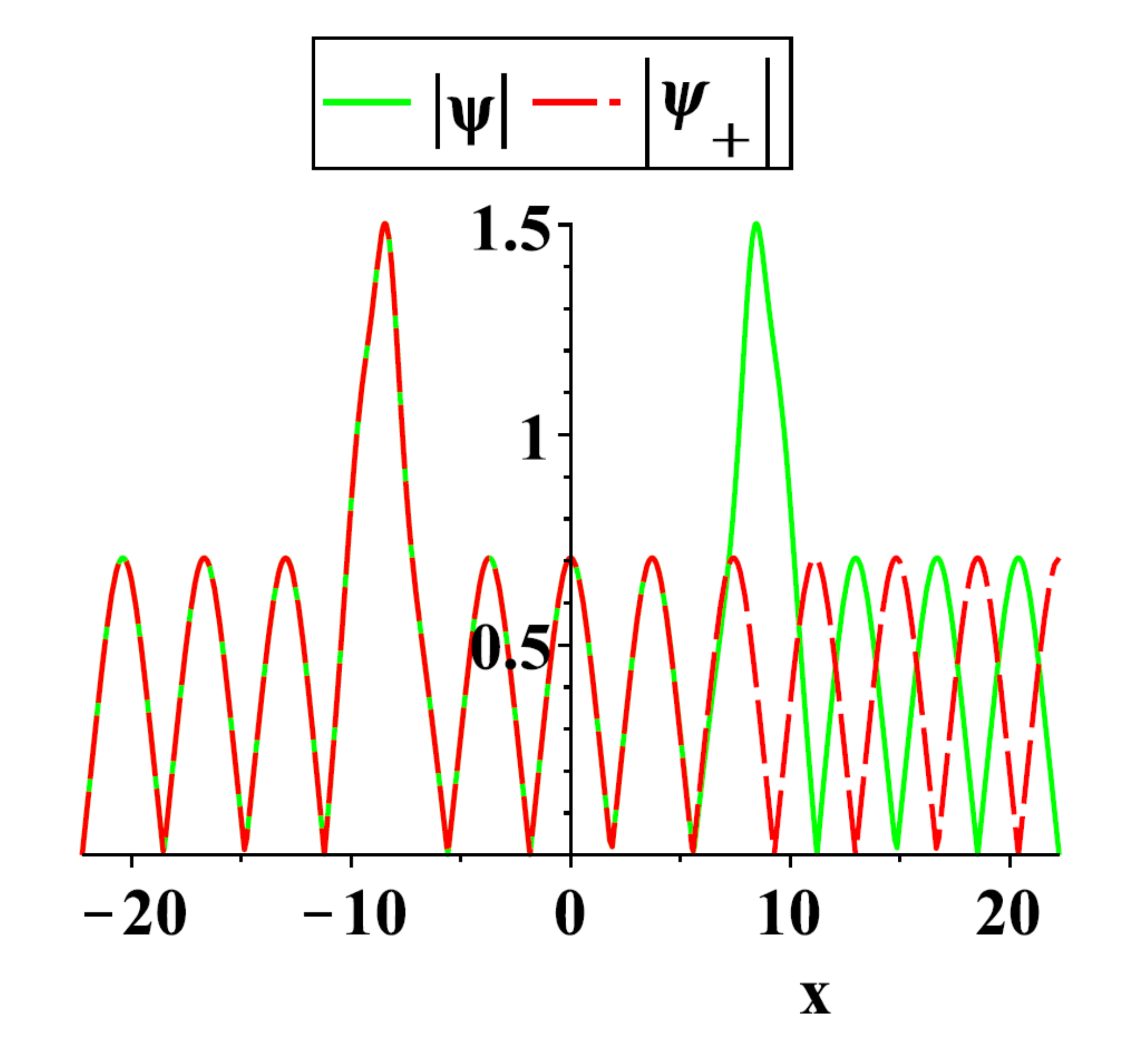}} \hfil
\subfigure[Asymptotic form for the right breather at $t=+10$]{
\includegraphics[width=.4\textwidth]{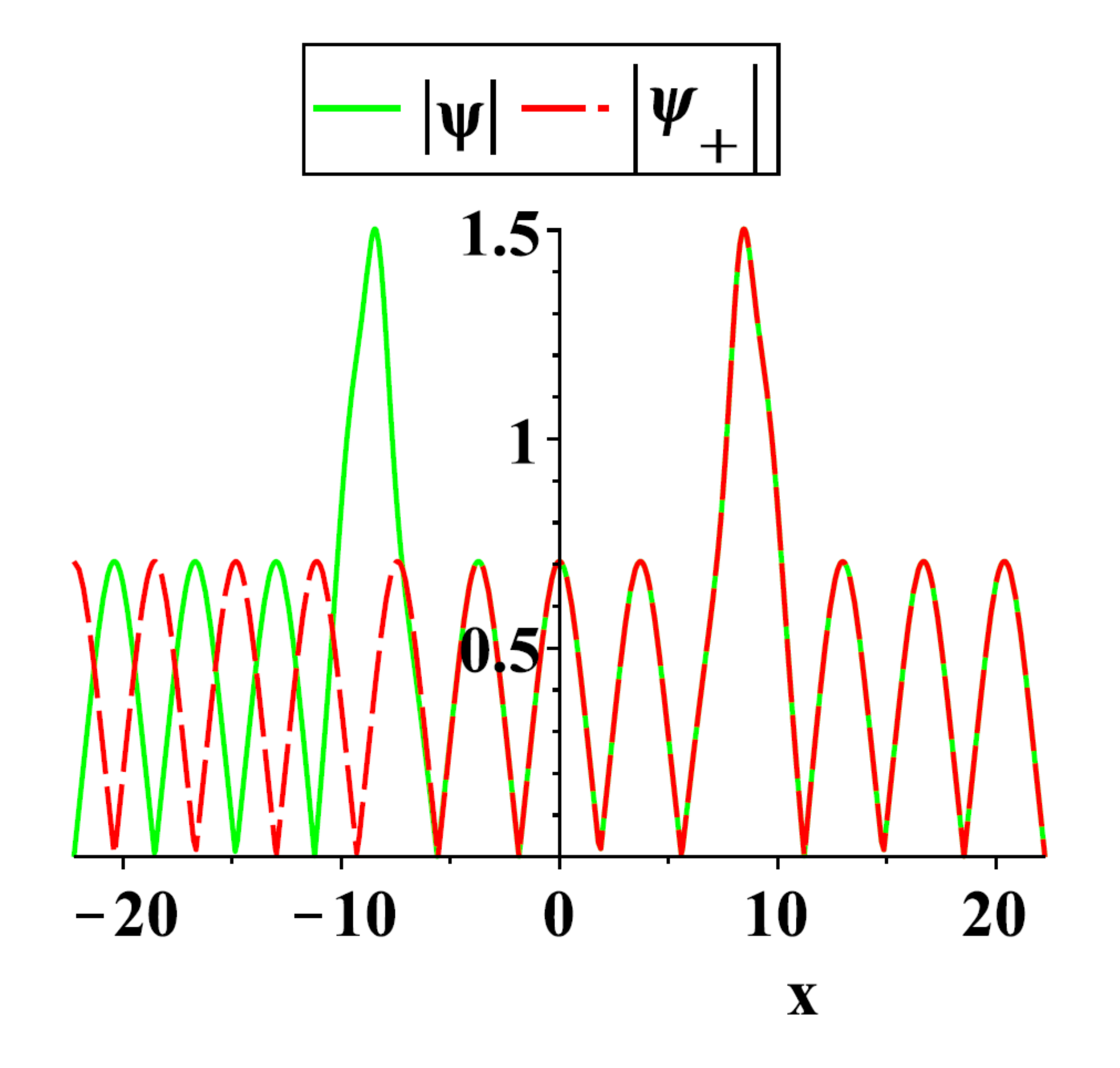}}
\caption{(color online): The asymptotic behaviors for the two-breather solution with cn background, where the solution behaves as an isolated one-breather solution in each region. Here, the same parameters as Fig.~\ref{fig:two-breathers}(a) are used.  (a) The green solid line  $|\psi|$ means the full exact two breather solutions at time $t=-10$, and the red dashed line $|\psi_-|$ represents the asymptotic single breather with negative velocity at time $t=-10$ ($\psi_{1}^{[-\infty]}$ in Theorem~\ref{thm4}). (b) $|\psi|$ is the same as (a), and the red dashed line $|\psi_-|$ represents the single breather with positive velocity at time $t=-10$ ($\psi_{2}^{[-\infty]}$ in Theorem~\ref{thm4}). (c) $|\psi|$ represents the two-breather solution at $t=+10$, and $|\psi_+|$ represents the single breather with negative velocity at time $t=+10$ ($\psi_{1}^{[+\infty]}$ in Theorem \ref{thm4}). (d) $|\psi|$ is the same as (c), and $|\psi_+|$ represents the single breather with positive velocity at time $t=+10$ ($\psi_{2}^{[+\infty]}$ in Theorem \ref{thm4}).}

\label{fig:two-breathers-}
\end{figure}

We illustrate the asymptotic analysis on the two cn breathers, i.e two breather solution with cn elliptic function background. It is seen that the asymptotic expression for $t\to-\infty$, which are obtained in Theorem \ref{thm4}, is consistent with two cn breathers [Fig.~\ref{fig:two-breathers-}(a) and (b)].
The asymptotic behaviors for $t \to +\infty$ also show good agreement [Fig.~\ref{fig:two-breathers-}(c) and (d)].

As another case, we consider the two-breather solution on the non-trivial phase background. By taking the parameters $k=\frac{1}{\sqrt{2}}$, $l=\frac{K(1/2)}{4}$ and $\alpha=1$, we obtain $\nu_1=-{\rm dn}^2(K(1/2)+2\ii l)$, $\nu_2=-\frac{1}{2}{\rm cn}^2(K(1/2)+2\ii l)$, $\nu_3=\frac{1}{2}{\rm sn}^2(K(1/2)+2\ii l)$ and the background solution. On the other hand, taking the parameters $\alpha_1=\alpha_2=1$, $z_1=-\frac{9K(1/2)}{10}+\frac{K'(1/2)\ii}{4}$, $z_2=-0.5+1.2335\ii$, we can plot two breathers (Fig.~\ref{fig:two-breathers} (b)). The maximum value is about $4.198$ located at the origin. The interaction can be obtained in the above and be verified as the cn case too.

\section{Modulational instability and Akhmediev breathers on the elliptic function background}\label{sec5}
Based on the linearized NLSE \eqref{eq:linear-nlse}, we analyze the spectral stability for the elliptic-function solution of NLSE. As shown previously, the elliptic-function solution can be represented as an abstract form $\psi=u(x)\ee^{\ii s_2t}$. Let us assume that the perturbed solution has the form
\begin{equation}\label{eq:perturbation}
p(x,t)=\ee^{\ii s_2 t}(p_1(x)\ee^{\ii\Omega t}+p_2^*(x)\ee^{-\ii\Omega^* t}),
\end{equation}
then the linearized equation is converted into the following spectral problem:
\begin{equation}\label{eq:per-spect}
\begin{bmatrix}
\frac{1}{2}\partial_x^2+2|u(x)|^2-s_2& u(x)^2\\
u(x)^{*2}&\frac{1}{2}\partial_x^2+2|u(x)|^2-s_2\\
\end{bmatrix}\begin{bmatrix}
p_1(x)\\
p_2(x)
\end{bmatrix}=\Omega \begin{bmatrix}
p_1(x)\\
-p_2(x)
\end{bmatrix}.
\end{equation}
With the aid of the stationary Lax pair, we can solve the linear spectral problem with stationary periodic solution exactly. Thus the spectrum for the spectral problem can be determined.

As shown in Sec.~\ref{subsec:linearizednlse},  $p(x,t)=\phi_1(x,t;\lambda)\phi_2(x,t;\lambda)+\varphi_1^*(x,t;\lambda)\varphi_2^*(x,t;\lambda)$ solves the linearized perturbed equation. Together with the vector solution for the Lax pair, we can set them as the form of equation \eqref{eq:perturbation}.  It follows that
\begin{equation}\label{eq:roots-1}
\begin{split}
p_1(x)=&\frac{1}{\vartheta_4^2(\frac{\alpha\,x}{2K})}\frac{\vartheta_4^2(\frac{-\alpha x+\ii(z-l)}{2K})}{\vartheta_4^2(\frac{\ii(z-l)}{2K})} \ee^{2[\alpha Z(\ii(z-l))+\ii\lambda]x}, \\
p_2(x)=&-\frac{1}{\vartheta_4^2(\frac{\alpha\,x}{2K})}\frac{\vartheta_4^2(\frac{\alpha x+K+\ii K'-\ii(z-l)}{2K})}{\vartheta_4^2(\frac{K+\ii K'-\ii(z+l)}{2K})} \ee^{2[\alpha Z(\ii(z-l))+\ii\lambda]x+2\alpha[Z(2\ii l+K)+\frac{\ii \pi}{2K}]x}, \\
\end{split}
\end{equation}
with $\Omega=2y;$ or
\begin{equation}\label{eq:roots-2}
\begin{split}
p_1(x)&=-\frac{1}{\vartheta_4^2(\frac{\alpha\,x}{2K})}\frac{\vartheta_1^2(\frac{-\alpha\, x+K+\ii K'-\ii (z+l)}{2K})}{\vartheta_4^2(\frac{K+\ii K'-\ii (z+l)}{2K})}{\rm e}^{2[\alpha Z(K+\ii K'-\ii (z+l))+\ii \lambda]x},\\
p_2(x)&= \frac{1}{\vartheta_4^2(\frac{\alpha\,x}{2K})}\frac{\vartheta_1^2(\frac{\ii (z-l)+\alpha\,x}{2K})}{\vartheta_4^2(\frac{\ii (z-l)}{2K})}{\rm e}^{2[\alpha Z(K+\ii K'-\ii (z+l))+\ii \lambda]x+2\alpha [Z(2\ii l+K)+\frac{\ii \pi}{2K}]x},  \\
\end{split}
\end{equation}
with $\Omega=-2y.$ We have four different roots $\lambda=\xi_i$, $i=1,2,3,4$ for fixed $\Omega.$ By the transformation $\lambda(z)=\mu(\ii(z-l)/\alpha)$, we know that this establishes the corresponding between the Riemann surface and the torus. Thus we have four different points $\mu(\ii (z_i-l)/\alpha)=\xi_i.$ Inserting the four different roots $\xi_i$s into equations \eqref{eq:roots-1} and \eqref{eq:roots-2}, we obtain four independent solutions for the linearized spectral problem \eqref{eq:per-spect}. So the spectral stability can be obtained by the analysis of solutions $p_1(x)$, $p_2(x)$, the bound properties of which will determine the spectrum of spectral problem \eqref{eq:per-spect}. Moreover, the coefficient of the exponential function in $p_1(x)$ and $p_2(x)$, which may be interpreted as a crystal momentum, determines the spectrum: i.e. $\textrm{Re}(\alpha Z(\ii(z-l))+\ii\lambda)=0$ or $\textrm{Re}(\alpha Z(K+\ii K'-\ii(z+l))+\ii\lambda)=0$. We know the determinant of solution $\Phi(x,t;\lambda)$ is a constant, so that $\textrm{Re}(\alpha Z(\ii(z-l))+\alpha Z(K+\ii K'-\ii(z+l))+2\ii\lambda)=0$. Then the spectral condition becomes $\textrm{Re}(Z(\ii(z-l))-Z(K+\ii K'-\ii(z+l)))=0.$ Following the method proposed by Deconinck and Segal \cite{DeconinckS17}, the spectrum can be obtained numerically by the tangent vector field
\begin{equation}\label{eq:tangent}
  v_{T}(z)=\left(\textrm{Im}[\ii({\rm dn}(\ii (z-l))+{\rm dn}(K+\ii K'-\ii (z+l)))], \textrm{Re}[\ii({\rm dn}(\ii (z-l))+{\rm dn}(K+\ii K'-\ii (z+l)))]\right)
\end{equation}
and the starting point $z=\frac{1}{2}(K'\pm \ii K)$ and $z=\frac{1}{2}(-K'\pm \ii K)$. By the formula $\lambda(z)=\mu(\ii(z-l)/\alpha)$, this determines the spectrum in the $\lambda$ plane. By the theory of Riemann surfaces, we know that the Abel map
\begin{equation}\label{eq:abel}
z-l=2\alpha\,K\mathbf{A}(\lambda)=\frac{\alpha}{2}\int_{0}^{\lambda}\frac{{\rm d}\mu}{\sqrt{(\mu-\lambda_1)(\mu-\lambda_1^*)(\mu-\lambda_2)(\mu-\lambda_2^*)}} \end{equation}
establishes the conformal mapping between the genus $1$ Riemann surface $\mathbb{S}$ and the rectangular region $\{z| -K'\leq {\rm Im}(z-l)\leq K',\,\,\,\, -K\leq {\rm Re}(z-l)\leq K\}$, by fixed the integral path in the Riemann surface, i.e., the relation \eqref{eq:mu(iz)}. Through the above analysis, the spectrum with respect to $\lambda$ can be obtained by the relation \eqref{eq:mu(iz)}. Also, the spectrum with respect to $\Omega=\pm2y$ can be obtained via \eqref{eq:dmu(iz)}.

In this section, our main goal is not to analyze the stability properties for the NLSE, but to construct a special solution which can describe the modulational instability. In the plane wave background, this type of solution is called Akhemediev breather, so we also use the same name in the case of the elliptic backgrounds.

\begin{figure}[tbh]
\centering
\subfigure[unstable spectrum in $z$ plane]{%
\includegraphics[width=.32\textwidth]{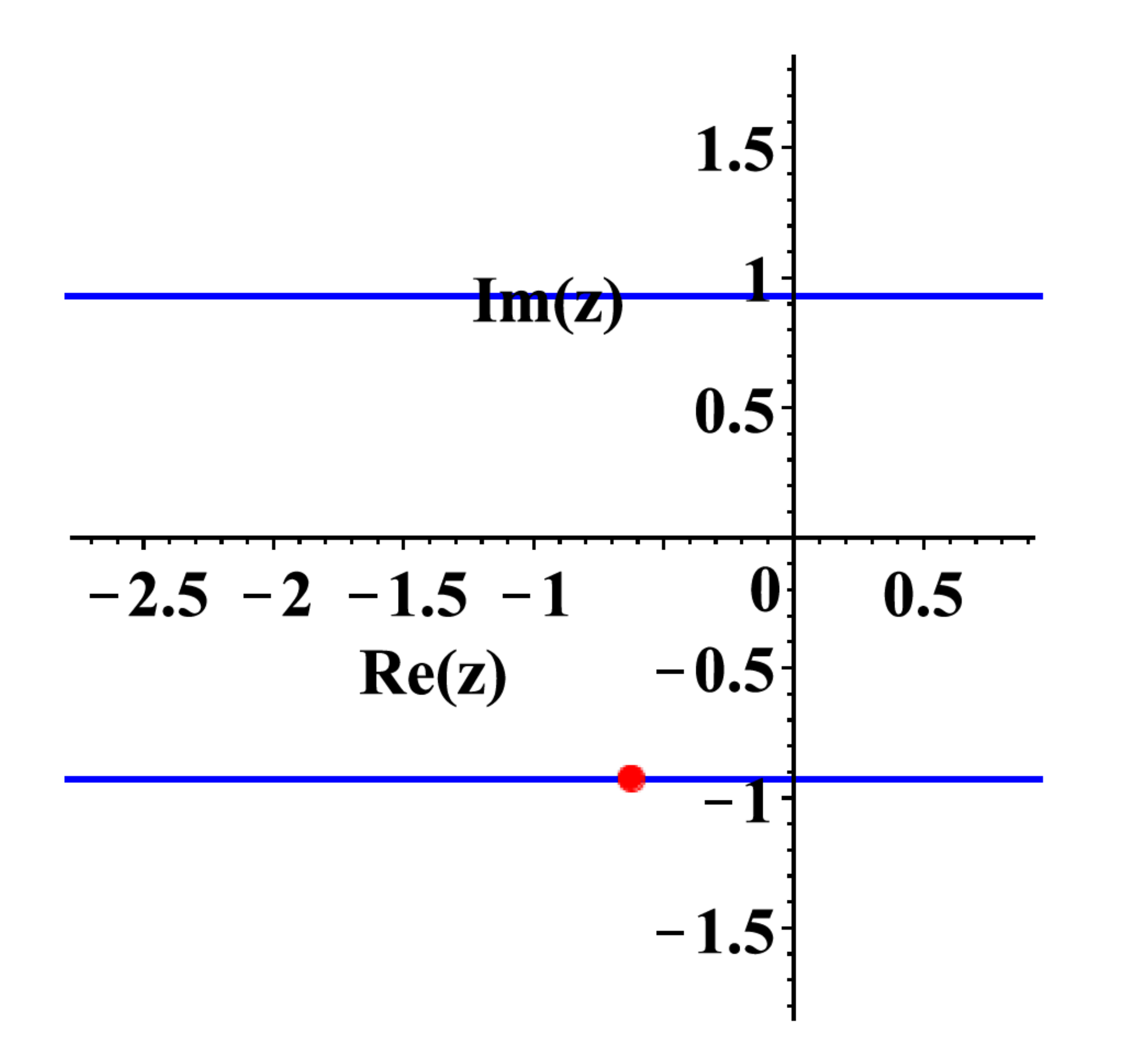}}
\subfigure[unstable spectrum in $\lambda$ plane]{%
\includegraphics[width=.32\textwidth]{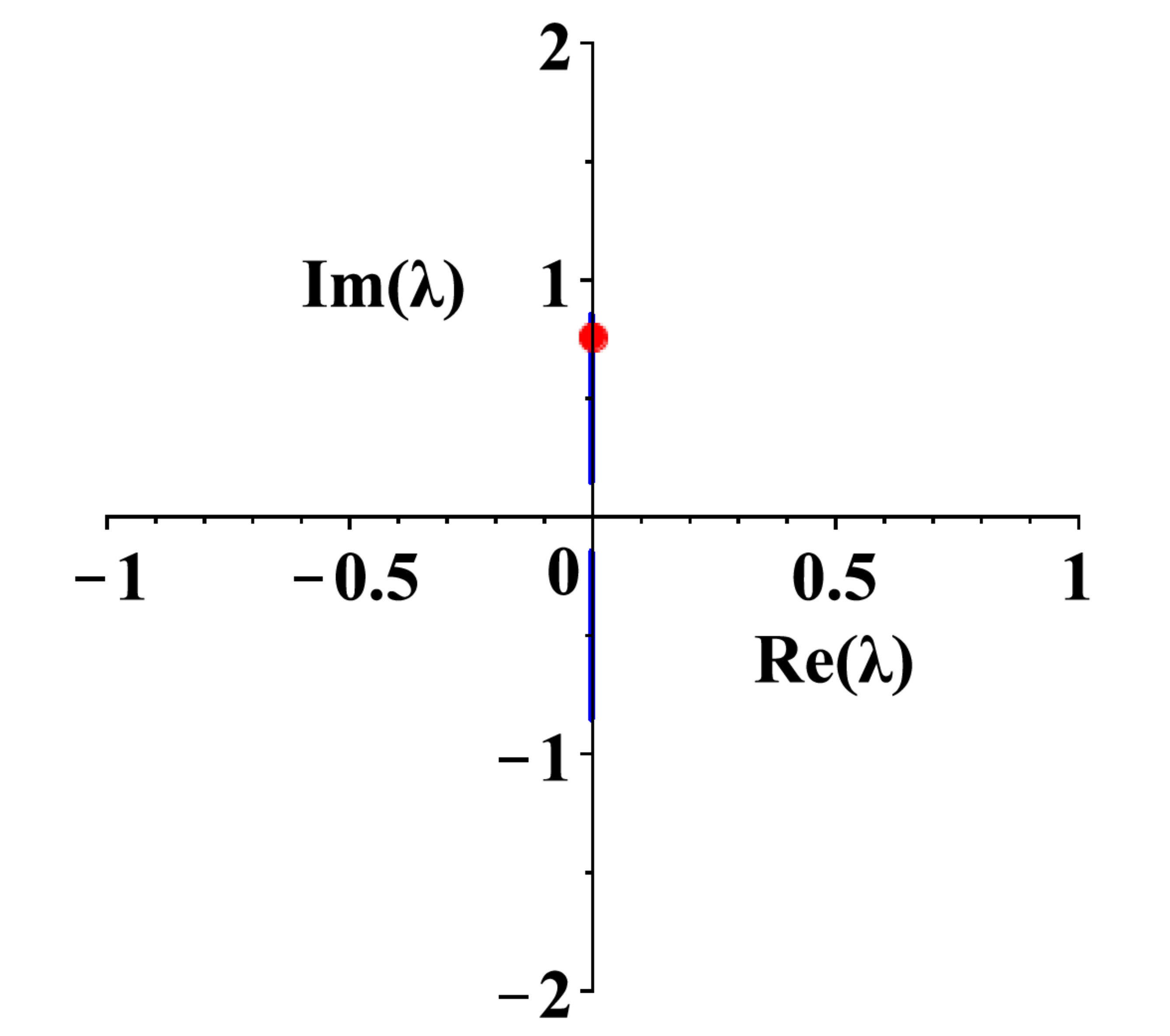}}
\subfigure[unstable spectrum in $\Omega$ plane]{%
\includegraphics[width=.32\textwidth]{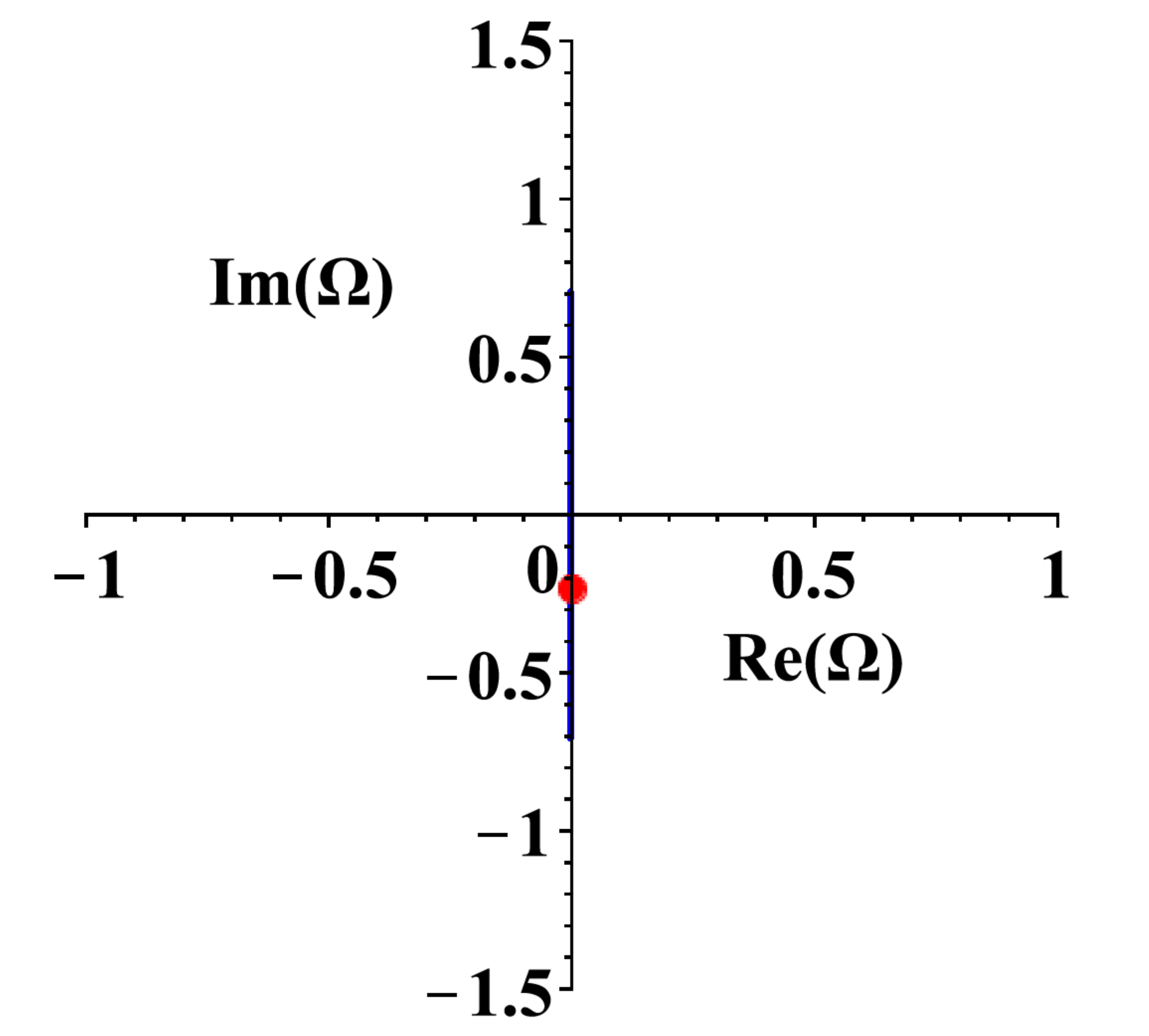}}
\caption{(color online): Unstable dn spectra in $z,\,\lambda,\,$ and $\Omega$ plane. The red point represents the parameter used for the plot of the Akhmediev dn breather in Fig.~\ref{fig:AB-dn}.}
\label{fig:spec1}
\end{figure}

Firstly, we consider the ${\rm dn}$ case by fixing the parameters $k=\frac{1}{\sqrt{2}}$, $\alpha=1$, $l=\frac{K'}{2}$. The resultant unstable spectra are shown in Fig.~\ref{fig:spec1}. This figure shows that there are two lines $z=s\pm \frac{K'}{2}$, $(-K\leq z-l\leq K)$ in $z$ plane. The unstable spectrum in $\lambda$ plane consists of two line segments located on the imaginary axis. In $\Omega$ plane, it is represented by one line segment on the imaginary axis.

\begin{figure}[tbh]
\centering
\subfigure[unstable spectrum in $z$ plane]{%
\includegraphics[width=.32\textwidth]{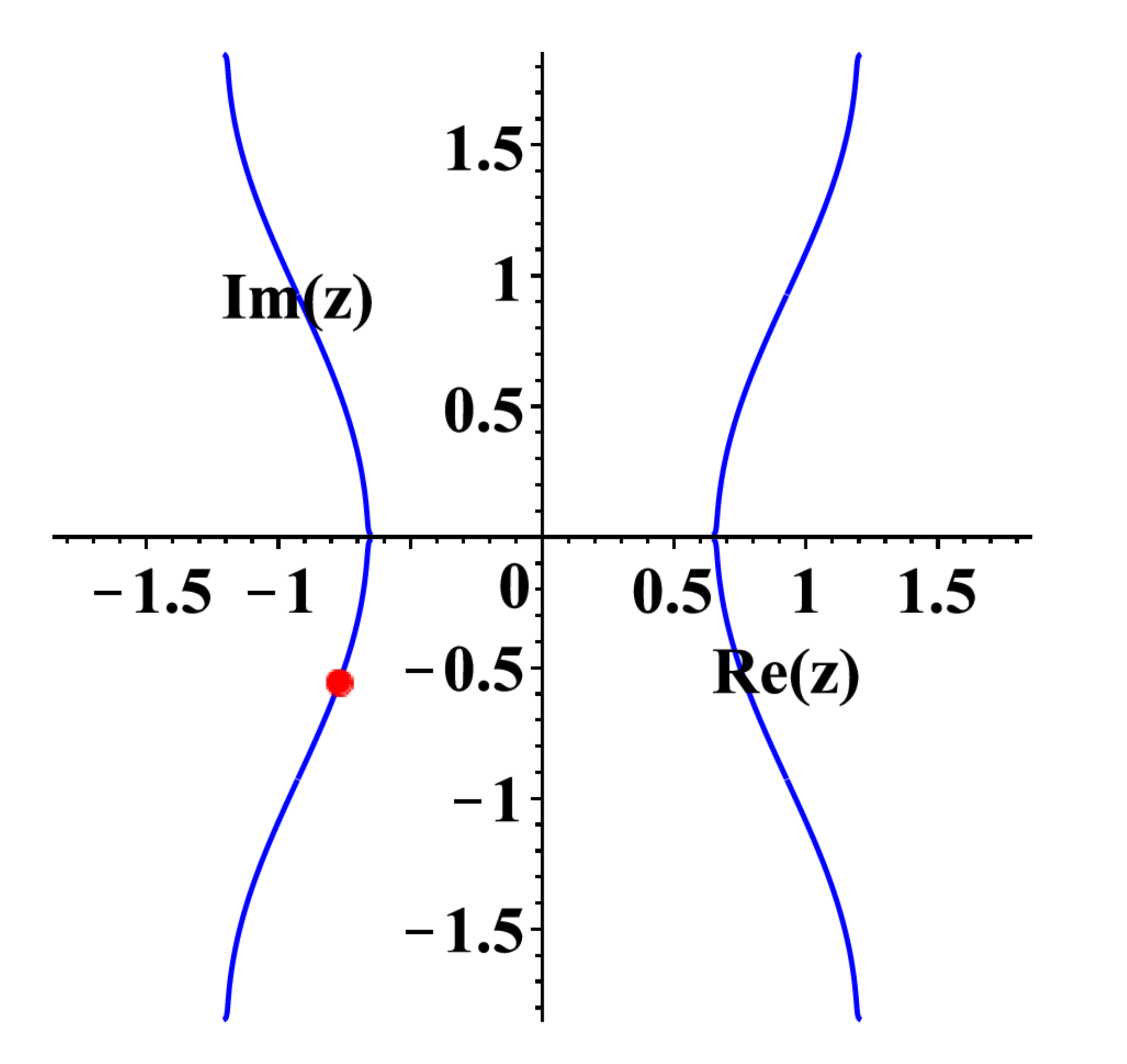}}
\subfigure[unstable spectrum in $\lambda$ plane]{%
\includegraphics[width=.32\textwidth]{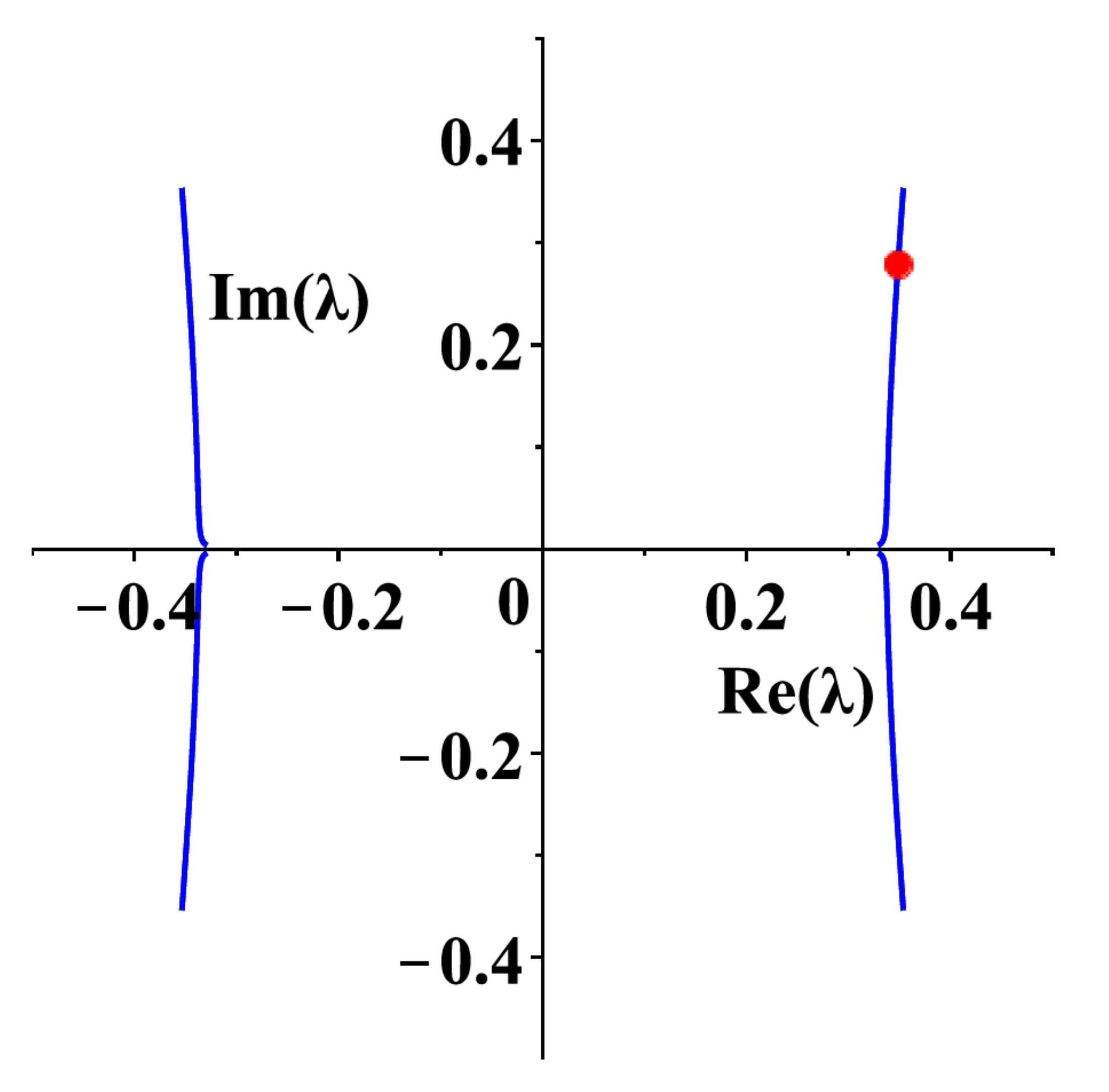}}
\subfigure[unstable spectrum in $\Omega$ plane]{%
\includegraphics[width=.32\textwidth]{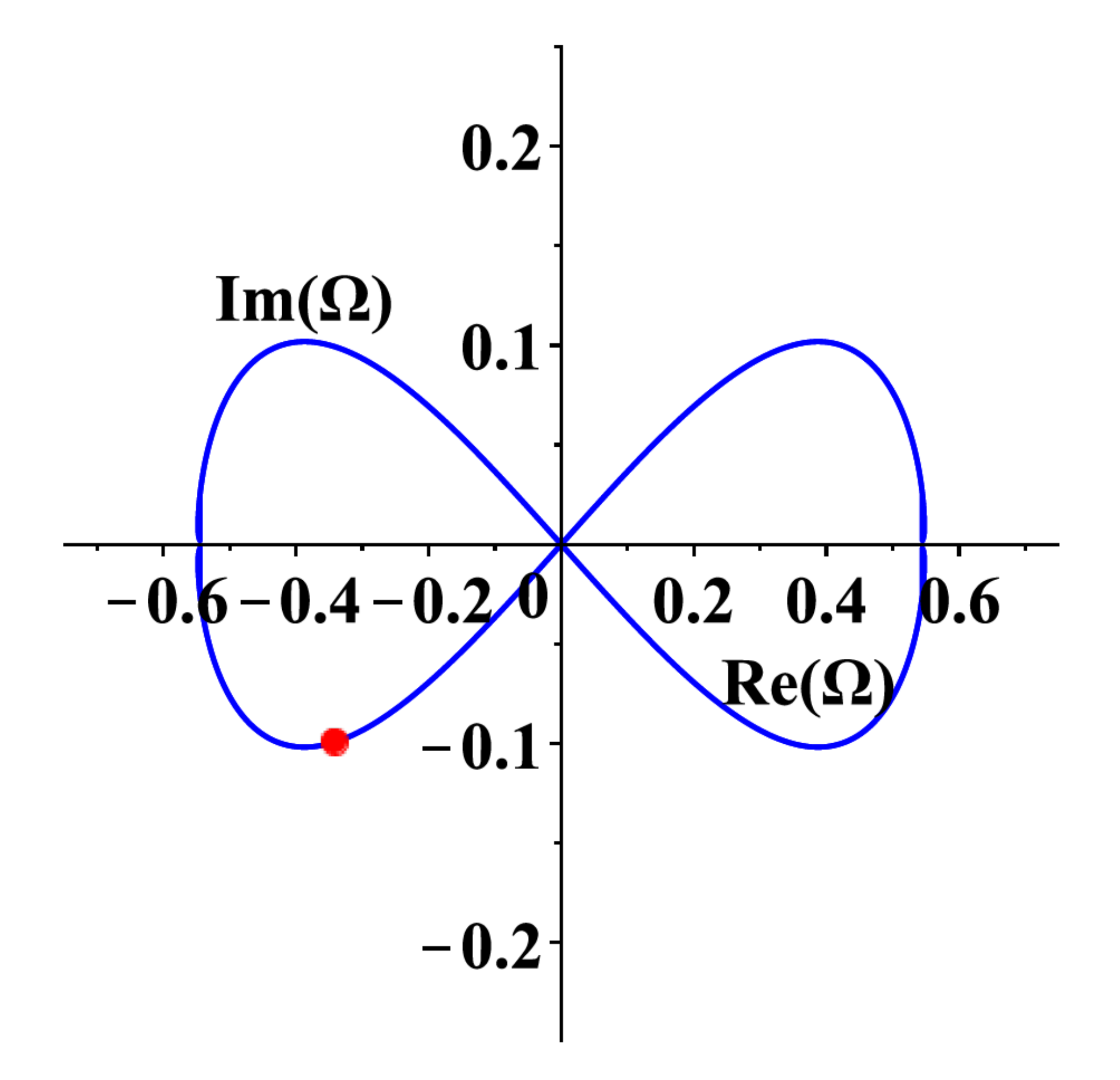}}
\caption{(color online): Unstable cn spectra in $z,\,\lambda,\,$ and $\Omega$ plane. The red point represents the parameter used for the plot of the Akhmediev cn breather in Fig.~\ref{fig:AB-cn-gene}(a).}

\label{fig:spec2}
\end{figure}

For the second case, we consider the ${\rm cn}$ case ($\nu_1=-1/2$, $\nu_2=0$ and $\nu_3=1/2$, i.e. $l=0$, $\alpha=1$ and $m=k^2=1/2$).  Here, we demonstrate the above-mentioned method in Ref.~\cite{DeconinckS17} to derive the unstable spectrum.
Firstly, we have $K(k)$ and $K'(k)$. From the four starting points $z=\frac{\ii}{2}(K(k)\pm \ii K'(k)), \frac{\ii}{2}(-K(k)\pm \ii K'(k))$, the curve can be replaced by the tangent line with the vector $\pm v_{T}(z)$ [equation \eqref{eq:tangent}] for the sufficiently small segment, where $\pm$ determines the direction. Combining the eight segments, we obtain the unstable spectrum with respect to $z$. Furthermore, using the formulas \eqref{eq:mu(iz)} and \eqref{eq:dmu(iz)}, we have the figure (Fig. \ref{fig:spec2}). It shows that there are two waist type curves in $z$ plane. Moreover, there are two almost vertical curves in $\lambda$ plane, and one figure-eight curve in $\Omega$ plane.

\begin{figure}[tbh]
\centering
\subfigure[unstable spectrum in $z$ plane]{%
\includegraphics[width=.32\textwidth]{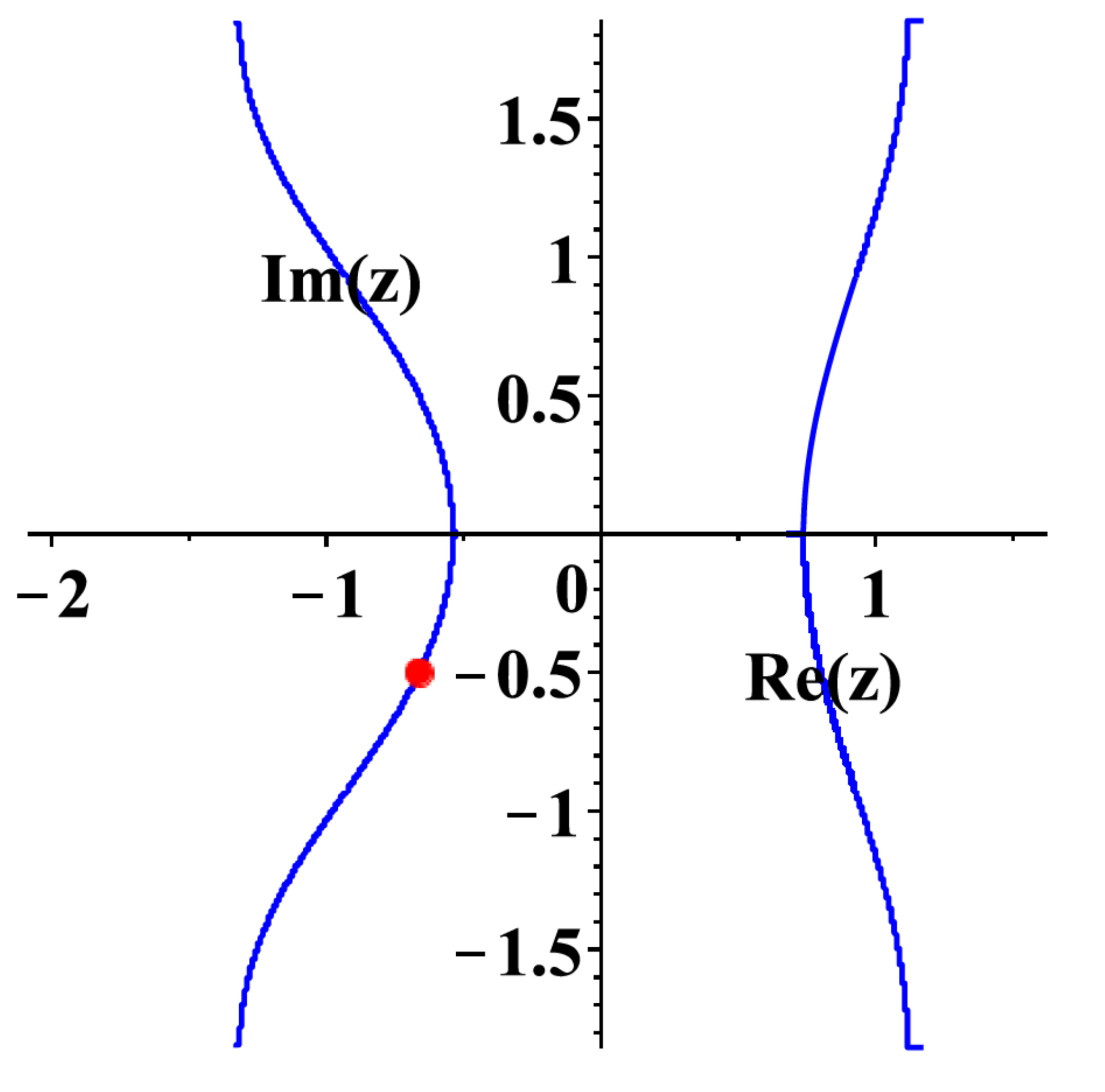}}
\subfigure[unstable spectrum in $\lambda$ plane]{%
\includegraphics[width=.32\textwidth]{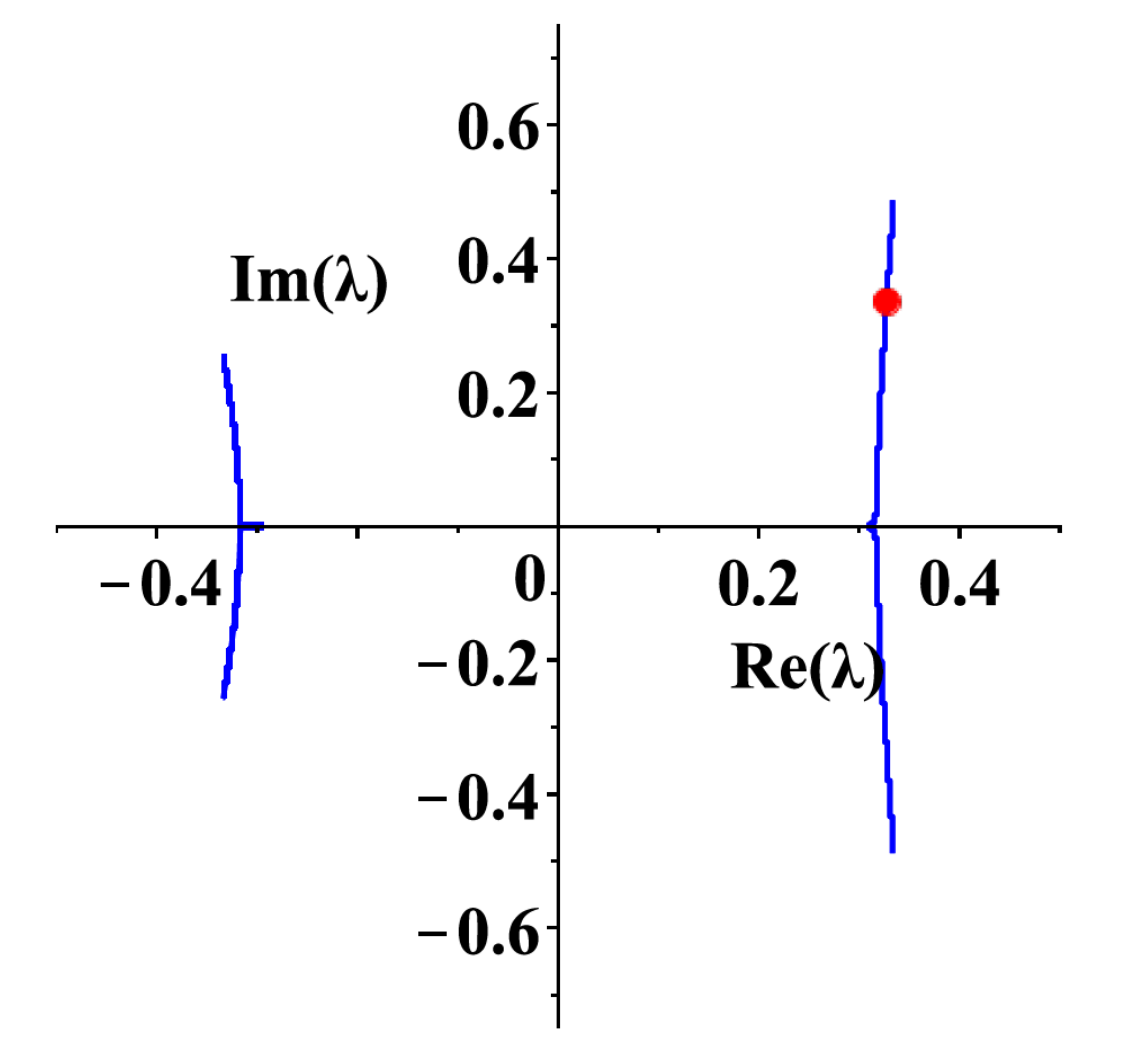}}
\subfigure[unstable spectrum in $\Omega$ plane]{%
\includegraphics[width=.32\textwidth]{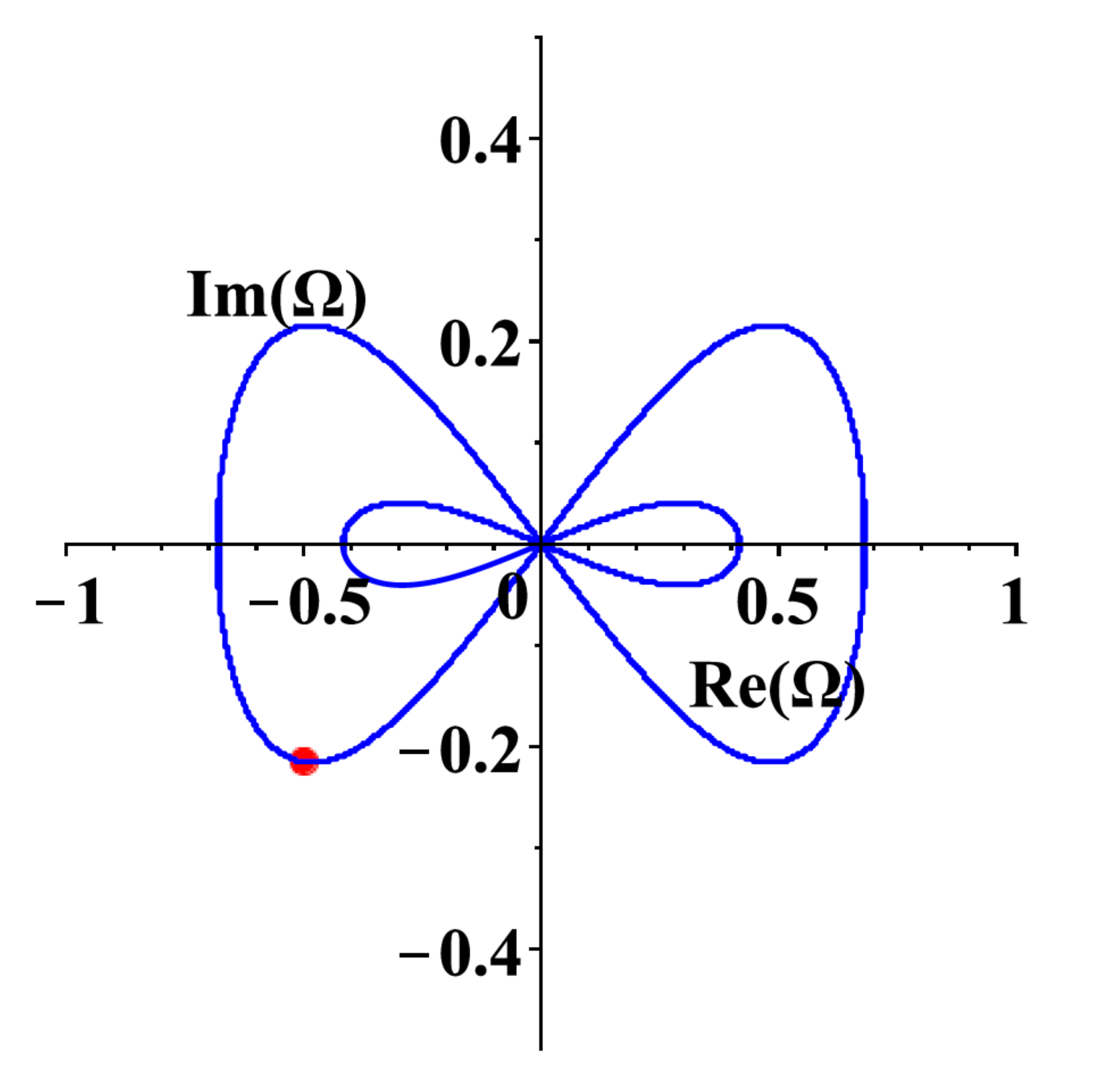}}
\caption{(color online): Unstable  spectra in $z,\,\lambda,\,$ and $\Omega$ plane for non-trivial phase-modulating background. The red point represents the parameter used for the plot of the Akhmediev breather in Fig.~\ref{fig:AB-cn-gene}(b).}
\label{fig:spec3}
\end{figure}

In a similar way, we obtain the figure for the non-trivial phase case (Fig.~\ref{fig:spec3}). It shows that there are two non-symmetric waist type curves in $z$ plane. Moreover, there are two non-symmetric almost vertical curves in $\lambda$ plane, and two figure-eight curves in $\Omega$ plane. For the general discussion on the unstable spectrum on the elliptic function solution of NLSE, we can refer to the reference \cite{DeconinckS17}. Here we just provide some special cases to illustrate the relationship between the unstable spectrum and the exact breather solutions in the following.

An interesting application of the above unstable spectra is that we can construct the exact solution to illustrate the modulation dynamics to describe the unstable process. We can obtain such solutions by setting the breather parameter $z_1$ in one-breather solution [equation \eqref{eq:n-breather} with $n=1$] to be values on the lines in Figs.~\ref{fig:spec1}-\ref{fig:spec3}. Below we provide a few examples.

\begin{figure}[tbh]
\centering
\subfigure[Akhmediev dn breather]{%
\includegraphics[width=.45\textwidth]{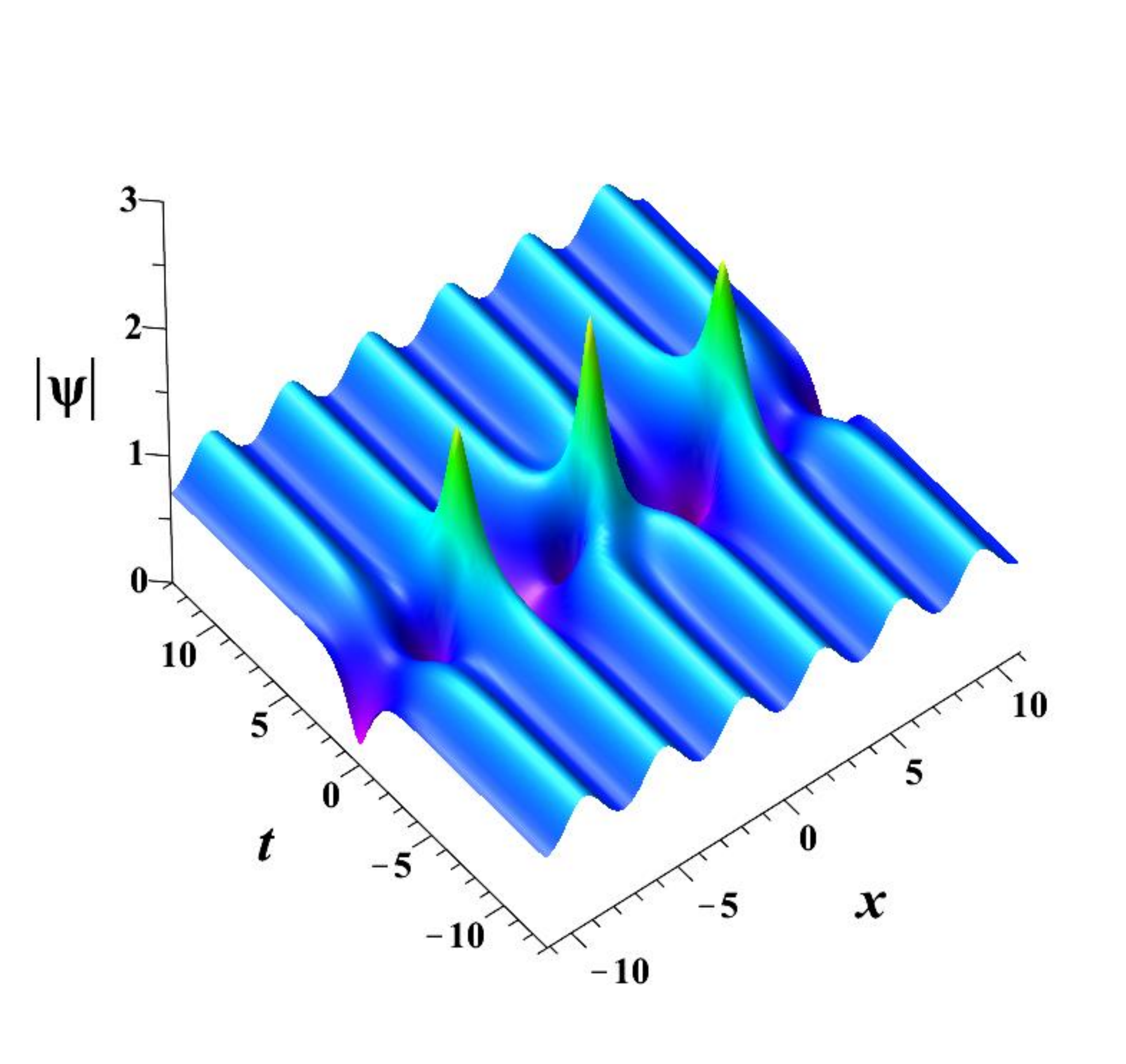}} \hfil
\subfigure[asymptotic behavior]{%
\includegraphics[width=.45\textwidth]{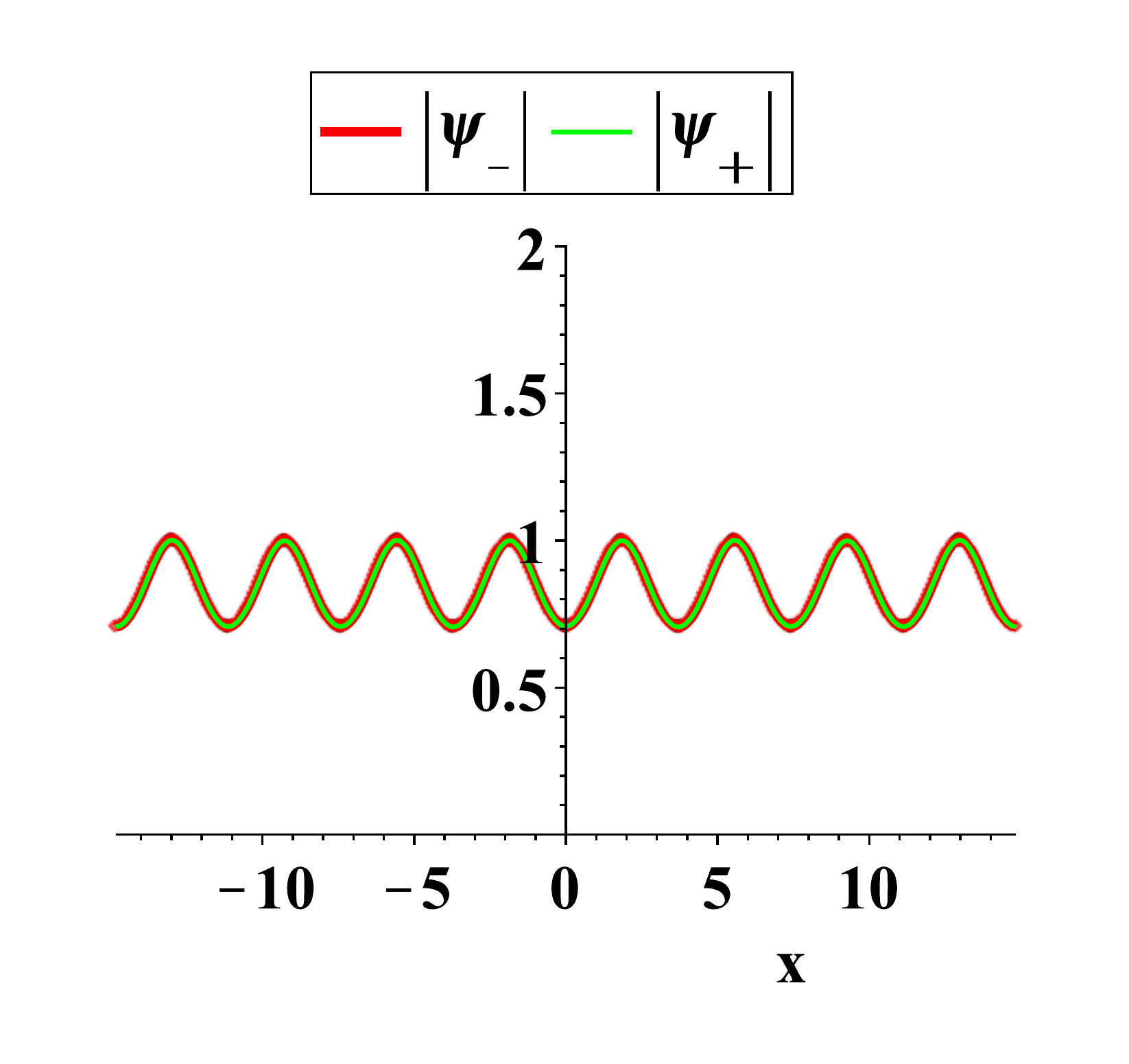}}
\caption{(color online): Akhmediev breathers (a) Akhmediev dn breather.  (b) The asymptotic behaviors. $\psi_\pm$ represent the different asymptotic behaviors at $t=\pm\infty$.}
\label{fig:AB-dn}
\end{figure}

Let the parameters be $k=\frac{1}{\sqrt{2}}$, and $\alpha=1$, $l=\frac{K'}{2}$, which infer $s_2=\frac{3}{4}$. Choosing the parameter $z_1=-0.527-\frac{K(1/2)}{2}\ii\approx -0.527-0.927\ii$, $\alpha_1=-1$, we obtain the Akhmediev dn breather (Fig \ref{fig:AB-dn} (a)). The maximum value is about $2.384$ located at the origin. From the figure \ref{fig:AB-dn} (a), we see that there are lots of peaks located on the line $t=0$, but the value of peaks is less than $2.384$, which can be obtained by the analysis of formula \eqref{eq:backlund} and is distinct with the Akhmediev breather in the plane wave background. In this setting, it is special that there is no shift of crest under the special parameters setting (Fig \ref{fig:AB-dn} (b)), by the above analysis \eqref{eq:breather-asy-1} and \eqref{eq:breather-asy-2} for the single breather solution. Here we should point out that this phenomenon with no shift on the crest is a general rule in the dn background. Through the figure \ref{fig:AB-dn} (a) and the analysis, we show that this exact breather solution describes the unstable dynamical behavior induced by the unstable spectrum.

\begin{figure}[tbh]
\centering
\subfigure[Akhmediev cn breather]{%
\includegraphics[width=.45\textwidth]{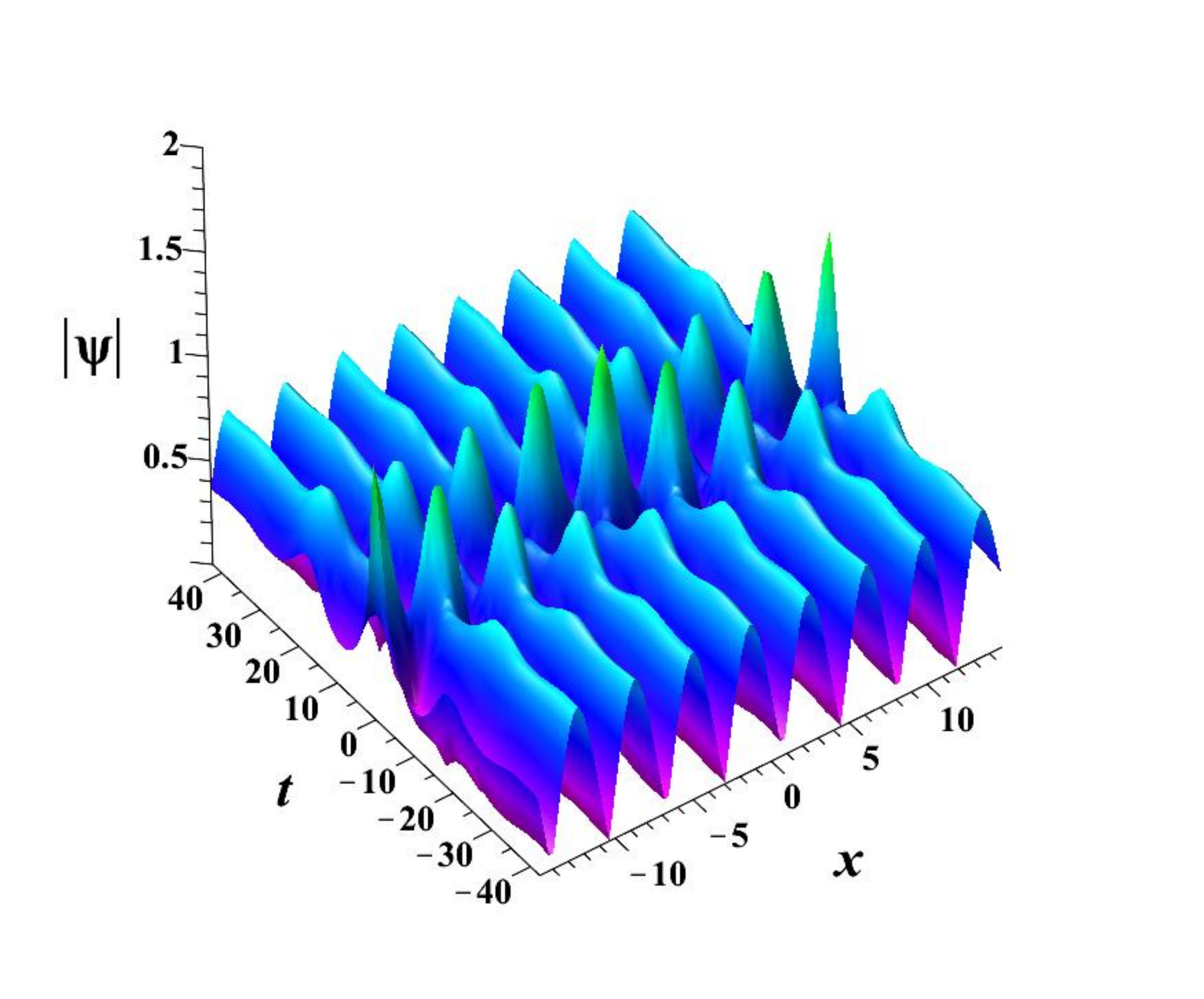}} \hfil
\subfigure[Akhmediev non-trivial phase breather]{%
\includegraphics[width=.45\textwidth]{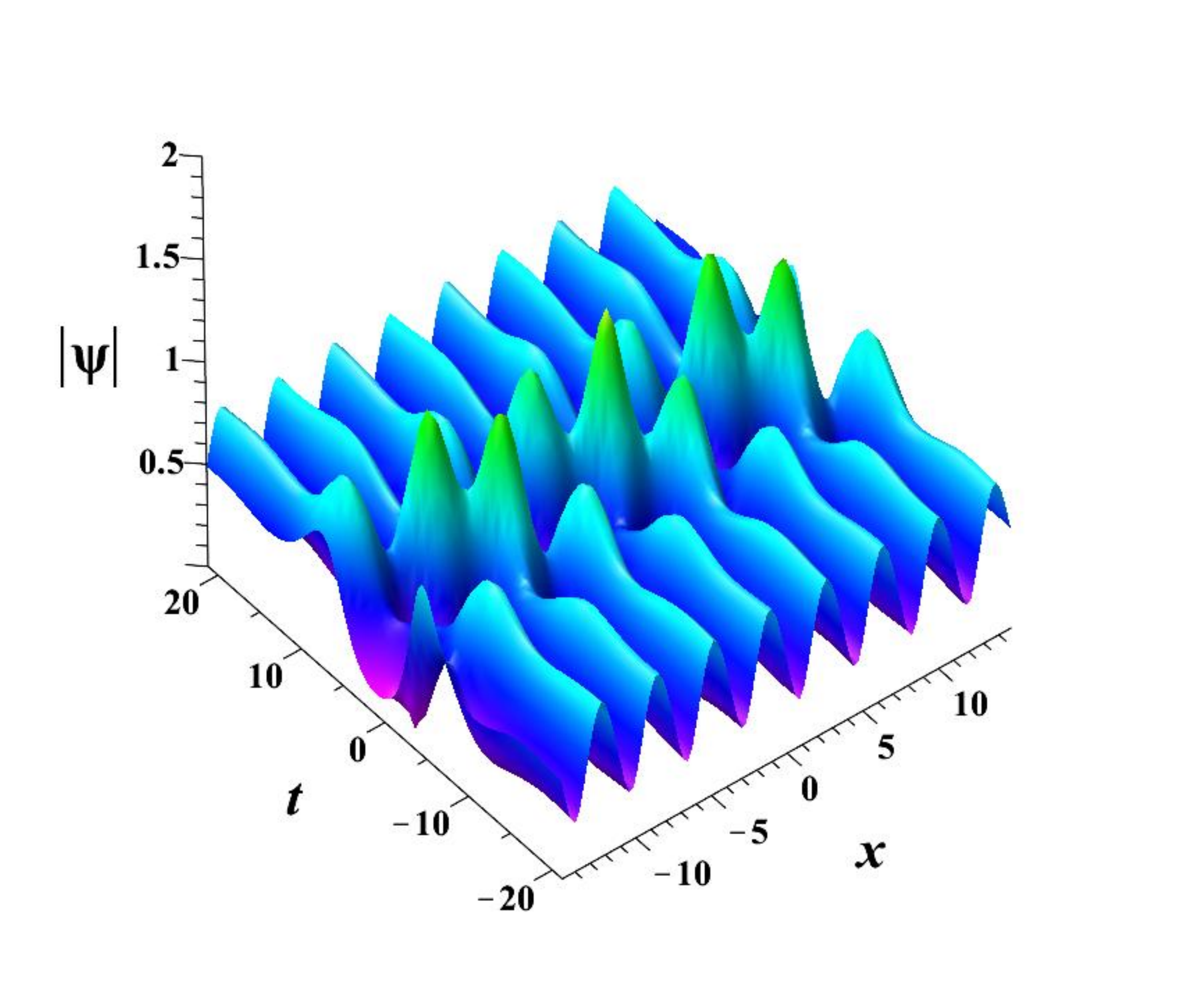}}
\caption{(color online): Akhmediev breathers (a) Akhmediev cn breather. (b) Akhmediev non-trivial phase breather.}
\label{fig:AB-cn-gene}
\end{figure}

Let us set $k=\frac{1}{\sqrt{2}}$, $l=0$, and $\alpha=1$, which implies $s_2=0$. Choosing the parameters $z_1\approx -0.771-0.560\ii$, $\alpha_1=-1$, we arrive at the Akhmediev cn breather, which admits the phase and crest shift simultaneously (Fig \ref{fig:AB-cn-gene} (a)). The maximum value of the peak is about $1.264$ located at origin. As the analysis in the above, we know that the other peaks' value is less than $1.264$. In this setting, we can obtain the shift of crest and phase through above general analysis.
The dynamics of breather describes the unstable process of the cn solution.

Finally, let us take the parameters $k=\frac{1}{\sqrt{2}}$, $\alpha=1$, and $l=\frac{K(1/2)}{8}$,  which yield  $\nu_1=-{\rm dn}^2(K(1/2)+2\ii l)$, $\nu_2=-\frac{1}{2}{\rm cn}^2(K(1/2)+2\ii l)$, $\nu_3=\frac{1}{2}{\rm sn}^2(K(1/2)+2\ii l)$ and $s_2=\frac{1}{2}(\nu_1+\nu_2+\nu_3)$. Choosing parameters $z_1=-0.663-0.503\ii$, $\alpha_1=-1$, we plot Fig.~\ref{fig:AB-cn-gene} (b). The maximum value of breather is about $1.414$ located at origin, which is greater than the other peaks located on the line $t=0$. This breather describes the unstable dynamics for the elliptic-function background with non-trivial phase modulation.

\section{Rogue waves on the elliptic function background}\label{sec6}

In this section, we consider the rogue waves, which are located at the branch points.  In this case, above two vector solutions \eqref{eq:fund-sol} merge into one vector solution. Thus we need to look for another linear independent vector solution. Here we would like to utilize the limit technique to deal with this problem. Due to the theta function representations include the parameters by Abel map, we still use the original parameter to derive the formulas of rogue waves. Suppose we expand the parameter $y(\lambda)$ at the branch point $\lambda_1$: $\lambda=\lambda_1+\epsilon_1^2$, here $\epsilon_1$ is a small parameter, then we have the expansions:
\begin{equation}\label{eq:expan-y(lambda)}
\begin{split}
y(\lambda)&=\epsilon_1y_1\sqrt{\left(1+\frac{\epsilon_1^2}{\lambda_1-\lambda_2}\right)\left(1+\frac{\epsilon_1^2}{\lambda_1-\lambda_3}\right)\left(1+\frac{\epsilon_1^2}{\lambda_1-\lambda_4}\right)}\\
&=y_1\epsilon_1\sum_{i=0}^{\infty}y_1^{[i]}\epsilon_1^{2i},
\end{split}
\end{equation}
where
\[
y_1=\sqrt{(\lambda_1-\lambda_2)(\lambda_1-\lambda_3)(\lambda_1-\lambda_4)},\,\,\,\, y_1^{[i]}\equiv\left(\sum_{l_2+l_3+l_4=i,} \prod_{j=2}^{4}\frac{\binom{\frac{1}{2}}{l_j}}{(\lambda_1-\lambda_j)^{l_j}}\right).
\]
It follows that
\begin{equation}\label{eq:expan-beta1}
\begin{split}
\beta_1(\lambda)&=2(\lambda_1+\epsilon_1^2)^2+s_2-2y(\lambda)    \\
            &=2\lambda_1^2+s_2+2\left[\epsilon_1^2(2\lambda_1+\epsilon_1^2)-y_1\epsilon_1\sum_{i=0}^{\infty}y_1^{[i]}\epsilon_1^{2i} \right]   \\
\end{split}
\end{equation}
and
\begin{equation}\label{eq:expan-C1}
\begin{split}
C_1(\lambda)&=2\lambda \beta_1(\lambda)-s_3      \\
   &=2\lambda_1^3+s_2\lambda_1-s_3+2\left[\epsilon_1^2[(3\lambda_1^2+s_2/2)+3\lambda_1\epsilon_1^2+\epsilon_1^4]-y_1\epsilon_1\left(\lambda_1y_1^{[0]}+\sum_{i=1}^{\infty}(\lambda_1y_1^{[i]}+y_1^{[i-1]})\epsilon_1^{2i}\right)\right].
\end{split}
\end{equation}
Moreover, the elements in the vector solution can be expanded in the following:
\begin{equation}\label{eq:expan-nubeta1}
\begin{split}
(\nu(x)-\beta_1(\lambda))^{\pm1/2}&=[W_1(x)]^{\pm1/2}(1+K_1(\epsilon_1)+K_2(\epsilon_1))^{\pm1/2}     \\
   &=[W_1(x)]^{\pm1/2}\sum_{i=0,}^{\infty}\sum_{j=0}^{i}\binom{\pm\frac{1}{2}}{i}\binom{i}{j}K_1^{i-j}K_2^j\\
   &=[W_1(x)]^{\pm1/2}\sum_{j=0,}^{\infty}\sum_{i=0}^{\infty} \binom{\pm\frac{1}{2}}{i+j}\binom{i+j}{j}K_1^{i}K_2^j,
\end{split}
\end{equation}
where
\[ W_1(x)=\nu(x)-(2\lambda_1^2+s_2),\,\,\,\,\,
K_1(\epsilon_1)=\frac{-2\epsilon_1^2(2\lambda_1+\epsilon_1^2)}{W_1(x)},\,\,\,\,
K_2(\epsilon_1)=\frac{2y_1\epsilon_1{\displaystyle\sum_{i=0}^{\infty}y_1^{[i]}\epsilon_1^{2i} }}{W_1(x)}.
\]
We need the combinatorial formulas:
\begin{equation*}
\begin{split}
\sum_{i=0}^{\infty} \binom{\pm\frac{1}{2}}{i+j}\binom{i+j}{j}K_1^{i}=&\sum_{i=0}^{\infty} \binom{\pm\frac{1}{2}}{i+j}\binom{i+j}{j}\left(\frac{-2}{W_1(x)}\right)^i\epsilon_1^{2i}(2\lambda_1+\epsilon_1^2)^i \\
=&\sum_{i=0,}^{\infty}\sum_{k=0}^{i} \binom{\frac{1}{2}}{i+j}\binom{i+j}{j}\binom{i}{k}\left(\frac{-2}{W_1(x)}\right)^i(2\lambda_1)^{i-k}\epsilon_1^{2(k+i)} \\
=&\sum_{k=0,}^{\infty}\sum_{i=2k}^{\infty} \binom{\pm\frac{1}{2}}{i-k+j}\binom{i-k+j}{j}\binom{i-k}{k}\left(\frac{-2}{W_1(x)}\right)^{i-k}(2\lambda_1)^{i-2k}\epsilon_1^{2i}\\
=&\sum_{i=0}^{\infty} \sum_{k=0}^{[i/2]}\binom{\pm\frac{1}{2}}{i-k+j}\binom{i-k+j}{j}\binom{i-k}{k}\left(\frac{-2}{W_1(x)}\right)^{i-k}(2\lambda_1)^{i-2k}\epsilon_1^{2i}
\end{split}
\end{equation*}
and
\begin{equation*}
\begin{split}
K_2^{j}=&\left(\frac{2y_1}{W_1(x)}\right)^j\epsilon_1^j\left(\sum_{i=0}^{\infty}y_1^{[i]}\epsilon_1^{2i}\right)^j \\
=&\left(\frac{2y_1}{W_1(x)}\right)^j\epsilon_1^j\sum_{n=0}^{\infty}j!\left(\sum_{\sum_{m=0}^nms_m=n,}^{\sum_{m=0}^{n}s_m=j} \prod_{m=0}^{n} \frac{\left(y_1^{[m]}\right)^{s_m}}{s_m!}\right)\epsilon_1^{2n}.
\end{split}
\end{equation*}
Let us introduce the notation
\[
 G^{\pm}_{i,j}\equiv\sum_{k=0}^{[i/2]}\binom{\pm\frac{1}{2}}{i-k+j}\binom{i-k+j}{j}\binom{i-k}{k}\left(\frac{-2}{W_1(x)}\right)^{i-k}(2\lambda_1)^{i-2k}
\]
and
\[
H_{n,j}=j!\left(\sum_{\sum_{m=0}^nms_m=n,}^{\sum_{m=0}^{n}s_m=j} \prod_{m=0}^{n} \frac{\left(y_1^{[m]}\right)^{s_m}}{s_m!}\right),
\]
then we arrive at
\begin{equation}\label{eq:expan-nubeta2}
\begin{split}
(\nu(x)-\beta_1(\lambda))^{\pm1/2}&=[W_1(x)]^{\pm1/2}\sum_{j=0}^{\infty} \left(\frac{2y_1}{W_1(x)}\right)^j\epsilon_1^j\sum_{k=0}^{\infty}\left( \sum_{i+n=k}G^{\pm}_{i,j}H_{n,j} \right)\epsilon_1^{2k}\\
&=[W_1(x)]^{\pm1/2}\sum_{k=0,}^{\infty}\sum_{j=0}^{\infty} \left(\frac{2y_1}{W_1(x)}\right)^j\left( \sum_{i+n=k}G^{\pm}_{i,j}H_{n,j} \right)\epsilon_1^{2k+j}\\
&=[W_1(x)]^{\pm1/2}\sum_{k=0,}^{\infty}\sum_{j=2k}^{\infty} \left(\frac{2y_1}{W_1(x)}\right)^{j-2k}\left( \sum_{i+n=k}G^{\pm}_{i,j-2k}H_{n,j-2k} \right)\epsilon_1^{j}\\
&=[W_1(x)]^{\pm1/2}\sum_{j=0}^{\infty}\left[\sum_{k=0}^{[j/2]} \left(\frac{2y_1}{W_1(x)}\right)^{j-2k}\left( \sum_{i+n=k}G^{\pm}_{i,j-2k}H_{n,j-2k} \right)\right]\epsilon_1^{j}\\
&\equiv [W_1(x)]^{\pm1/2}\sum_{j=0}^{\infty}\nu_{1\pm}^{[j]} \epsilon_1^{j},
\end{split}
\end{equation}
where $[\cdot]$ represents the floor function.
In a similar fashion, we get the expansion for $1/(\nu(x)-\beta_1(\lambda))$:
\begin{equation}\label{eq:expan-nubeta3}
(\nu(x)-\beta_1(\lambda))^{-1}=[W_1(x)]^{-1}\sum_{j=0}^{\infty}R_j\epsilon_1^{j},
\end{equation}
where we have defined
\begin{align}
R_j &\equiv \sum_{k=0}^{[j/2]} \left(\frac{2y_1}{W_1(x)}\right)^{j-2k}\left( \sum_{i+n=k}G'_{i,j-2k}H_{n,j-2k} \right), \\
 G'_{i,j}&\equiv\sum_{k=0}^{[i/2]}(-1)^{i-k+j}\binom{i-k+j}{j}\binom{i-k}{k}\left(\frac{-2}{W_1(x)}\right)^{i-k}(2\lambda_1)^{i-2k}.
\end{align}
With the help of equations \eqref{eq:expan-C1} and \eqref{eq:expan-nubeta3}, it follows that
\begin{equation}\label{eq:expan-nubeta4}
\begin{split}
\frac{C_1}{(\nu(x)-\beta_1(\lambda))}&=\frac{C_1}{W_1(x)}\sum_{j=0}^{\infty}R_j\epsilon_1^{j}\\
&=\frac{1}{W_1(x)}\sum_{j=0}^{\infty}\left[(2\lambda_1^3+s_2\lambda_1-s_3)R_j-2\lambda_1y_1y_1^{[0]} R_{j-1}\chi_{[j\geq 1]}+
2\left(3\lambda_1^2+\frac{s_2}{2}\right)R_{j-2}\chi_{[j\geq 2]}\right. \\ &\,\,\,\,\,\,\, \left.-2y_1\left(\sum_{i=0}^{[(j-3)/2]}(\lambda_1y_1^{[i+1]}+y_1^{[i]})R_{j-3-2i}\right)\chi_{[j\geq 3]}+6\lambda_1 R_{j-4}\chi_{[j\geq 4]}+2R_{j-6}\chi_{[j\geq 6]}\right]\epsilon_1^j\\
&\equiv\sum_{j=0}^{\infty}I_j(x) \epsilon_1^j,
\end{split}
\end{equation}
where \[\chi_{[j\geq s]}=\left\{\begin{matrix}1,&j\geq s \\
0,&j<s.
\end{matrix}\right.
\]
Then we obtain the expansion for the phase factor:
\begin{equation}
\begin{split}
 \theta_1(\lambda)&=\ii \int_{0}^{x}\frac{C_1{\rm d}s}{\nu(s)-\beta_1}+\ii \lambda x+\ii (\frac{s_2}{2}+y)t\\
 &=\ii \left[\int_0^xI_0(s) {\rm d}s+\lambda_1x+\frac{s_2}{2}t+\sum_{j=1}^{\infty}\left(\int_0^xI_j(s) {\rm d}s\right) \epsilon_1^j+x\epsilon_1^2+y_1\sum_{i=0}^{\infty}y_1^{[i]}t\epsilon_1^{2i+1} \right]\\
 &\equiv \sum_{i=0}^{\infty} \Theta_1^{[i]} \epsilon_1^i.
 \end{split}
\end{equation}

\begin{remark}\label{rem4}
Here we point out that the integrations $\int_0^x I_j(s) {\rm d}s,$ $(i\geq 1)$ do not possess integrals of the third kind. In general, the integration $\int_0^x I_j(s) {\rm d}s$
can be expressed in terms of elliptic functions and elliptic integrals of the second and third kind via the formulas \eqref{eq:v01},\eqref{eq:v2} and \eqref{eq:vm3}. It is hard to give a claim that the integrations $\int_0^x I_j(s) {\rm d}s,$ $(i\geq 1)$ do not have the elliptic integral of the third  by the pure algebraic calculations. It is much easier to give a proof by analyzing the singularity point at the point $x=\xi \mod \Lambda\in \{2mK+n\ii \,K' |m,n\in \mathbb{Z}\}$ when $\nu(\xi)-(2\lambda_1^2+s_2)=0$. By using the conformal transformation \eqref{eq:mu(iz)} and \eqref{eq:dmu(iz)}, the integration
\[
\ii \int_0^x\frac{C_1\,{\rm d} s}{\nu(s)-\beta_1}=\frac{\ii }{2}\int_0^{\alpha\,x}\frac{\frac{{\rm d}}{{\rm d}z} {\rm sn}^2(\ii (z-l))\,{\rm d} s}{{\rm sn}^2(\ii (z-l))-{\rm sn}^2(s)}
\]
has the logarithmic singularity at $x=\xi\mod \Lambda$. Expanding the above integration at $\lambda=\lambda_1$ is equivalent to the expansion at $z=z_{s}\equiv\frac{K'+\ii K}{2}=-\ii (\alpha\,\xi+z_0).$ Firstly, we have the following two expansions:
\[
{\rm sn}^2(\ii (z-l))-{\rm sn}^2(s)=\sum_{i=1}^{\infty}\left[ \left(\frac{{\rm d}^i}{i!{\rm d}z^i} {\rm sn}^2(\ii (z-l))\right)|_{z=z_s}[(z-z_s)^i-(\ii \alpha\xi-\ii  s)^i]\right],
\]
and
\[
\frac{{\rm d}}{{\rm d}z} {\rm sn}^2(\ii (z-l))=\sum_{i=1}^{\infty} \left(\frac{{\rm d}^i}{(i-1)!{\rm d}z^i} {\rm sn}^2(\ii (z-l))\right)|_{z=z_s}(z-z_s)^{i-1}.
\]
Thus we have
\[
\int\frac{\ii }{2}\frac{\frac{{\rm d}}{{\rm d}z} {\rm sn}^2(\ii (z-l)) {\rm d}s}{{\rm sn}^2(\ii (z-l))-{\rm sn}^2(s)}=\frac{1}{2}\ln(s-\alpha\xi)+\frac{O(z-z_s)}{O(s-\alpha\xi)},
\]
which infers that the logarithmic singularity can not appear at the higher-order term of $(z-z_s).$ Moreover, the higher-order terms of $(z-z_s)$ do not involve elliptic integrals of the third kind since the leading order term involves the logarithmic singularity. From the conformal transformation
\[ \alpha\int_{0}^{\lambda}\frac{{\rm d}\mu}{\sqrt{(\mu-\lambda_1)(\mu-\lambda_1^*)(\mu-\lambda_2)(\mu-\lambda_2^*)}}=2(z-l).\]
we can deduce that, at the critical point $z=z_s$, we have the expansion $\lambda(z)-\lambda_1=(z-z_s)^2H^2(z-z_s)$, where $H(z-z_s)$ is an analytic function with $H(0)\neq 0.$ So $\epsilon_1=(z-z_s)H(z-z_s)$  and
\[
\int\frac{C_1 {\rm d}s}{\nu(s)-\beta_1}=\frac{1}{2}\ln(s-\alpha\xi)+\frac{O(\epsilon_1)}{O(s-\alpha\xi)},
\]
which implies that there appears no elliptic integral of the third kind in the coefficients of higher-order terms of $\epsilon_1$.
\end{remark}

By the Schur polynomial, we have
\[
\exp[\theta_1]=\exp[\Theta_1^{[0]}]\sum_{i=0}^{\infty}E_1^{[i]}\epsilon_1^i,
\]
where
\[
E_1^{[i]}=\sum_{\sum_{k=1}^{m}kl_k=i}\left(\prod_{j=1}^{m}\frac{(\Theta_1^{[j]})^{l_j}}{l_j!}\right).
\]
Summarizing, we obtain the expansion for the first component of vector solution:
\[
\sqrt{\nu(x)-\beta_1(\lambda)} \exp(\theta_1)=
 \sqrt{W_1(x)} \exp[\Theta_1^{[0]}] \sum_{i=0}^{\infty} \phi_1^{[i]} \epsilon_1^{i},\,\,\,\,\, \phi_1^{[i]} =\sum_{j=0}^{i} E_1^{[j]}\nu_{1+}^{[i-j]}.
\]
To derive the expansion for the second component, we need to expand the following expression:
\[
r_1=\frac{\ii h}{f-y}=\frac{2\psi^*\lambda-\ii \psi^*_x}{\nu(x)-\beta_1(\lambda)}=\frac{2\nu(x)\lambda_1-\ii (\nu_x(x)/2+\ii q)+2\nu(x)\epsilon_1^2}{\sqrt{\nu(x)}(\nu(x)-\beta_1(\lambda))}\exp\left[\ii q\int_{0}^{x}\frac{{\rm d}s}{\nu(s)}-\ii s_2t\right].
\]
In fact, the second component of the vector solution has the expansion
\begin{equation*}
\begin{split}
&\frac{2\nu(x)\lambda_1-\ii (\nu_x(x)/2+\ii q)+2\nu(x)\epsilon_1^2}{\sqrt{\nu(x)}}\frac{\exp(\theta_1)}{\sqrt{\nu(x)-\beta_1(\lambda)}}\exp\left[\ii q\int_{0}^{x}\frac{{\rm d}s}{\nu(s)}-\ii s_2t\right]\\
=&\exp\left[\ii q\int_{0}^{x}\frac{{\rm d}s}{\nu(s)}-\ii s_2t\right] \frac{\sqrt{\nu(x)}\exp[\Theta_1^{[0]}]}{\sqrt{W_1(x)}} \sum_{i=0}^{\infty} \varphi_1^{[i]} \epsilon_1^{i},
\end{split}
\end{equation*}
where
\[
\varphi_1^{[i]}=\frac{2\nu(x)\lambda_1-\ii (\nu_x(x)/2+\ii q)}{\nu(x)}\sum_{j=0}^{i} E_1^{[j]}\nu_{1-}^{[i-j]}+2\chi_{[i\geq 2]}\sum_{j=0}^{i-2} E_1^{[j]}\nu_{1-}^{[i-2-j]}.
\]
In a similar procedure, we obtain the expansion at $\lambda=\lambda_2$:
\begin{equation*}
\begin{split}
\sqrt{\nu(x)-\beta_1(\lambda)} \exp(\theta_1)&=
 \sqrt{W_1(x)} \exp[\Theta_2^{[0]}] \sum_{i=0}^{\infty} \phi_2^{[i]} \epsilon_2^{i}, \\
r_1\sqrt{\nu(x)-\beta_1(\lambda)} \exp(\theta_1) &=\exp\left[\ii q\int_{0}^{x}\frac{{\rm d}s}{\nu(s)}-\ii s_2t\right] \frac{\sqrt{\nu(x)}\exp[\Theta_2^{[0]}]}{\sqrt{W_1(x)}} \sum_{i=0}^{\infty} \varphi_2^{[i]} \epsilon_2^{i},
 \end{split}
\end{equation*}
where $\Theta_2^{[i]}$, $\phi_2^{[i]}$ and $\varphi_2^{[i]}$ are $\Theta_1^{[i]}$, $\phi_1^{[i]}$ and $\varphi_1^{[i]}$ by exchanging the variables $\lambda_1$ to $\lambda_2.$ By the Remark \ref{rem4} and the integration formulas \eqref{eq:v01}, \eqref{eq:v2} and \eqref{eq:vm3}, we know that the functions $\phi_k^{[i]}$, $\varphi_k^{[i]}$ ($k=1,2; i\in \mathbb{Z}^+$) are the rational function of $\sn$, $\cn$, $\dn$ and the second type of elliptic integration $E(u)$.

Summarizing the above, we can obtain the following theorem of rogue waves:
\begin{theorem}\label{thm5}
The general rogue wave formulas obtained by the high order Darboux transformation (in the  Appendix~\ref{sec:appA}) can be represented as
\begin{equation}
\psi_{n_1,n_2}=\sqrt{\nu(x)}\left[\frac{\det(\mathbf{K}_n)}{\det(\mathbf{M}_n)}\right]\exp\left[-\ii q\int_{0}^{x}\frac{ {\rm d}s}{\nu(s)}+\ii  s_2 t\right],
\end{equation}
where $\mathbf{M}_n=\nu(x)\mathbf{X}_n^{\dag}\mathbf{D}_n\mathbf{X}_n+\mathbf{Y}_n^{\dag}\mathbf{D}_n\mathbf{Y}_n$, $\mathbf{K}_n=\mathbf{M}_n-2\mathbf{X}_{n,1}^{\dag}\mathbf{Y}_{n,1}$, $\mathbf{X}_{n,1}=[\mathbf{X}_{1}^{[1]},\mathbf{X}_{1}^{[2]}]$, $\mathbf{Y}_{n,1}=[\mathbf{Y}_{1}^{[1]},\mathbf{Y}_{1}^{[2]}]$ and
\[ \mathbf{D}_n=\begin{bmatrix}
\mathbf{D}^{[1,1]} & \mathbf{D}^{[1,2]}\\
\mathbf{D}^{[2,1]} & \mathbf{D}^{[2,2]}\\
\end{bmatrix},\,\,\,\,\,
\mathbf{D}^{[l,k]}=\left(\binom{i+j-2}{i-1}\frac{(-1)^{j-1}}{(\lambda_k-\lambda_l^*)^{i+j-1}}\right)_{1\leq i\leq n_l,0\leq j\leq n_k},\]
\[
 \mathbf{X}_n=\begin{bmatrix}
\mathbf{X}^{[1]} & 0\\
0 &\mathbf{X}^{[2]}  \\
\end{bmatrix},\,\,\,\,\, \mathbf{Y}_n=\begin{bmatrix}
\mathbf{Y}^{[1]} & 0\\
0 &\mathbf{Y}^{[2]}  \\
\end{bmatrix},  \]
and $l,k=1,2$, $n_1+n_2=n,$
\[
\mathbf{X}^{[l]}=\begin{bmatrix}
\varphi_l^{[1]}&\varphi_l^{[3]}& \varphi_l^{[5]}&\cdots &\varphi_l^{[2n_l-1]}  \\[8pt]
0 &\varphi_l^{[1]}& \varphi_l^{[3]}&\cdots &\varphi_l^{[2n_l-3]} \\
 \vdots&\vdots&\vdots&&\vdots\\
0 &0&0&\cdots &\varphi_l^{[1]}   \\
\end{bmatrix},\,\,\,\,
\mathbf{Y}^{[l]}=(\nu(x)-(2\lambda_l^2+s_2))\begin{bmatrix}
\phi_l^{[1]}&\phi_l^{[3]}& \phi_l^{[5]}&\cdots &\phi_l^{[2n_l-1]}  \\[8pt]
0 &\phi_l^{[1]}& \phi_l^{[3]}&\cdots &\phi_l^{[2n_l-3]} \\
 \vdots&\vdots&\vdots&&\vdots\\
0 &0&0&\cdots &\phi_l^{[1]}   \\
\end{bmatrix},
\]
and $\mathbf{X}_1^{[l]}$ and $\mathbf{Y}_1^{[l]}$ represent the first row of $\mathbf{X}^{[l]}$ and $\mathbf{Y}^{[l]}$, respectively.
\end{theorem}
The general expressions for the high order rogue waves are very complicated. However, the extremal value at the point $(x,t)=(0,0)$ can be obtained through the determinant formula. Since the NLSE possesses the scaling symmetry, we set the amplitude for the background be $1$: i.e., $\nu_3=1$  without loss of generality. Then, the vector fundamental solution at $(x,t)=(0,0)$ can be written as
\[
\begin{bmatrix}
\phi_1(0,0;\lambda) \\[5pt]
\varphi_1(0,0;\lambda) \\
\end{bmatrix}=\begin{bmatrix}
\sqrt{\nu_3-2(\lambda^2+s_2/2-y)} \\[5pt]
\ii \sqrt{\nu_3-2(\lambda^2+s_2/2+y)} \\
\end{bmatrix}.
\]
At the branch points $\lambda=\lambda_i+\epsilon_i^2$, we know that $y(-\epsilon_i)=-y(\epsilon_i)$, $i=1,2$. So we have the expansion
\[
\begin{bmatrix}
\phi_1(0,0;\lambda) \\[5pt]
\varphi_1(0,0;\lambda) \\
\end{bmatrix}=\sum_{j=0}^{\infty} \begin{bmatrix}
\phi_1^{[j]}(0,0;\lambda_i) \\[5pt]
\ii (-1)^{i}\phi_1^{[j]}(0,0;\lambda_i) \\
\end{bmatrix}\epsilon_i^{j},\,\,\,\,\, i=1,2.
\]
Moreover, we have
\[
\psi_n(0,0)=\frac{\det(\mathbf{M}_n-\ii \mathbf{X}_{n,1}^{\dag}\mathbf{X}_{n,1})}{\det(\mathbf{M}_n)},
\]
where $\mathbf{M}_n=\mathbf{X}_n^{\dag}\mathbf{D}_n\mathbf{X}_n$, $\mathbf{X}_{1,n}=[\phi_1^{[1]}(\lambda_1), \phi_1^{[3]}(\lambda_1),\cdots,\phi_1^{[2n_1-1]}(\lambda_1),\phi_1^{[1]}(\lambda_2), \phi_1^{[3]}(\lambda_2),\cdots,\phi_1^{[2n_2-1]}(\lambda_2)],$ and
\[
\mathbf{X}_n=\begin{bmatrix}
\mathbf{X}^{[1]} & 0\\
0 &\mathbf{X}^{[2]}  \\
\end{bmatrix},\,\,\,\,\,\, \mathbf{X}^{[i]}=\begin{bmatrix}
\phi_1^{[1]}(\lambda_i)& \phi_1^{[3]}(\lambda_i)&\cdots &\phi_1^{[2n_1-1]}(\lambda_i) \\
0&\phi_1^{[1]}(\lambda_i)&\cdots&\phi_1^{[2n_1-3]}(\lambda_i) \\
&&& \\
0&0&0&\phi_1^{[1]}(\lambda_i) \\
\end{bmatrix},\,\,\,\,\, i=1,2.
\]
By the formula of the Cauchy determinant and taking its special limit, we find \[
\psi_n(0,0)=\frac{\det(\widehat{\mathbf{D}}_n)}{\det(\mathbf{D}_n)}=1+2(n_1{\rm Im}(\lambda_1)+n_2{\rm Im}(\lambda_2)),
\]
where
\[
 \widehat{\mathbf{D}}_n=\begin{bmatrix}
 \widehat{\mathbf{D}}^{[1,1]} &  \widehat{\mathbf{D}}^{[1,2]}\\
 \widehat{\mathbf{D}}^{[2,1]} &  \widehat{\mathbf{D}}^{[2,2]}\\
\end{bmatrix},\,\,\,\,\,
 \widehat{\mathbf{D}}^{[l,k]}=\left(\binom{i+j-2}{i-1}\frac{(-1)^{i-1}}{(\lambda_k-\lambda_l^*)^{i+j-1}}-\ii \delta_{1,1}\right)_{1\leq i\leq n_l,0\leq j\leq n_k},\]
and $1\leq l,k\leq 2$, $\delta_{1,1}=1$ if $i=j=1$, otherwise $\delta_{1,1}=0.$

Here we stress that the high order rogue waves given previously do not possess the free parameters. Actually, since the fundamental solution can involve lots of free parameters, it is easy to bring the free parameters into the formulas of rogue waves as in reference \cite{GuoLL12}.

Now we consider the several explicit examples of the rogue waves. Firstly, we consider the first order rogue wave. We fix the parameters $\lambda_1=b+c\ii $, $\lambda_2=-b+d\ii $, $\lambda_3=\lambda_1^*$ and $\lambda_4=\lambda_2^*$.
Using the above expansion, we can obtain that
\begin{equation}\label{eq:phi1-varphi1}
\begin{split}
\varphi_1&=\sqrt{u_1(x)}p(x)\exp\left(\theta_1(x,t)\right), \,\,\,\,\,\, \theta_1(x,t)=\ii \int_{0}^{x}\frac{{d}s}{u_1(s)}+\ii \lambda_1x+\ii \frac{s_2}{2}t\\
\phi_1&=\sqrt{\nu(x)}\sqrt{u_1(x)}\frac{(\lambda_1^2+(s_2-\nu(x))/2)p(x)+1}{\lambda_1\nu(x)+(2q+\ii \nu_x(x))/4}\exp\left(\theta_1(x,t)+\ii \theta_0(x,t)\right),
\end{split}
\end{equation}
where $\theta_0(x,t)=q\int_0^x\frac{{d}s}{\nu(s)}-s_2t$, $p(x)=\frac{1}{u_1(x)}+\ii t +J(x)+k_1$, $u_1(x)=\nu(x)-(2\lambda_1^2+s_2)$, $C_1=2\lambda_1(2\lambda_1^2+s_2)-s_3$, $k_1\in \mathbb{C}$, and the integration
\begin{equation*}
\begin{split}
J(x)&=-4\ii \lambda_1\int_{0}^{x}\frac{{d}s}{u_1(s)}-2\ii C_1\int_{0}^{x}\frac{{d}s}{u_1^2(s)}\\
&=A_1\left(\frac{E}{K}\alpha\,x+Z(\alpha\,x)\right)+A_2x+A_3\frac{{\rm sn}(\alpha\,x){\rm cn}(\alpha\,x){\rm dn}(\alpha\,x)}{u_1(x)},
\end{split}
\end{equation*}
where $\alpha=\sqrt{4b^2+(c+d)^2}$, $A_1=\frac{-\alpha}{2c A_0}$, $A_2=\frac{1}{2c}$, $A_3=\frac{2\alpha d}{A_0}$, $A_0=(\lambda_1-\lambda_2)(\lambda_1-\lambda_2^*)$.
Thus the rogue wave solution can be obtained as
\begin{equation}\label{eq:rw-sol}
\psi_1=\sqrt{\nu(x)}\left(1-\frac{\ii \,{\rm Im}(\lambda_1)[4\lambda_1\nu(x)+(2q+\ii \nu_x(x))][(\lambda_1^{*2}+(s_2-\nu(x))/2)|p(x)|^2+p(x)]}{|\lambda_1\nu(x)+(2q+\ii \nu_x(x))/4|^2|p(x)|^2+\nu(x)|(\lambda_1^2+(s_2-\nu(x))/2)p(x)+1|^2}\right) {\rm e}^{-\ii \theta_0(x,t)}.
\end{equation}
The parameter $k_1$ determines not only the location of the rogue wave but also its shape and dynamics, since it is influenced by the periodic modulation of the background. The rogue waves on cn background and dn background at upper branch point were given in \cite{ChenP18}.
If we switch the parameter $\lambda_1$
to $\lambda_2=-b+d\ii $, then we obtain another family of parameters $A_1=\frac{-\alpha}{2dA_0}$, $A_2=\frac{1}{2d}$, $A_3=\frac{2\alpha c}{A_0}$, $A_0=(\lambda_2-\lambda_1)(\lambda_2-\lambda_1^*)$, which will determine another family of rogue wave solutions arising from the different branch point.

Here if we intend to prove the asymptotic behavior when $x,t\to \infty$, we need to use the theta-function representation mentioned previously. It is easy to prove that the rogue wave solution approaches the background solution by the shift of phase and crest (dn,cn or non-trivial phase function). Since the calculation is similar to Sec.~\ref{sec4}, here we omit the detailed asymptotic analysis.

Using Theorem \ref{thm5}, we can also obtain the two-rogue-wave solution:
\begin{equation}\label{eq:two-rw}
\psi_2=\sqrt{\nu(x)}\left[\frac{\det(\mathbf{K}_2)}{\det(\mathbf{M}_2)}\right]\exp\left[-\ii q\int_{0}^{x}\frac{ {\rm d}s}{\nu(s)}+\ii  s_2 t\right],
\end{equation}
where
\[ \mathbf{M}_2=\begin{bmatrix}
\frac{|\varphi_1|^2+|\phi_1|^2}{\lambda_1-\lambda_1^*} &\frac{\varphi_1^*\varphi_2+\phi_1^*\phi_2}{\lambda_2-\lambda_1^*}  \\[5pt]
\frac{\varphi_1\varphi_2^*+\phi_1\phi_2^*}{\lambda_1-\lambda_2^*} & \frac{|\varphi_2|^2+|\phi_2|^2}{\lambda_2-\lambda_2^*} \\
\end{bmatrix},\,\,\,\,\,\,\, \mathbf{K_2}=\mathbf{M}_2-\frac{2{\rm e}^{\ii \theta_0(x,t)}}{\sqrt{\nu(x)}}\begin{bmatrix}
\phi_1^*\\[5pt]
\phi_2^* \\
\end{bmatrix}\begin{bmatrix}
\varphi_1,&
\varphi_2
\end{bmatrix},
 \]
 and $\phi_i$ and $\varphi_i$ are given by equations \eqref{eq:phi1-varphi1}.

\subsubsection{\textbf{Example 1: The rogue waves and two rogue waves on the dn solution background}} Since the NLSE possesses the scaling symmetry, we can fix the amplitude of background. So we choose the parameters $b=0$, $c=\frac{1}{2}(1+\kappa)$ and $d=\frac{1}{2}(1-\kappa)$, $(|\kappa|<1)$, then $\nu_1=0$, $\nu_2=\kappa^2$ and $\nu_3=1$. Moreover, $m=1-\kappa^2$, $q=-s_3=0$, $s_2=\frac{1}{2}(1+\kappa^2)$. It follows that the elliptic function solution is
\[
\psi={\rm dn}(x|1-\kappa^2){\rm e}^{\frac\ii {2}(1+\kappa^2)t}.
\]
By fixing the background solution, two different families of rogue wave solution can be obtained by the formula \eqref{eq:rw-sol} depending on the choice of different branch points. When $\kappa\to 0$, we have $\psi\to \exp(\ii t/2)$. So when $\kappa$ is small and approaches zero, the rogue waves would exhibit the similar scenario as the classical rogue waves on the plane wave background. Since there exist two different types of rogue waves, it is shown that while one of the two is similar to the classical one, the other one is different from the classical one, which locates on the lower branch point. When $\kappa$ is small enough, these rogue waves vanish and become the plane waves. On the other hand, when $\kappa\to 1$, then $\psi\to {\rm sech}(x)$. In this case, there still exist two types of rogue waves. It is shown that the rogue waves seem a second-order soliton in the center part. In this case, two types of rogue waves have the similar structure.

\begin{figure}[tb]
\centering
\begin{tabular}{cc}
\includegraphics[width=.45\textwidth]{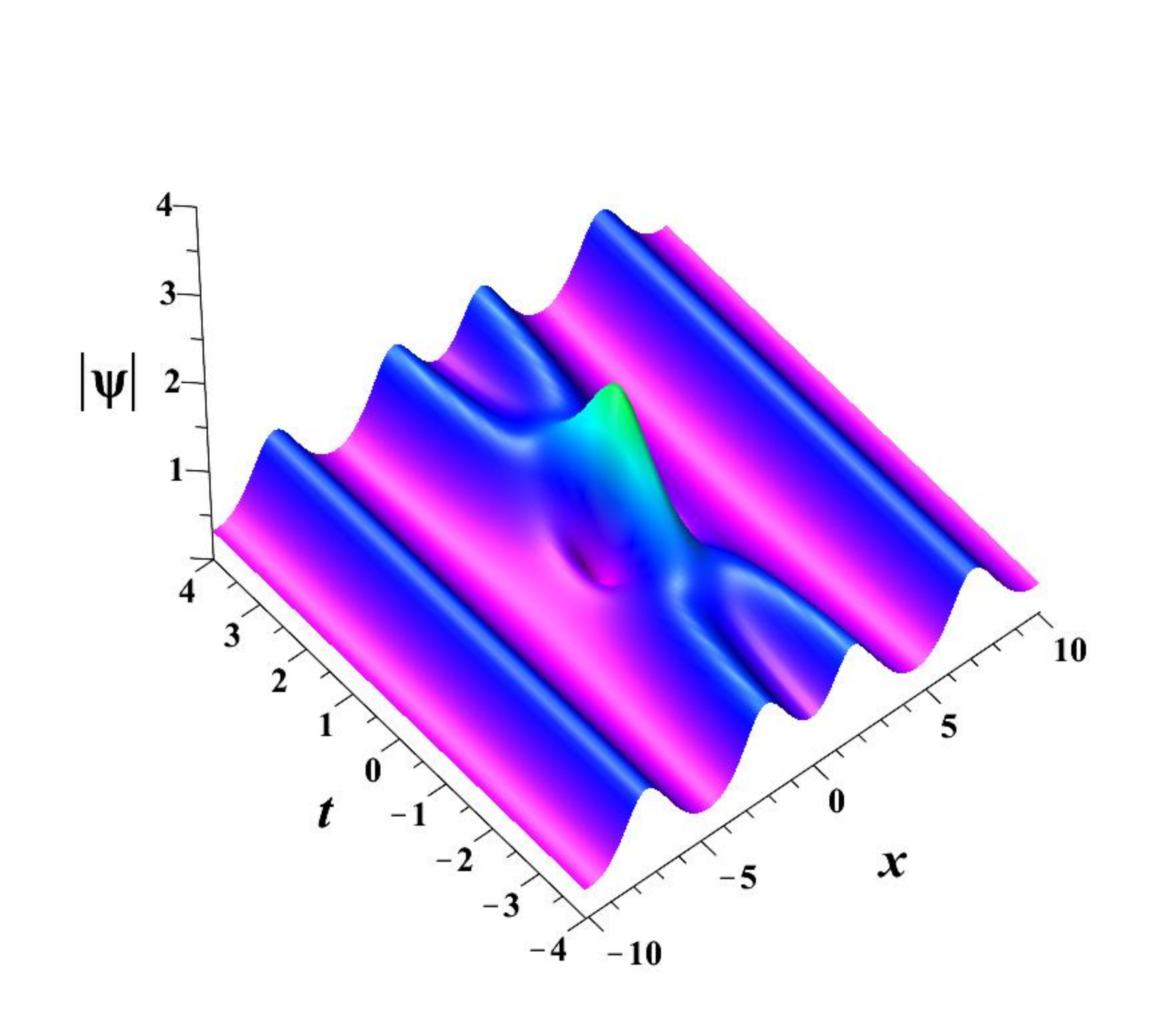} & \includegraphics[width=.47\textwidth]{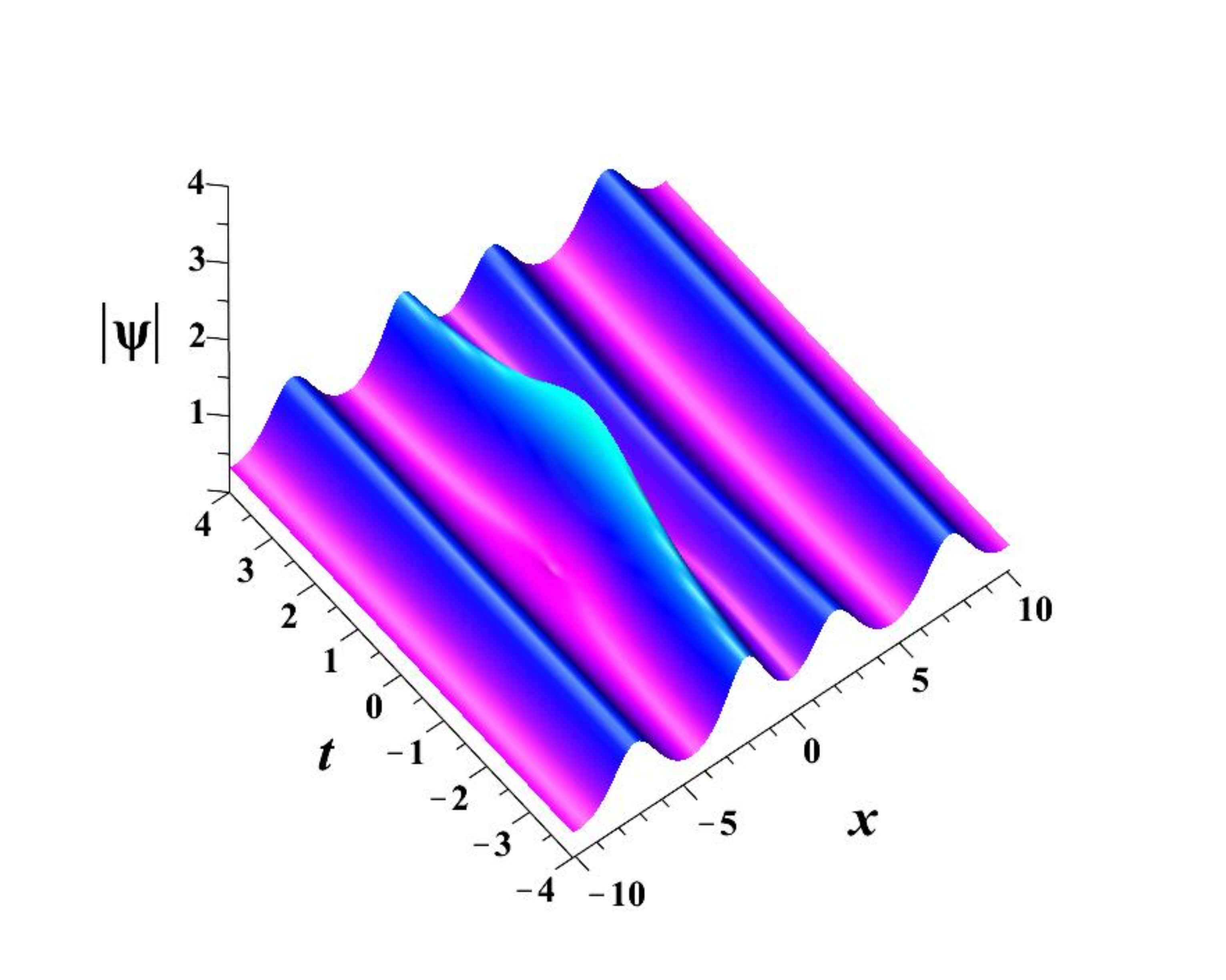} \\
{\footnotesize (a) dn rogue wave} &  {\footnotesize (b) dn rogue wave at another position} \\
\includegraphics[width=.44\textwidth]{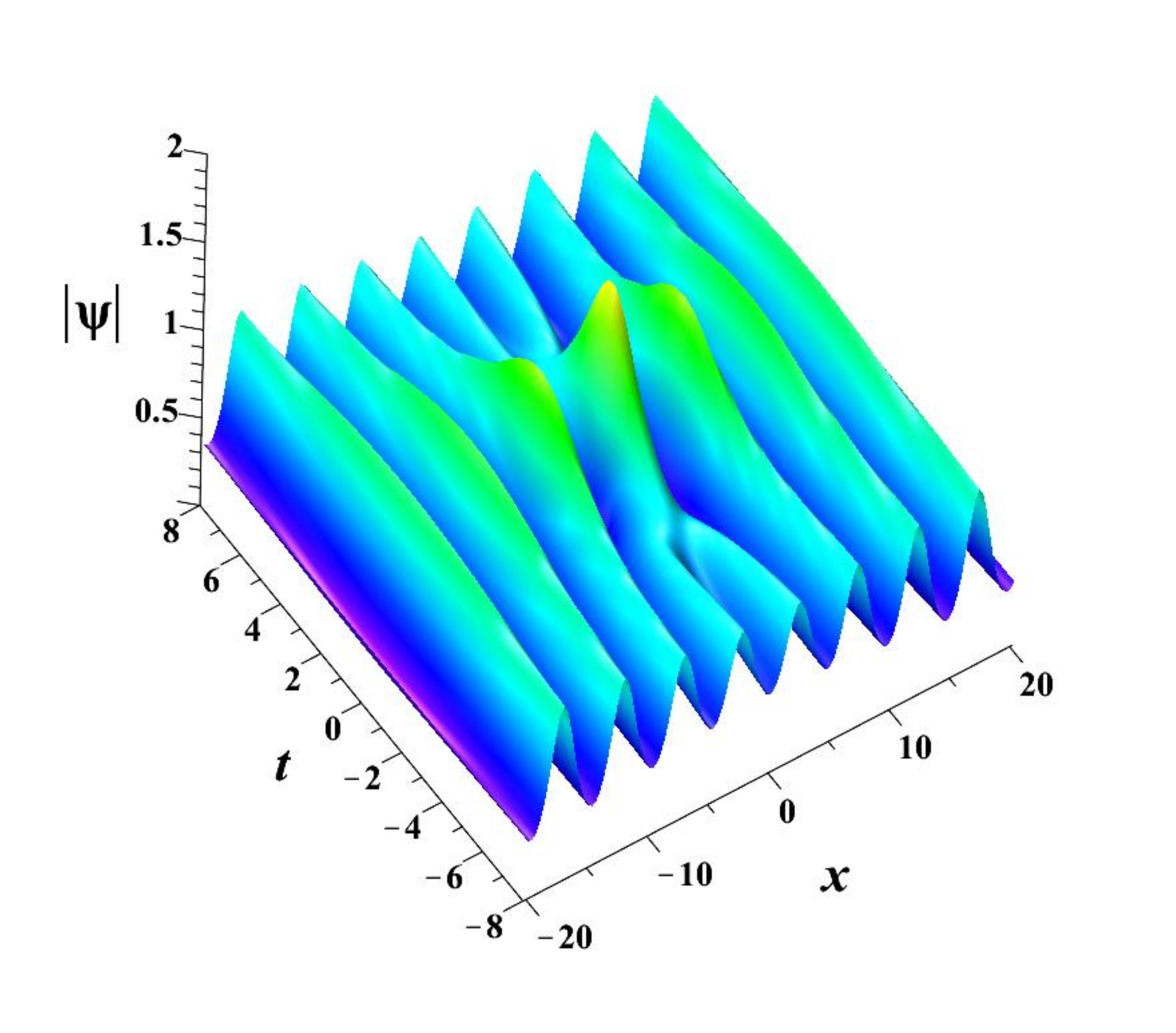} & \includegraphics[width=.5\textwidth]{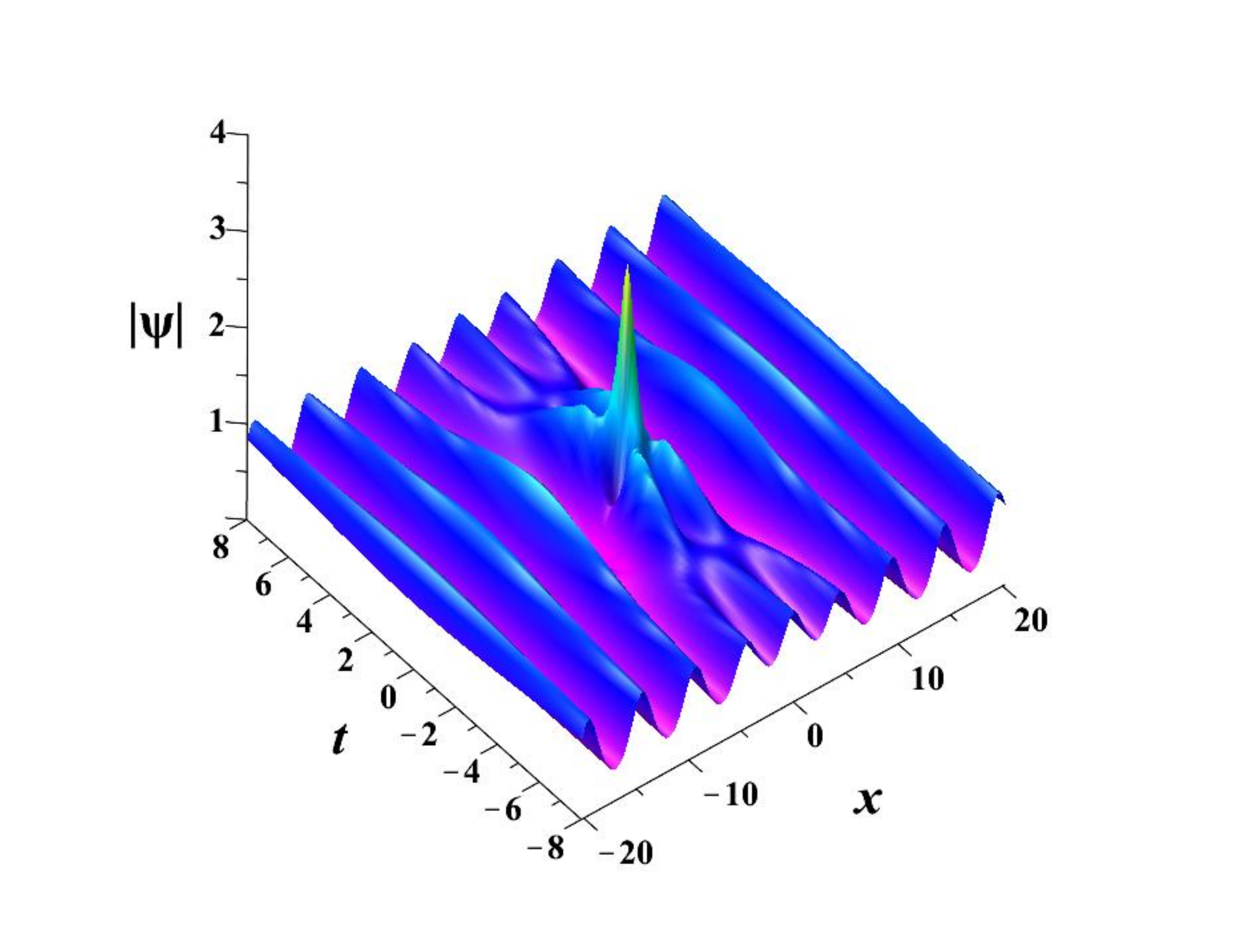} \\
{\footnotesize (c) dn rogue wave for another branch point} &  {\footnotesize (d) two-dn rogue wave}
\end{tabular}
\caption{(color online): Rogue wave solutions on the dn background. (a) The dn rogue wave with $\lambda_1=\frac{2}{3}\ii$ and $k_1=0$. (b) The dn rogue wave with $\lambda_1=\frac{2}{3}\ii$ and $k_1=2$. (c) The dn rogue wave with $\lambda_2=\frac{1}{3}\ii$ and $k_1=0$. (d) The two-dn rogue wave.}
\label{fig:dn-rw}
\end{figure}

Choosing the parameters $\kappa=\frac{1}{3}$, $\lambda_1=\frac{2}{3}\ii$, $k_1=0$ and inserting them into formula \eqref{eq:rw-sol}, we obtain the rogue wave solution on the dn background (Fig. \ref{fig:dn-rw} (a)). The maximum value of the rogue wave is $\frac{7}{3}$. It is seen that the central part of the dn rogue wave under this parameter looks likes a second-order soliton. For the other part, it is still likely the dn solution. If we change the parameter with $k_1=2$, the shape of the rogue wave changes  (Fig. \ref{fig:dn-rw} (b)). The maximum value of this rogue wave is about $1.89$, which is less than $7/3$. Meanwhile the emergent peak is deviated from the line $x=0.$

Choosing another type parameters $\kappa=\frac{1}{3}$, $\lambda_2=\frac{1}{3}\ii$, $k_1=0$ and plugging them into formula \eqref{eq:rw-sol}, we obtain another type of rogue waves. The maximum value is about $\frac{5}{3}$ (Fig. \ref{fig:dn-rw}(c)). In \cite{ChenP18}, the authors do not give this case since the integration is singular. In the same way as above, if the parameter $k_1$ varies, the shape also changes. Here we do not show the figure.

It is interesting that the two rogue waves can exist simultaneously, which can be achieved by the two-rogue wave formula \eqref{eq:two-rw}.
Taking the parameters $\kappa=\frac{1}{3}$, $\lambda_1=\frac{2}{3}\ii$, $\lambda_2=\frac{1}{3}\ii$, $k_1=k_2=0$ and inserting them into the formula of two rogue waves \eqref{eq:two-rw}, we obtain the two rogue waves with the maximum value $3$ at origin (Fig. \ref{fig:dn-rw}(d)). In this parameter setting, the two rogue waves attain the maximum value. If we change the parameters $k_1$ and $k_2$, we  obtain two types of rogue wave situated at different locations.

\subsubsection{\textbf{Example 2: The rogue waves and two rogue waves on the cn solution background}} Since the NLSE possesses the scaling symmetry, we can fix the amplitude of background. So we choose the parameters $c=d=1/2$, then $\nu_1=-4b^2$, $\nu_2=0$ and $\nu_3=1$. Moreover, $m=1/(1+4\,b^2)$, $q=-s_3=0$, $s_2=\frac{1}{2}-2b^2$. It follows that the elliptic-function solution is
\[
\psi={\rm cn}(\sqrt{1+4b^2}x|m){\rm e}^{\ii (\frac{1}{2}-2b^2)t}.
\]
By fixing the background solution, two different families of rogue wave solutions originating from two branch points can be obtained by the formula \eqref{eq:rw-sol}. In this case, we find that two kinds of rogue waves move to the left and right. When $\kappa\to 0$, we have $\psi\to \exp(\ii t/2)$. So when $\kappa$ is small and is close to zero, the rogue waves exhibit the similar scenario as the classical rogue waves on the plane wave background. Since there exist two different types of rogue waves, it is shown that one of them is similar to the classical one, while the other one is different from the classical one, which locates on the lower branch point. When $\kappa$ is small enough, these rogue waves merge into the mere plane wave solution. On the other hand, when $\kappa\to 1$, i.e.,  $\psi\to {\rm sech}(x)$, there still exist two families of rogue waves. It is shown that the rogue waves looks like a second-order soliton in the center part, and under this case, two rogue waves have the similar structure.

\begin{figure}[tb]
\centering
\subfigure[cn rogue wave moving to the right]{%
\includegraphics[width=.47\textwidth]{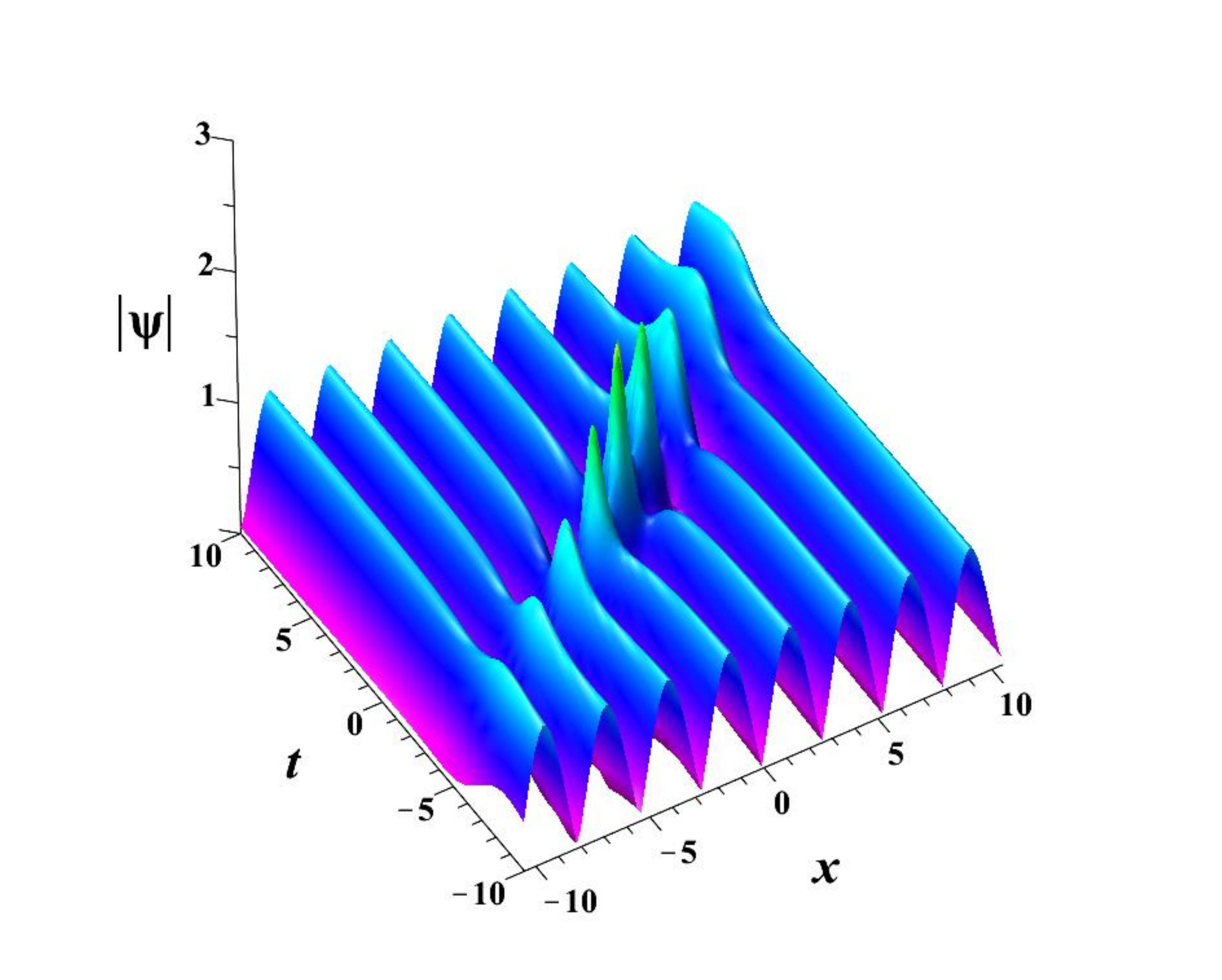}} \hfil
\subfigure[cn rogue wave moving to the left]{%
\includegraphics[width=.47\textwidth]{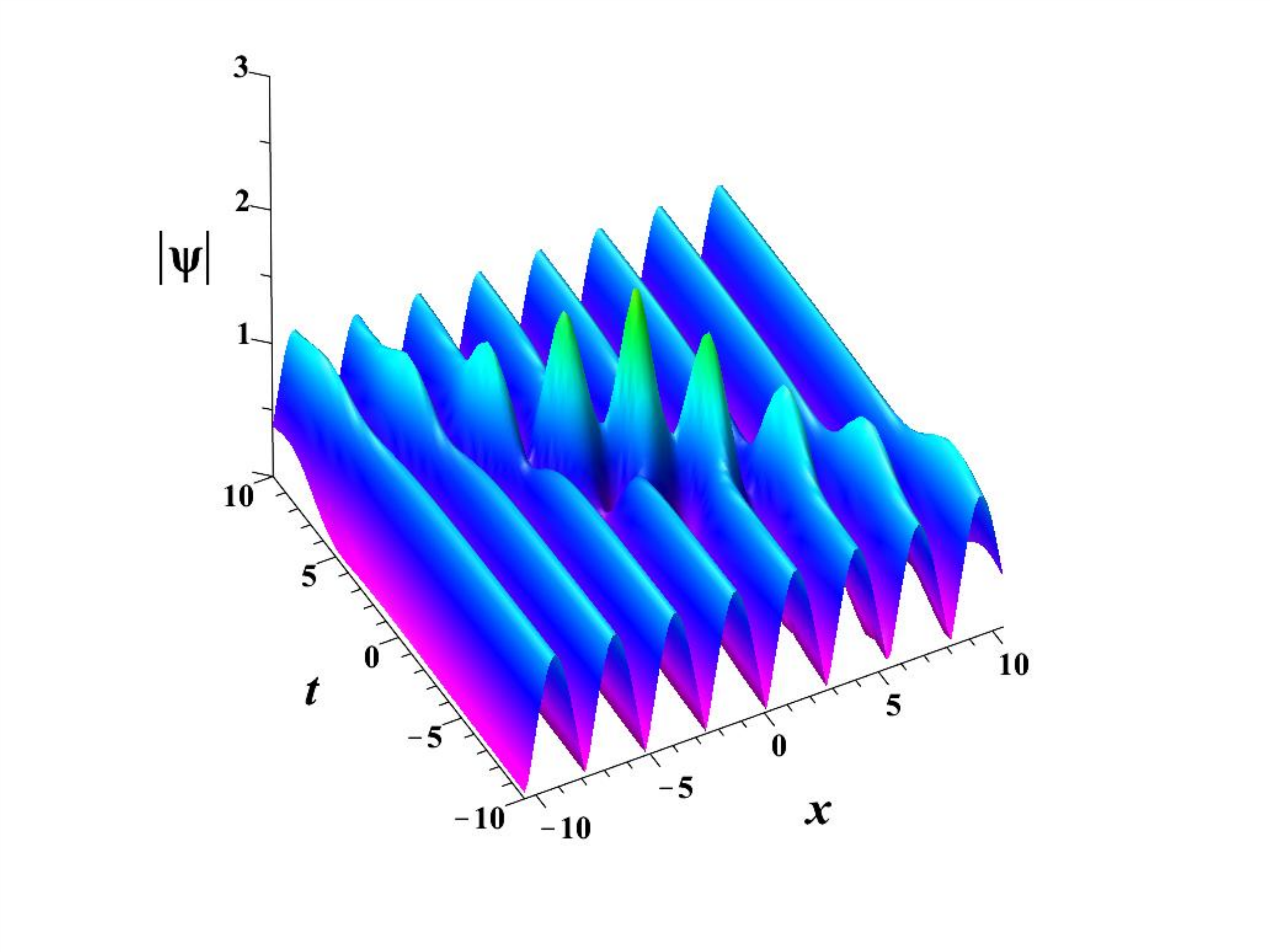}} \\
\subfigure[two-cn-rogue waves]{%
\includegraphics[width=.47\textwidth]{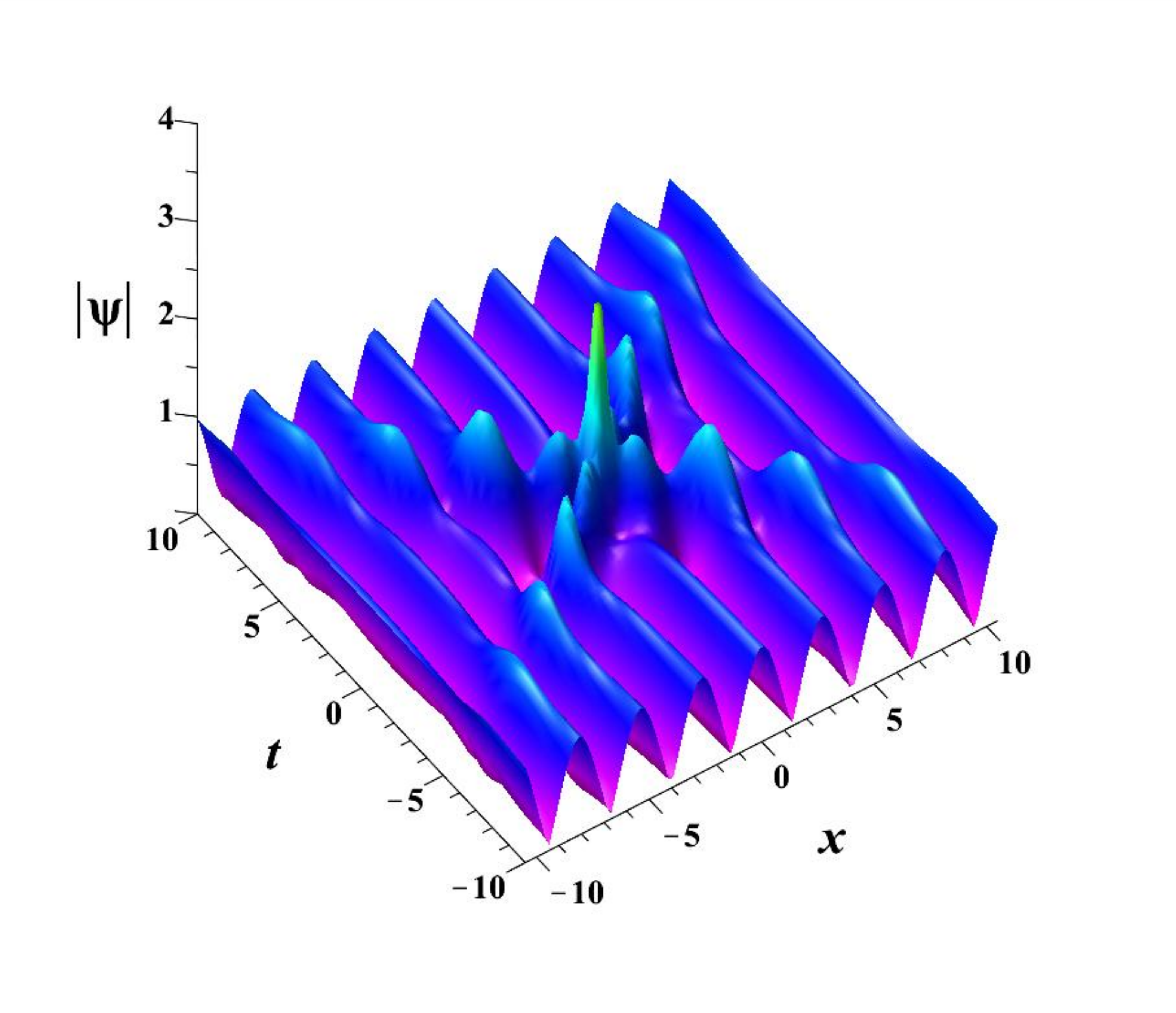}} \hfil

\caption{(color online): Rogue wave solutions with cn background. (a) The cn rogue wave with positive velocity. (b) The cn rogue wave with negative velocity. (c) The two-cn-rogue waves.}
\label{fig:cn-rw-1}
\end{figure}

Choosing the parameters $b=\frac{1}{2}$, $\lambda_1=-\frac{1}{2}+\frac{\ii}{2}$, $k_1=0$, we plot the figure for the rogue wave on the ${\rm cn}$ function background. The maximum value equals to 2 located at origin (Fig. \ref{fig:cn-rw-1} (a)). Also, if we change the parameters $\lambda_2=\frac{1}{2}+\frac{\ii}{2}$, $k_2=0$, we obtain another type of rogue waves. The maximum value equals to 2 located at origin (Fig. \ref{fig:cn-rw-1} (b)).

For the two rogue waves, we set the parameters $b=\frac{1}{2}$, $\lambda_1=-\frac{1}{2}+\frac{\ii}{2}$, $\lambda_2=\frac{1}{2}+\frac{\ii}{2}$, $k_1=k_2=0$, and then we find that the maximum value of two rogue waves equals to 3 located at origin (Fig. \ref{fig:cn-rw-1}(c)). If we change the parameters $k_1$ and $k_2$, two rogue waves change their locations.

\subsubsection{\textbf{Example 3: The rogue waves and two rogue waves on the general elliptic-function solution background}} Since the NLSE possesses the scaling symmetry, we can fix the amplitude of background. So we choose the parameters $b\neq 0$, $c=\frac{1}{2}(1+\kappa)$ and $d=\frac{1}{2}(1-\kappa),$ $(\kappa\neq0)$, then $\nu_1=-4b^2$, $\nu_2=\kappa^2$ and $\nu_3=1$. Moreover, $m=(1-\kappa^2)/(1+4b^2)$, $q=-s_3=-2b\kappa$, $s_2=-2b^2+\frac{1}{2}(1+\kappa^2)$. It follows that the elliptic function solution is
\[
\psi=\sqrt{1-(1-\kappa^2)\,{\rm sn}^2(\sqrt{1+4b^2}x|m)
}\exp\left(\ii s_2t-\frac{2b\kappa\ii }{\sqrt{1+4b^2}}\Pi({\rm am}(\sqrt{1+4b^2}x|m);1-\kappa^2|m)\right).
\]
By fixing the background solution, the two different families of the rogue wave solutions can be obtained by the formula \eqref{eq:rw-sol} corresponding to the two branch points. In this case, we obtain two rogue waves which have the asymmetry pairs.

\begin{figure}[tbh]
\centering
\subfigure[non-trivial phase rogue waves]{%
\includegraphics[width=.48\textwidth]{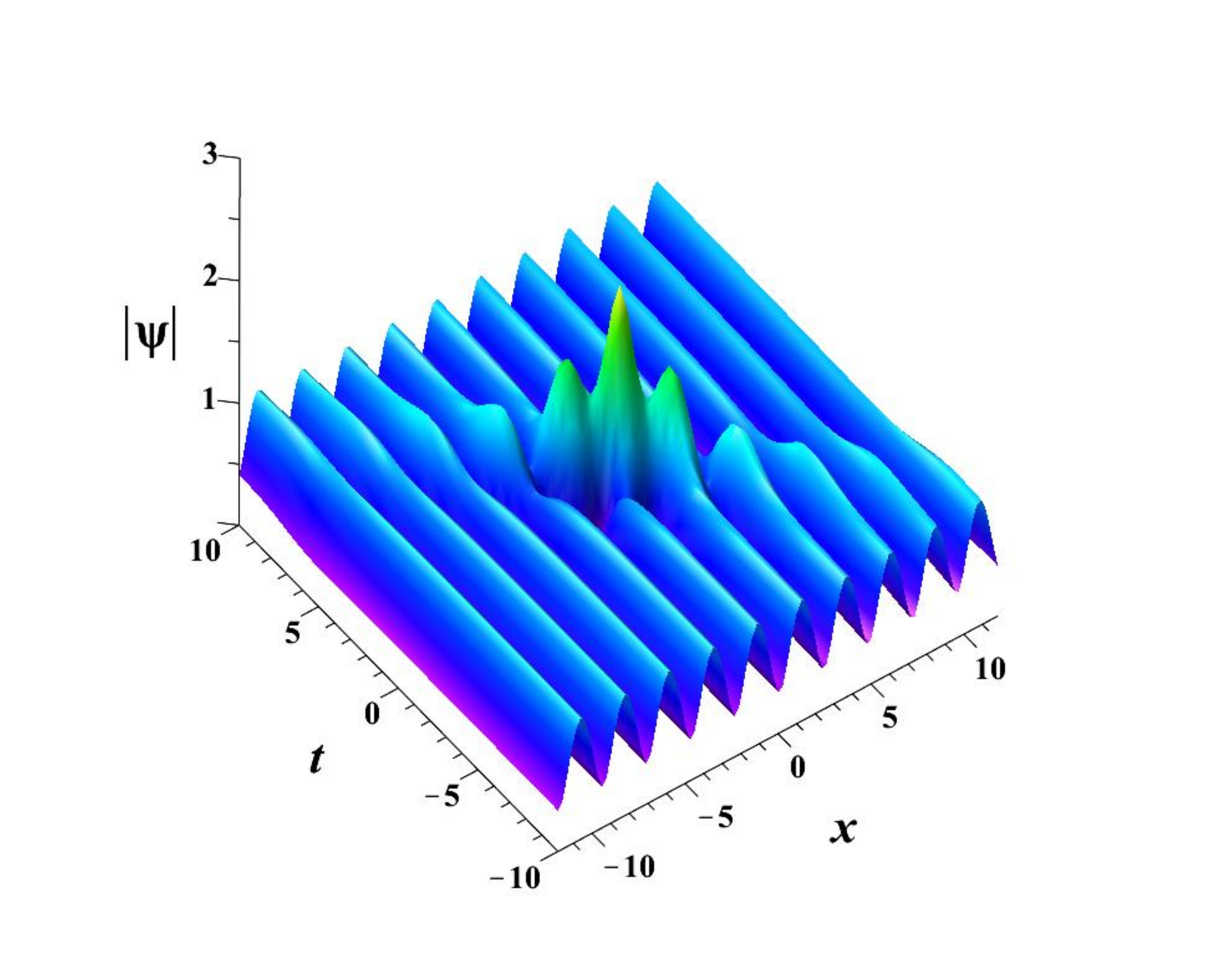}} \hfill
\subfigure[non-trivial phase rogue waves]{%
\includegraphics[width=.45\textwidth]{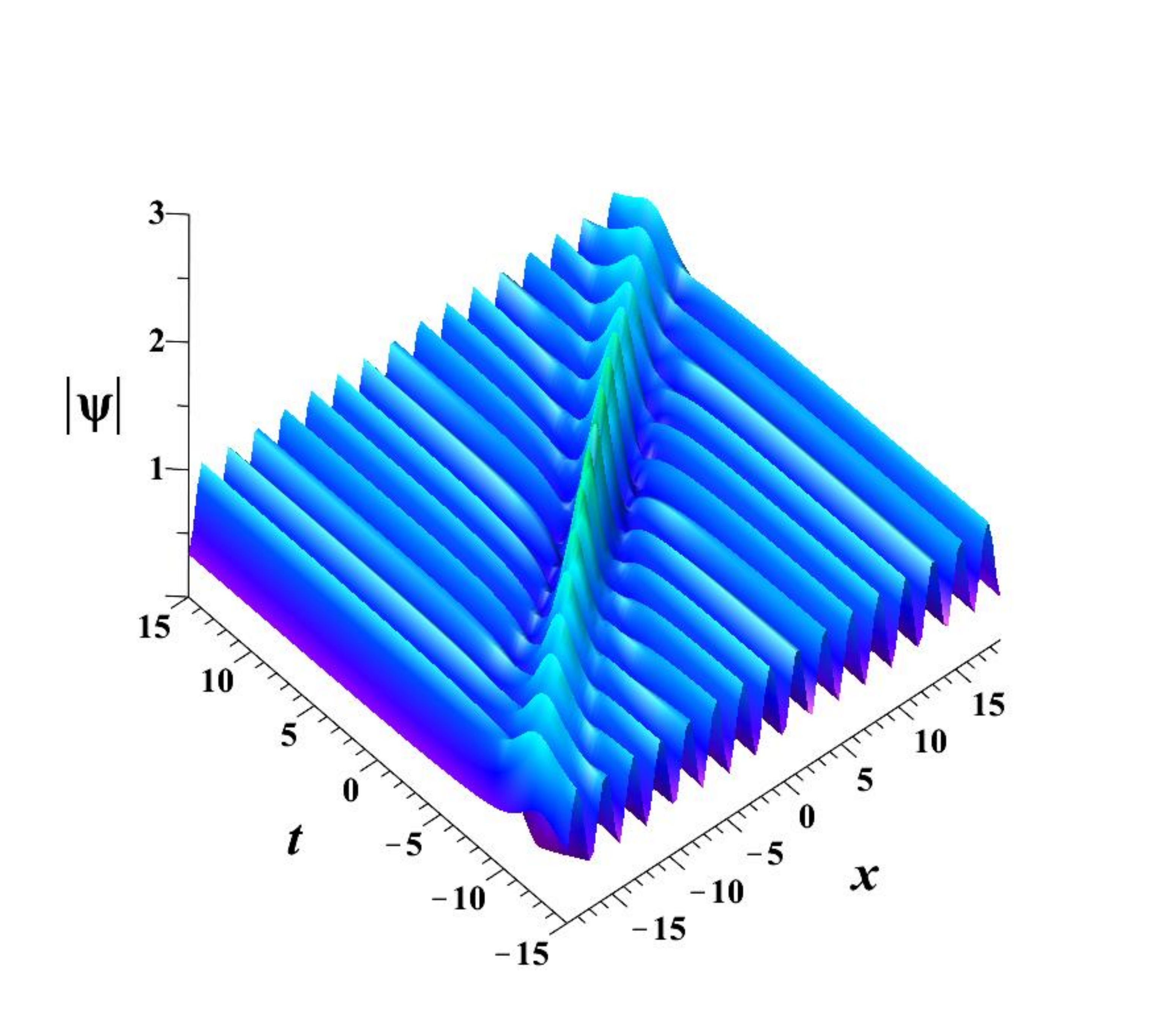}} \\
\subfigure[two-rogue waves]{%
\includegraphics[width=.48\textwidth]{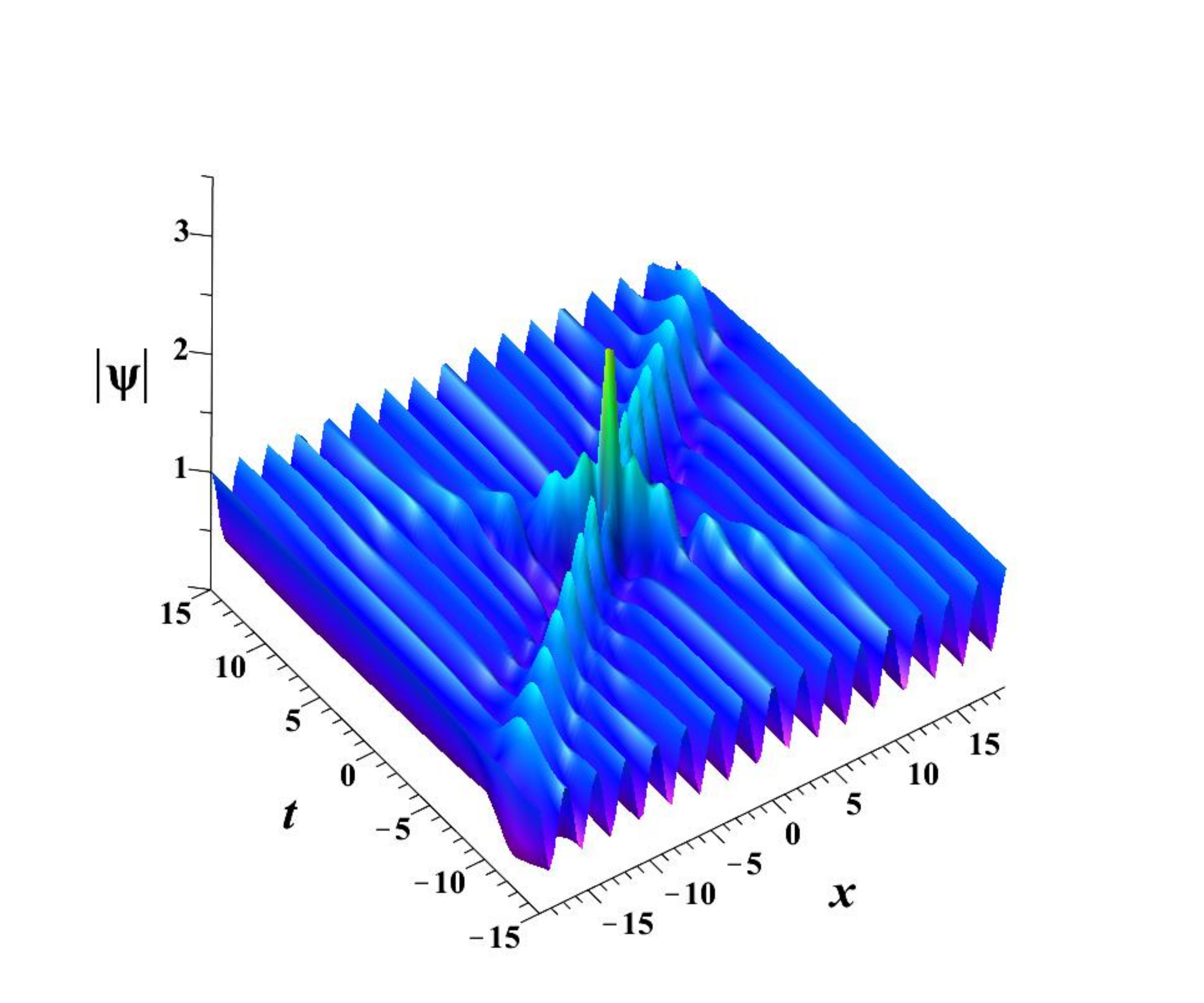}}
\caption{(color online): The rogue wave solutions with general elliptic-function solution background. (a) and (b) non-trivial phase rogue waves generated from the different branch points. (c) The two-rogue waves}
\label{fig:gene-rw}
\end{figure}

Inserting the parameters $b=\frac{\sqrt{5}}{4}$, $\kappa=\frac{1}{3}$, $\lambda_1=-\frac{\sqrt{5}}{4}+\frac{\ii}{3}$, $k_1=0$ into the formula \eqref{eq:rw-sol}, we obtain the rogue waves on the non-trivial phase background (Fig. \ref{fig:gene-rw}(a)). The maximum value is $\frac{5}{3}$ located at origin.

Similarly, inserting the parameters $b=\frac{\sqrt{5}}{4}$, $\kappa=\frac{1}{3}$,  $\lambda_2=\frac{\sqrt{5}}{4}+\frac{2\ii}{3}$, $k_1=0$ into the formula \eqref{eq:rw-sol}, we arrive at another type of rogue waves (Fig. \ref{fig:gene-rw}(b)). In this parameter setting, the maximum value is $\frac{7}{3}$.

Plugging the parameters $b=\frac{\sqrt{5}}{4}$, $\kappa=\frac{1}{3}$, $\lambda_1=-\frac{\sqrt{5}}{4}+\frac{\ii}{3}$, $\lambda_2=\frac{\sqrt{5}}{4}+\frac{2\ii}{3}$, $k_1=k_2=0$ into the formula \eqref{eq:rw-sol}, we obtain the two rogue waves, which
attain the maximum amplitude $3$ (Fig. \ref{fig:gene-rw}(c)).

\subsubsection{\textbf{Example 4: Multi high order rogue waves}}

For the general higher-order rogue waves, we have given the general formula above. In practice, since the expression $y_1$ involves the roots of complex number, we can absorb the parameter $y_1$ into parameter $\epsilon_i$ to make the expressions more compact. On the other hand, the general expression for the elements are very complicated, and hence we only provide the second-order and two-second order rogue waves in the ${\rm cn}$ background. For the systematic treatment of the dynamics of general higher-order rogue waves, we would like to exhibit them in the other works.

By the computer plotting, we exhibit the figures for the fundamental second-second order rogue waves (Fig. \ref{fig: sec-sec-rw}), in which the exact formulas are shown in Appendix~\ref{sec:appC}. Both figures show the same solution by different coordinates to display the dynamics clearly. From the left figure, it is seen that four solitons collide at the origin, where they show the maximum value $5$. When $x$ and $t$ become large enough, the rogue waves tend to the background solution ${\rm cn}(2x|\frac{1}{2}){\rm e}^{-\ii t},$ which can not be observed from the picture directly. The right figure shows the detail of the dynamics near the origin $(x,t)=(0,0)$, implying the more complicated profile than that of the plane-wave background.

\begin{figure}[tbh]
\centering
\includegraphics[width=.48\textwidth]{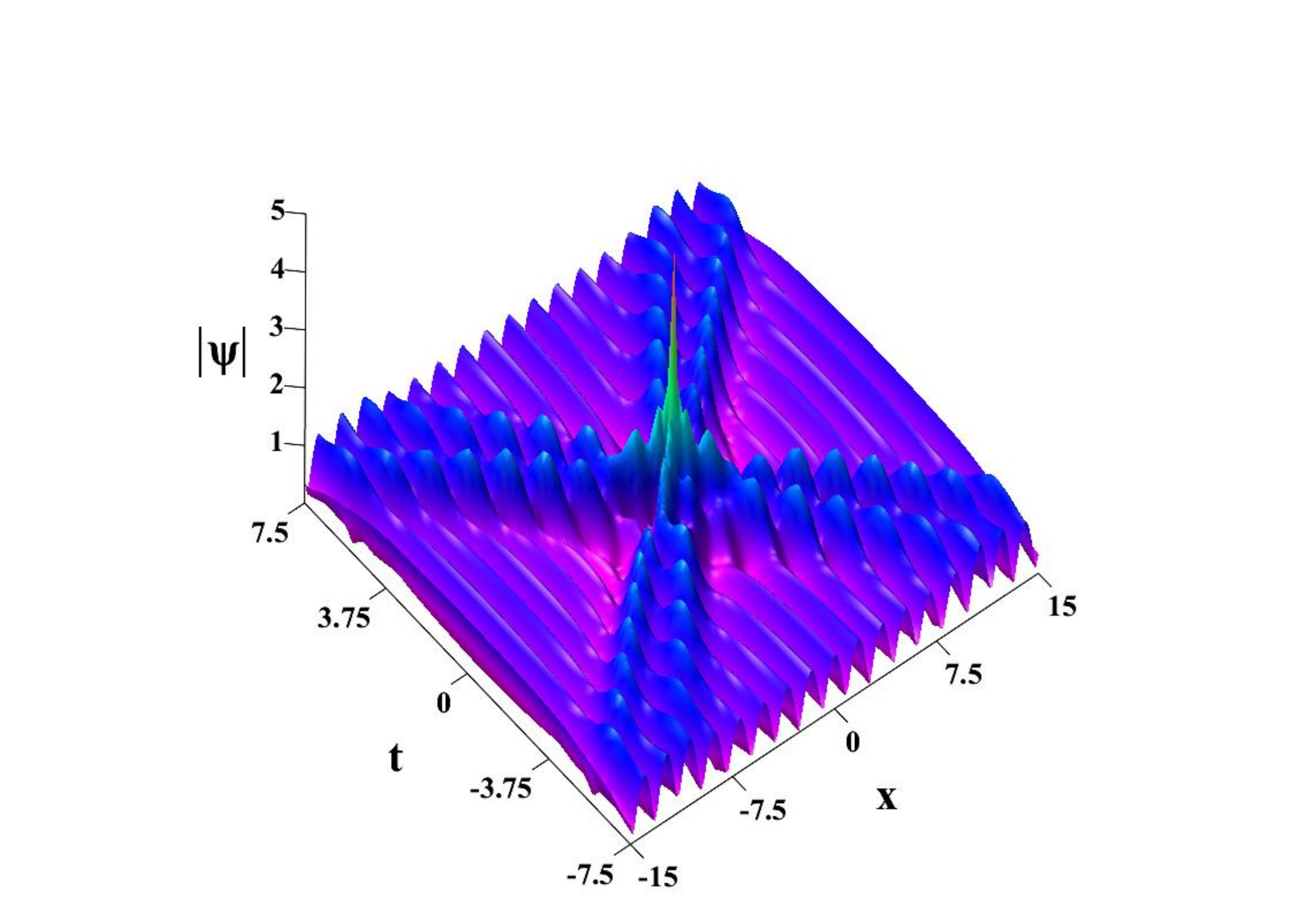} \hfil
\includegraphics[width=.48\textwidth]{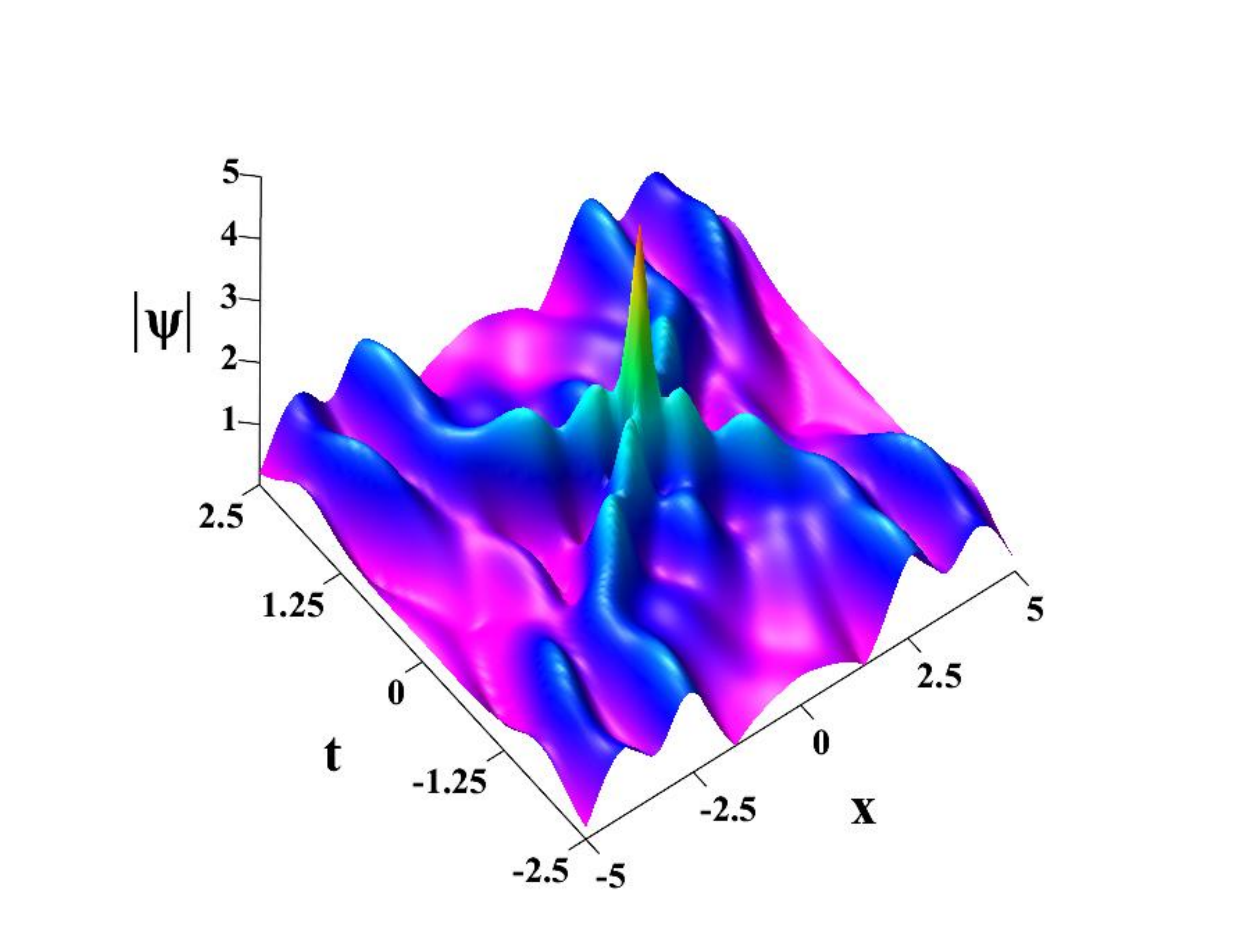}
\caption{(color online): The second-second order rogue waves. The left and right figures represent the same solution with different plot ranges.}
\label{fig: sec-sec-rw}
\end{figure}

\section{Conclusion and Discussion}\label{sec7}
In this work, we systematically constructed the breathers and rogue waves for the focusing NLSE on the elliptic function background by the Darboux transformation or its generalized version. What is more, we provided the asymptotic analysis of breathers with different velocity by using the addition formula of theta functions. To verify the asymptotic analysis, we plot some figures of interaction of two breathers. Furthermore, we establish the relation between modulational instability and the Akhmediev breather or rogue wave solutions on the elliptic function background by the linear stability analysis, which describe the dynamical destabilization process of modulational instability in the same way as the case of the plane-wave background. The method provided here can be easily extended to the other integrable model, such as the sine-Gordon equation, modified KdV equation, derivative nonlinear Sch\"odinger equation and so on.

For the physical significance, the solutions found in this work will be observed in physical experiments such as nonlinear fibre optics \cite{Kibler2010} with oscillating background and the rogue waves in the presence of the soliton trains in Bose-Einstein condensates \cite{Strecker2002,Khawaja2002,Carr2000}. Generalization of the solutions to the multi-component NLSE with modulated background will also be an interesting future issue, and there are several relevant works from both experiment and theory \cite{Hamner2011,Li2018,Kurosaki2007}.

The long-time asymptotics for the focusing NLSE with plane wave background without discrete spectrum tends to an elliptic function in a triangle region \cite{BiodiniM17}. It can be expected that the long-time asymptotics with discrete spectrum can be represented by the breathers in this work.
Meanwhile, the classical inverse scattering method on plane wave background can not involve the rogue waves and high order rogue waves directly; we need to perform a sensitive limiting procedure to obtain them. Recently, Bilman and Miller \cite{BilmanM17} proposed a new scheme called robust inverse scattering transform to deal with the spectral singularity with the aid of the Darboux transformation. Meanwhile, it is hopeful to be used in analyzing the long-time asymptotics for the rogue waves and high order rogue wave on the plane wave background. The rogue waves and high order rogue waves are the possible results for their long-time dynamics. For both of the above works, they need to be analyzed in the future work with the aid of Riemann-Hilbert representation \cite{KotlyarovS17,BertolaT17}.

\appendix
\renewcommand\thesection{\Alph{section}}
\makeatletter
\renewcommand{\theequation}{%
\thesection.\arabic{equation}}
\@addtoreset{equation}{section}
\makeatother

\section{Darboux transformation on the elliptic function background}\label{sec:appA}
In this appendix, we give the relative results on the Darboux transformation. The theory for the Darboux transformation is very mature.

\subsection*{Elementary Darboux transformation and its multi and higher order version}
The elementary Darboux transforation can be stated by the following theorem, which is essential to yield the multi-fold and higher order versions \cite{Cieslieski09,MatveevS91,GuoLL12}.

\begin{theorem}
Assume that there is an entire matrix function $\Phi(x,t;\lambda)$ for the Lax pair \eqref{eq:lax-nlse} with the potential function $\psi(x,t)\in \mathbf{L}^{\infty}(\mathbb{R}^2)\cup\mathbf{C}^{\infty}(\mathbb{R}^2)$, which is normalized at $(x,t)=(0,0)$: $\Phi(0,0;\lambda)=\mathbb{I}$, then
the Darboux transforation
\begin{equation}
\mathbf{T}_1=\mathbb{I}-\frac{\lambda_1-\lambda_1^*}{\lambda-\lambda_1^*}\mathbf{P}_1,\,\,\,\,\, \mathbf{P}_1=\frac{\Phi_1\Phi_1^{\dag}}{\Phi_1^{\dag}\Phi_1},
\end{equation}
where $\Phi_1=\Phi(x,t;\lambda_1)(1,c_1)^{\rm T}\equiv(\phi_1,\varphi_1)^{\rm T}$ is a special solution $\lambda=\lambda_1$, converts the Lax pair \eqref{eq:lax-nlse} into the Lax pair with same form \eqref{eq:lax-nlse} by replacing the potential function $\psi(x,t)$ with a new one $\psi_1(x,t)$
\begin{equation}
\psi_1(x,t)=\psi(x,t)+\frac{2(\lambda_1^*-\lambda_1)\varphi_1^*\phi_1}{|\phi_1|^2+|\varphi_1|^2},
\label{eq:backlund}
\end{equation}
and the normalized matrix solution for the new system is $\Phi[1]=\mathbf{T}_1(x,t;\lambda)\Phi(x,t;\lambda)\mathbf{T}_1(0,0;\lambda)^{-1}$, which is also an entire matrix function with respect to $\lambda$.
\end{theorem}

In the background of the elliptic function, we need to the entire fundamental matrix solution $\Phi^{\rm N}(x,t;\lambda)$ given in \eqref{eq:fund-sol-norm}.

Now we analyze the properties for the elementary Darboux transformation and B\"acklund transformation. Through the B\"acklund transformation and mean value inequality, we know that the amplitude of the new potential function ranges in the interval $[\min(|\psi|)-2\mathrm{Im}(\lambda_1),\max(|\psi|)+2\mathrm{Im}(\lambda_1)]$. Now if we need to choose the solution $q_1$ to attend the maximum value $\max(|\psi|)+2\mathrm{Im}(\lambda_1),$ then $\varphi_1/\phi_1=-\ii.$ For simplicity, we assume that $\psi_1(0,0)=1$ and $\varphi_1(0,0)=-\ii.$ For instance, if we choose the seed solution as the plane wave solution $\psi=\psi_{BC}=1$. Meanwhile, the rogue wave solutions is located on the branch point $\lambda=\ii,$ in which case the vector solution $\Psi_1$ is not a periodic solution, but a rational function. To make the solution attend the maximum at the origin, we should impose the initial condition for the vector solution as $\Phi_1(0,0)=k(1,\ii )^{\rm T}.$

Actually, even though the formula for the B\"acklund transformation is simple, the new generated function is not easy to analyze by the classical tools of mathematical analysis for the elliptic-function background. For the breathers on the elliptic function background, we define that the center of breather is located at the maximum value point, which we choose the solution $\Phi_1=\Phi(x,t;\lambda_1)(1,-\ii)^{T}$. In practice, the local minimum value can also be seen as the center of breather, i.e. $\Phi_1=\Phi(x,t;\lambda_1)(1,\ii)^{T}$. For the other parameter choices, the  determination of the center of breathers need the use of the numerical method.

Based on the above elementary Darboux transformation, we can iterate it to obtain the multi- and higher order ones. If we iterate the Darboux transformation with \textit{different} spectral parameters $\lambda_i$, $i=1,2,\cdots,n$, then we obtain the \textit{multi-fold} Darboux matrix. Otherwise, if we iterate it with \textit{the same} spectral parameter, then the \textit{higher order} Darboux matrix can be obtained. For the multi-fold Darboux matrix, it is purely algebraic procedure. On the other hand, for the higher order Darboux matrix, because the iteration is always performed on the same spectral point which is a moving singularity point for the higher order iteration, we should use the Laurent series expansion to continue the higher order iteration.

Firstly, we consider the multi-fold Darboux matrix and its B\"acklund transformation. Suppose we have $n$ different spectral parameters $\lambda_i$, $i=1,2,\cdots,n$ and its corresponding solutions $\Phi_i=\Phi(x,t;\lambda_i)(1,c_i)^{T}$, then the $n$ times of iteration can reduce as the $n$-fold Darboux matrix:
\begin{equation}\label{eq:n-fold}
\mathbf{T}[n](x,t;\lambda)=\mathbf{T}_n(x,t;\lambda)\mathbf{T}_{n-1}(x,t;\lambda)\cdots \mathbf{T}_{1}(x,t;\lambda)=\mathbb{I}-\mathbf{X}_n(x,t)\mathbf{M}_n(x,t)^{-1}\mathbf{D}_n(\lambda)^{-1}\mathbf{X}_n(x,t)^{\dag}
\end{equation}
where
\begin{equation*}
  \begin{split}
      \mathbf{M}_n(x,t)=&\left(\frac{\Phi_i^{\dag}\Phi_j}{\lambda_j-\lambda_i^*}\right)_{1\leq i,j\leq n},  \\
      \mathbf{X}_n(x,t)=&\left[\Phi_1,\Phi_2,\cdots,\Phi_n\right],\\
      \mathbf{D}_n(\lambda)=&{\rm diag}\left(\lambda-\lambda_1,\lambda-\lambda_2,\cdots,\lambda-\lambda_n\right).
  \end{split}
\end{equation*}
The corresponding B\"acklund transformation is
\begin{equation}\label{eq:backlund-n-fold}
\psi_n=\psi-2\mathbf{X}_{n,1}\mathbf{M}_n(x,t)^{-1}\mathbf{X}_{n,2}^{\dag}.
\end{equation}

The higher order Darboux matrix corresponds to the iterated Darboux matrix with the same spectral parameter, which is useful in constructing the higher order soliton and higher order rogue waves. In the elliptic-function background, there are two pairs of conjugated roots. Thus we consider the following higher order Darboux matrix with two different parameters $\lambda_1$ and $\lambda_2$.
To construct the higher order Darboux matrix, we expand the vector solution $\Phi(x,t;\lambda)$ at the points $\lambda=\lambda_1$ and $\lambda=\lambda_2$:
\begin{equation}
\Phi(x,t;\lambda)=\sum_{i=0}^{\infty}\Phi_1^{[i]}(\lambda-\lambda_1)^i,\,\,\,\,\,\,
\Phi(x,t;\lambda)=\sum_{i=0}^{\infty}\Phi_2^{[i]}(\lambda-\lambda_2)^i.
\end{equation}

Now we construct the higher order Darboux matrix and the corresponding B\"acklund transformation based on $\Phi$. We consider the one from vector solution $\Phi$:
 \begin{equation}
 \mathbf{T}[n_1,n_2](x,t;\lambda)=\mathbb{I}-\mathbf{X}\mathbf{M}^{-1}\mathbf{D}\mathbf{X}^{\dag},\,\,\,\,\,\mathbf{M}=\mathbf{Y}\mathbf{S}\mathbf{Y}^{\dag},
 \end{equation}
where
\begin{equation}
\begin{split}
\mathbf{X}_a=&\left[\Phi_a^{[0]},\Phi_a^{[1]},\cdots, \Phi_a^{[n_a-1]}\right],\,\,\,\,\,\,\mathbf{S}_{a,b}=\left(\frac{(-1)^{j+1}}{(\lambda_b-\lambda_a^*)^{i+j+1}}\binom{i+j-2}{i-1}\right)_{1\leq i\leq n_a,1\leq j\leq n_b}, \\ \\
\mathbf{Y}_a=&\begin{bmatrix}
\Phi_a^{[0]\dag}&0 & \cdots&0 \\
\Phi_a^{[1]\dag}&\Phi_a^{[0]\dag}&\cdots&0 \\
\vdots &\vdots &&\vdots \\
\Phi_a^{[n_a-1]\dag}&\Phi_a^{[n_a-1]\dag}&\cdots&\Phi_a^{[0]\dag} \\
\end{bmatrix},\,\,\,\,\,\,\mathbf{D}_a=\begin{bmatrix}
\frac{1}{(\lambda-\lambda_a^*)}&0&\cdots& 0\\
\frac{1}{(\lambda-\lambda_a^*)^2}&\frac{1}{(\lambda-\lambda_a^*)}&\cdots& 0\\
\vdots&\vdots& &\vdots\\
\frac{1}{(\lambda-\lambda_a^*)^{n}}&\frac{1}{(\lambda-\lambda_a^*)^{n-1}}&\cdots& \frac{1}{(\lambda-\lambda_a^*)}\\
\end{bmatrix}, \\
\mathbf{X}=&\left[\mathbf{X}_1,\mathbf{X}_2\right],\,\,\,\,\,\, \mathbf{Y}=\begin{bmatrix}
\mathbf{Y}_1&0\\
0&\mathbf{Y}_2 \\
\end{bmatrix},\,\,\,\,\,\,\,\mathbf{D}=\begin{bmatrix}
\mathbf{D}_1&0\\
0&\mathbf{D}_2 \\
\end{bmatrix},\,\,\,\,\,\,\,\,  \mathbf{S}=\begin{bmatrix}
\mathbf{S}_{1,1}&\mathbf{S}_{1,2} \\
\mathbf{S}_{2,1}&\mathbf{S}_{2,2} \\
\end{bmatrix}.
\end{split}
\end{equation}
The corresponding B\"acklund transformation can be represented as
\begin{equation}
\psi_{n_1,n_2}=\psi-2\mathbf{X}_1\mathbf{M}^{-1}\mathbf{X}_2^{\dag}.
\label{eq:high-RW}
\end{equation}

\subsection*{The trivial Darboux matrix and the translation}
We would like to discuss the solutions constructed from the Darboux transformation. Actually, even though the formula for the Darboux transformation itself is simple, it is not very easy to write down the formula in a compact form. In this subsection, we deal with this problem. Firstly, we give the proof that, in this case, the Darboux transformation is equivalent to the translation and phase gauge.

By the symmetry, we know that $(\ii \sigma_2)\Phi^{\rm T}(-\ii \sigma_2)$
satisfies the adjoint Lax pair. Thus the squared wave function can be constructed as the following way:
\[
\mathbf{L}(x,t;\lambda)=-\frac{1}{2}\Phi(\lambda)\ii \sigma_3 (\ii \sigma_2)\Phi^{\rm T}(-\ii \sigma_2),\,\,\,\,\,\, \Phi(\lambda)=(\Phi_1(\lambda),\Phi_2(\lambda)).
\]
On the other hand, we know the Darboux matrix
\[
\mathbf{T}(\lambda)=\mathbb{I}-\frac{\lambda_1-\lambda_1^*}{\lambda-\lambda_1^*}\mathbf{P}_1(x,t),\,\,\,\,\, \mathbf{P}_1(x,t)=\frac{\Phi_1(\lambda_1)\Phi_1(\lambda_1)^{\dag}}{\Phi_1(\lambda_1)^{\dag}\Phi_1(\lambda_1)} .
\]
Now we apply the Darboux matrix to construct a new squared wave function:
\[
\widehat{\mathbf{L}}(x,t;\lambda)=-\frac{1}{2} \mathbf{T}(\lambda)\Phi(\lambda) \ii \sigma_3 (\ii \sigma_2)\Phi^{\rm T}\mathbf{T}(\lambda)^{\rm T}(-\ii \sigma_2)\left(\frac{\lambda-\lambda_1^*}{\lambda-\lambda_1}\right).
\]
Here we just need to verify that there is no singularity at point $\lambda_1$ and $\lambda_1^*$. It is seen that
\[
\frac{\mathbf{T}(x,t;\lambda)\Phi_1(\lambda)}{\lambda-\lambda_1}
\]
has no singularity at point $\lambda=\lambda_1.$ So is the squared wave function $\widehat{\mathbf{L}}(x,t;\lambda)$. Similarly, we know that
\[
{\rm Res}_{\lambda=\lambda_1^*}\mathbf{T}(\lambda)\Phi_1(\lambda)=\mathbf{P}_1(x,t)\Phi_1(\lambda_1^*)=0,
\]
which infers that the squared wave function $\widehat{\mathbf{L}}(x,t;\lambda)$ has no singularity at point $\lambda=\lambda_1^*.$ Therefore the squared wave function $\widehat{\mathbf{L}}(x,t;\lambda)$ is an entire function. It is readily to see that the squared wave function $\widehat{\mathbf{L}}(x,t;\lambda)$ has the same asymptotic behavior as $\mathbf{L}(x,t;\lambda).$ Meanwhile, by virtue of properties of the Darboux matrix, we obtain that $\det(\mathbf{L})=\det(\widehat{\mathbf{L}}).$ This is to say, the matrices $\mathbf{L}$ and $\widehat{\mathbf{L}}$ share the same structure. So the coefficients satisfy the same differential relation.  Then the solutions transformed by the above special Darboux transformation is equivalent to the translation of coordinates and the phases. By direct calculation, the Darboux matrix can be rewritten as
\[
\mathbf{T}(\lambda)=\frac{\lambda-\lambda_1}{\lambda-\lambda_1^*}\left(\mathbb{I}-\frac{\lambda_1^*-\lambda_1}{\lambda-\lambda_1} (\ii \sigma_2)\mathbf{P}_1^*(-\ii \sigma_2)\right).
\]
So the transformed squared wave function $\widehat{\mathbf{L}}(\lambda)$ can be rewritten as
\[
\widehat{\mathbf{L}}(x,t;\lambda) =\mathbf{T}(\lambda) \mathbf{L}(x,t;\lambda) \left(\mathbb{I}-\frac{\lambda_1^*-\lambda_1}{\lambda-\lambda_1}\mathbf{P}_1\right)=-\ii \sigma_3\lambda^2+\widehat{\mathbf{L}}_1\lambda+\widehat{\mathbf{L}}_0,
\]
where
\[
\widehat{\mathbf{L}}_1=\mathbf{L}_1+\ii (\lambda_1^*-\lambda_1)[\sigma_3,\mathbf{P}_1]
\]
and
\[
\widehat{\mathbf{L}}_0=\mathbf{L}_0+(\lambda_1^*-\lambda_1)\left\{[\mathbf{P}_1,\mathbf{L}_1]+\ii (\lambda_1\sigma_3\mathbf{P}_1-\lambda_1^*\mathbf{P}_1\sigma_3)\right\}.
\]
The matrix functions $\widehat{\mathbf{L}}_1$ and $\widehat{\mathbf{L}}_0$ shares the same differential equation as  $\mathbf{L}_1$ and $\mathbf{L}_0$ as in equations \eqref{eq:l-x-part} and \eqref{eq:l-t-part}.

The aim for this subsection is to prove the shift of crest and phase, if we choose the special vector solution $\Phi_1$ with $c_1=0$ or $c_1=\infty.$ The new obtained solution is an elliptic function solution with the shift of crest and phase. The explicit shift of crest and phase can be calculated by the theta function given in section \ref{sec4}.

\section{The formula of theta function}\label{sec:appB}
Here we need to use the integration formulas: (Ref.~\cite{ByrdFriedman}, 336.00-336.03)
\begin{equation}\label{eq:v01}
V_0=u,\quad V_1=\int \frac{{\rm d}u}{1-\alpha^2 {\rm sn}^2u}=\Pi(\varphi,\alpha^2,k),\,\,\, \varphi={\rm am} \,u,
\end{equation}
and
\begin{equation}\label{eq:v2}
\begin{split}
V_2=&\int \frac{{\rm d}u}{(1-\alpha^2 {\rm sn}^2u)^2}=\frac{1}{2(\alpha^2-1)(k^2-\alpha^2)}
\left[\alpha^2E(u)+(k^2-\alpha^2)u+(2\alpha^2(1+k^2)-\alpha^4-3k^2)\Pi(\varphi,\alpha^2,k)\right. \\
&\left.-\frac{\alpha^4 {\rm sn}\,u\, {\rm cn}\,u\, {\rm dn}\,u}{1-\alpha^2{\rm sn}^2u} \right]
\end{split}
\end{equation}
and
\begin{equation}\label{eq:vm3}
\begin{split}
V_{m+3}&=\frac{1}{2(m+2)(1-\alpha^2)(k^2-\alpha^2)}\left[(2m+1)k^2 V_m+ 2(m+1)(\alpha^2k^2+\alpha^2-3k^2)V_{m+1}\right.\\
&\left.+(2m+3)(\alpha^4-2\alpha^2(1+k^2)+3k^2)V_{m+2}+\frac{\alpha^4 {\rm sn}\,u\, {\rm cn}\,u\, {\rm dn}\,u}{(1-\alpha^2{\rm sn}^2u)^{m+2}}\right].
\end{split}
\end{equation}

The elliptic integral of the second kind is given by
\[
E(u):=\int_{0}^{u}{\rm dn}^2v\,{\rm d}v= Z(u)+\frac{E(m)}{K(m)}u,
\]
where $\vartheta_4(u)$ and $Z(u)$ are the theta function and the Jacobi zeta function introduced in \eqref{eq:thetadef2} and \eqref{eq:jacobizeta}.

The elliptic integral of the third kind is given by (see Appendix B of Ref.~\cite{Takahashi16}  and  17.4.35 of Ref.~\cite{AbramowitzStegun} for derivation.)
\begin{equation}\label{eq:appBPi}
\Pi(\varphi,\alpha^2,k)=\int_{0}^{u}\frac{{\rm d}v}{1-\alpha^2\,{\rm sn}^2v}=u+{\rm sc}\,a\, {\rm nd}\,a
\left(\frac{1}{2}\ln \frac{\vartheta_4(\frac{\pi}{2K}(u-a))}{\vartheta_4(\frac{\pi}{2K}(u+a))}+uZ(a)\right),\,\,\, {\rm sn}(a)=\alpha/k.
\end{equation}

\noindent\textbf{Theta functions and related formulas:}

We define the theta function by \cite{KharchevZ15}
\begin{equation}
\vartheta_{a,b}(u|\tau):=\sum_{n\in \mathbb{Z}} \ee^{\ii\pi\tau(n+a)^2+2\pi\ii(n+a)(n+b)},
\end{equation}
and
\begin{equation}\label{eq:thetadef2}
\vartheta_3(u|\tau)=\vartheta_{0,0}(u|\tau),\,\,\,\, \vartheta_4(u|\tau)=\vartheta_{0,\frac{1}{2}}(u|\tau),\,\,\,\,
\vartheta_2(u|\tau)=\vartheta_{\frac{1}{2},0}(u|\tau),\,\,\,\, \vartheta_1(u|\tau)=-\vartheta_{\frac{1}{2},\frac{1}{2}}(u|\tau).
\end{equation}
We also write $\vartheta_j(u,q)=\vartheta_j(u|\tau)$ with the nome $q=\ee^{\ii\pi\tau}$, where $\tau=\ii K'/K$, $K=K(m)$, $K'=K(1-m)$. The notation $\vartheta_j=\vartheta_j(0|\tau)$ is always used. The Jacobi elliptic functions in terms of theta functions are
\[
\sn(2Ku)=\frac{\vartheta_3}{\vartheta_2} \frac{\vartheta_1(u)}{\vartheta_4(u)},\,\,\,\,\, \cn(2Ku)=\frac{\vartheta_4}{\vartheta_2} \frac{\vartheta_2(u)}{\vartheta_4(u)},\,\,\,\,\,
\dn(2Ku)=\frac{\vartheta_4}{\vartheta_3} \frac{\vartheta_3(u)}{\vartheta_4(u)}.
\]
The elliptic parameters are given by $m=\frac{\vartheta_2^4}{\vartheta_3^4}$ and $m'=1-m=\frac{\vartheta_4^4}{\vartheta_3^4}$.

We introduce the Jacobi zeta function:
\begin{equation}\label{eq:jacobizeta}
Z(u|m)=\frac{1}{2K}\frac{\vartheta_4'(\frac{u}{2K})}{\vartheta_4(\frac{u}{2K})}=\frac{\dd}{\dd u}\ln\vartheta_4(\frac{u}{2K}).
\end{equation}

The three-term Weierstrass addition identities (or Fay's identities) are given by \cite{KharchevZ15}
\begin{equation}\label{eq:addition}
\begin{split}
  &\vartheta_1(a+b)\vartheta_2(a-b)\vartheta_3(c+d)\vartheta_4(c-d)- \vartheta_1(a-c)\vartheta_2(a+c)\vartheta_3(c-b)\vartheta_4(c+b)\\
= &\vartheta_1(b+d)\vartheta_2(b-d)\vartheta_3(a+c)\vartheta_4(a-c),\\
&\vartheta_1(u+a)\vartheta_1(u-a)\vartheta_4(v+b)\vartheta_4(v-b)-\vartheta_1(v+a)\vartheta_1(v-a)\vartheta_4(u+b)\vartheta_4(u-b)\\
=&\vartheta_1(u+v)\vartheta_1(u-v)\vartheta_4(a+b)\vartheta_4(a-b), \\
&\vartheta_2(u+a)\vartheta_2(u-a)\vartheta_4(v+b)\vartheta_4(v-b)+\vartheta_1(u+v)\vartheta_1(u-v)\vartheta_3(a+b)\vartheta_3(a-b)\\
=&\vartheta_2(v+a)\vartheta_2(v-a)\vartheta_4(u+b)\vartheta_4(u-b), \\
&\vartheta_2(u+v)\vartheta_2(u-v)\vartheta_2(a+b)\vartheta_2(a-b)-\vartheta_3(u+a)\vartheta_3(u-a)\vartheta_3(v+b)\vartheta_3(v-b)\\
=&-\vartheta_4(v+a)\vartheta_4(v-a)\vartheta_4(u+b)\vartheta_4(u-b).
\end{split}
\end{equation}
The shift formulas by half-periods are:
\begin{equation}\label{eq:shift}
\begin{split}
\vartheta_1\left(u+\frac{1+\tau}{2}\right)&={\rm e}^{-\pi \ii (u+\tau/4)}\vartheta_3(u),\quad \vartheta_2\left(u+\frac{1+\tau}{2}\right)=-\ii {\rm e}^{-\pi\ii (u+\tau/4)}\vartheta_4(u),\\
\vartheta_3\left(u+\frac{1+\tau}{2}\right)&=\ii {\rm e}^{-\pi\ii (u+\tau/4)}\vartheta_1(u),\quad \vartheta_4\left(u+\frac{1+\tau}{2}\right)={\rm e}^{-\pi\ii (u+\tau/4)}\vartheta_2(u).
\end{split}
\end{equation}

\section{The second-second order rogue waves}\label{sec:appC}
In this appendix, we give a detailed calculation on the construction of second-second order rogue waves.

Choosing the parameters $b=\frac{\sqrt{3}}{2}$,  $c=d=0.5$, and $\lambda=\lambda_i+\frac{\ii \epsilon_i^2}{a_i^2}$, $a_1=\sqrt{-3-\ii \sqrt{3}}$, $a_2=\sqrt{-3+\ii \sqrt{3}}$, then we have
\[
\theta_1(\lambda)=\left\{ \begin{matrix} \theta_1^{[0]}+J_1(x)\epsilon_1+J_2(x)\epsilon_1^2+J_3(x)\epsilon_1^3+O(\epsilon_1^4),&\lambda \in O(\lambda_1,\delta)\\[8pt]
\theta_2^{[0]}+(J_1^*(x)+2\ii t)\epsilon_2+J_2^*(x)\epsilon_2^2+\left(J_3^*(x)-\frac{11\,\ii +\sqrt{3}}{24}t\right)\epsilon_2^3+O(\epsilon_2^4), &\lambda \in O(\lambda_2,\delta)
\end{matrix}\right.
\]
where
\begin{equation*}
\begin{split}
J_1(x)=&x+\ii t-\left(\frac{1}{2}-\frac{\sqrt{3}}{6}\ii \right)\left(P_4(x)+\frac{P_2(x)}{P_1(x)}\right), \,\,\,\,\,\,\,
J_2(x)=-\left(\frac{1}{2}-\frac{\sqrt{3}}{6}\ii \right)\frac{P_2(x)}{P_1(x)^2}, \\
J_3(x)=&\left(\frac{1}{48}-\ii \frac{7\sqrt{3}}{144}\right)x+\frac{\sqrt{3}-11\ii }{48}t-\left(\frac{1}{24}-\ii \frac{\sqrt{3}}{108}\right)\left(P_4(x)+\frac{P_2(x)}{P_1(x)}\right)\\&-\left(\frac{1}{9}+\ii \frac{\sqrt{3}}{9}\right)\frac{P_2(x)}{P_1(x)^2}-\left(\frac{2}{3}-\frac{2\sqrt{3}}{9}\ii \right)\frac{P_2(x)}{P_1(x)^3}, \\
\end{split}
\end{equation*}
and
\begin{equation*}
\begin{split}
P_1(x)&=\ii \sqrt{3}-{\rm cn}^2(2x|0.5),\\
P_2(x)&={\rm sn}^2(2x|0.5){\rm cn}^2(2x|0.5){\rm dn}^2(2x|0.5),\\
P_4(x)&=\frac{2E}{K}x+Z(2x|0.5).
\end{split}
\end{equation*}

Then the vector solutions can be expanded as
\begin{equation*}
\sqrt{\nu(x)-\beta_1(\lambda)}\exp(\theta_1)=\left\{\begin{matrix}
\sqrt{P_1(x)}\exp(\theta_1^{[0]})\left(1+\varphi_1^{[1]}\epsilon_1+\varphi_1^{[2]}\epsilon_1^2+\varphi_1^{[3]}\epsilon_1^3+O(\epsilon_1^4)\right),&\lambda \in O(\lambda_1,\delta) \\[8pt]
\sqrt{P_1^*(x)}\exp(\theta_2^{[0]})\left(1+\varphi_2^{[1]}\epsilon_2+\varphi_2^{[2]}\epsilon_2^2+\varphi_2^{[3]}\epsilon_2^3+O(\epsilon_2^4)\right),&\lambda \in O(\lambda_2,\delta)  \\
\end{matrix}\right.
\end{equation*}
and
\begin{equation*}
r_1(\lambda)\sqrt{\nu(x)-\beta_1(\lambda)}\exp(\theta_1)=\left\{
\begin{matrix}
\frac{{\rm cn}(2x|0.5)\exp(\theta_1^{[0]}+\ii t)}{\sqrt{P_1(x)}}\left(\phi_1^{[0]}+\phi_1^{[1]}\epsilon_1+\phi_1^{[2]}\epsilon_1^2+\phi_1^{[3]}\epsilon_1^3+O(\epsilon_1^4)\right),\!\!\!&\lambda \!\in\! O(\lambda_1,\delta) \\[8pt]
\frac{{\rm cn}(2x|0.5)\exp(\theta_2^{[0]}+\ii t)}{\sqrt{P_1^*(x)}}\left(\phi_2^{[0]}+\phi_2^{[1]}\epsilon_2+\phi_2^{[2]}\epsilon_2^2+\phi_2^{[3]}\epsilon_2^3+O(\epsilon_2^4)\right),\!\!\!&\lambda \!\in\! O(\lambda_2,\delta). \\
\end{matrix}\right.
\end{equation*}
We just list the odd order coefficients for the above expansions, which would be used to construct the high order rogue waves:
\begin{equation*}
\begin{split}
\varphi_1^{[1]}&=1-J_1(x)P_1(x),\,\,\,\,\,\,\, \phi_1^{[1]}=r_1^{[0]}\left(J_1+\frac{1}{P_1}\right), \\
\varphi_1^{[3]}&=-\left(J_1J_2+\frac{1}{6}J_1^3+J_3\right)P_1+\frac{\sqrt{3}\, \ii }{3}J_1+\frac{1}{2}J_1^2+J_2-\frac{11+\ii \sqrt{3}}{48}+\frac{\frac{\ii
\sqrt{3}}{3}+\frac{1}{2}J_1}{P_1}+\frac{1}{2P_1^2}, \\
\phi_1^{[3]}&=r_1^{[0]}\left[J_3+J_1J_2+\frac{1}{6}J_1^3+\frac{\frac{\sqrt{3}\ii }{3}J_1+\frac{1}{2}J_1^2+J_2-\frac{11+\ii \sqrt{3}}{48}}{P_1}+\frac{\frac{3}{2}J_1+\ii \sqrt{3}}{P_1^2}+\frac{5}{2P_1^3}\right]+r_1^{[2]}\left(J_1+\frac{1}{P_1}\right), \end{split}
\end{equation*}
and
\begin{equation*}
\begin{split}
\varphi_2^{[1]}=&1-(J_1^*+2\ii t)P_1^*,\,\,\,\,\,\,\, \phi_2^{[1]}=r_2^{[0]}\left(J_1^*+2\ii t+\frac{1}{P_1^*}\right), \\
\varphi_2^{[3]}=&-\left[(J_1^*+2\ii t)J_2^*+\frac{1}{6}(J_1^*+2\ii t)^3+\left(J_3^*-\frac{11\,\ii +\sqrt{3}}{24}t\right)\right]P_1^*-\frac{\sqrt{3}\, \ii }{3}(J_1^*+2\ii t)+\frac{1}{2}(J_1^*+2\ii t)^2+J_2^*\\
&-\frac{11-\ii \sqrt{3}}{48}+\frac{\frac{-\ii
\sqrt{3}}{3}+\frac{1}{2}(J_1^*+2\ii t)}{P_1^*}+\frac{1}{2P_1^{*2}}, \\
\phi_2^{[3]}=&r_2^{[0]}\left[\left(J_3^*-\frac{11\,\ii +\sqrt{3}}{24}t\right)+(J_1^*+2\ii t)J_2^*+\frac{1}{6}(J_1^*+2\ii t)^3+\frac{-\frac{\sqrt{3}\ii }{3}(J_1^*+2\ii t)+\frac{1}{2}(J_1^*+2\ii t)^2+J_2^*-\frac{11-\ii \sqrt{3}}{48}}{P_1^*}\right. \\
 &\left.+\frac{\frac{3}{2}(J_1^*+2\ii t)-\ii \sqrt{3}}{P_1^{*2}}+\frac{5}{2P_1^{*3}}\right]+r_2^{[2]}\left((J_1^*+2\ii t)+\frac{1}{P_1^*}\right),
\end{split}
\end{equation*}
and
\begin{equation*}
\begin{split}
r_1^{[0]}&=\frac{2\ii \,P_2(x)}{{\rm cn}(2x|0.5)}+(\sqrt{3}+\ii ){\rm cn}(2x|0.5),\,\,\,\,r_1^{[2]}=-\left(\frac{\ii }{2}+\frac{\sqrt{3}}{6}\right){\rm cn}(2x|0.5),\\
r_2^{[0]}&=\frac{2\ii \,P_2(x)}{{\rm cn}(2x|0.5)}-(\sqrt{3}-\ii ){\rm cn}(2x|0.5),\,\,\,\,\,\, r_2^{[2]}=-\left(\frac{\ii }{2}-\frac{\sqrt{3}}{6}\right){\rm cn}(2x|0.5).
\end{split}
\end{equation*}

The $(2,2)$-rogue waves can be represented as
\begin{equation}\label{eq:second-second-rw}
\psi_{(2,2)}={\rm cn}(2x|0.5)\left[\frac{\det(\mathbf{K}_4)}{\det(\mathbf{M}_4)}\right]{\rm e}^{-\ii t}
\end{equation}
where
\[
\mathbf{M}_4=\mathbf{X}_4^{\dag}\mathbf{D}_4\mathbf{X}_4+\mathbf{Y}_4^{\dag}\mathbf{D}_4\mathbf{Y}_4,\,\,\,\,\,\, \mathbf{K}_4=\mathbf{M}_4-2\mathbf{X}^{\dag}_{4,1}\mathbf{Y}_{4,1}
\]
and
\[
\mathbf{X}_4=\begin{bmatrix}
\varphi_1^{[1]}&\varphi_1^{[3]}&0&0\\
0&\varphi_1^{[1]}&0&0\\
0&0& \varphi_2^{[1]}&\varphi_2^{[3]}\\
0&0&0&\varphi_2^{[1]}\\
\end{bmatrix},\,\,\,\,\, \mathbf{Y}_4=\begin{bmatrix}
\phi_1^{[1]}&\phi_1^{[3]}&0&0\\
0&\phi_1^{[1]}&0&0\\
0&0& \phi_2^{[1]}&\phi_2^{[3]}\\
0&0&0&\phi_2^{[1]}\\
\end{bmatrix},\,\,\,\,\,\,\, \begin{matrix}
\mathbf{X}_{4,1}=\left[\varphi_1^{[1]},\varphi_1^{[2]},\varphi_2^{[1]},\varphi_2^{[2]}\right] \\[8pt]
\mathbf{Y}_{4,1}=\left[\phi_1^{[1]},\phi_1^{[2]},\phi_2^{[1]},\phi_2^{[2]}\right] \\
\end{matrix}
\]
and
\[
\mathbf{D}_4=\begin{bmatrix}
\frac{1}{\lambda_1-\lambda_1^*}& \frac{-\ii }{(\lambda_1-\lambda_1^*)^2a_1} &  \frac{1}{\lambda_2-\lambda_1^*} & \frac{-\ii }{(\lambda_2-\lambda_1^*)^2a_2} \\
\frac{-\ii }{(\lambda_1-\lambda_1^*)^2a_1^*} & \frac{-2}{(\lambda_1-\lambda_1^*)^3|a_1|^2} &  \frac{-\ii }{(\lambda_2-\lambda_1^*)^2a_1^*} & \frac{-2}{(\lambda_2-\lambda_1^*)^3a_2a_1^*} \\
\frac{1}{\lambda_1-\lambda_2^*}& \frac{-\ii }{(\lambda_1-\lambda_2^*)^2a_1} &  \frac{1}{\lambda_2-\lambda_2^*} & \frac{-\ii }{(\lambda_2-\lambda_1^*)^2a_2} \\
\frac{-\ii }{(\lambda_1-\lambda_2^*)^2a_2^*}& \frac{-2}{(\lambda_1-\lambda_2^*)^3a_1a_2^*} &  \frac{-\ii }{(\lambda_2-\lambda_2^*)^2a_2^*} & \frac{-2}{(\lambda_2-\lambda_2^*)|a_2|^2} \\
\end{bmatrix}.
\]

\section*{Acknowledgments}
B.-F.  acknowledges the partial support by NSF under Grant No. DMS-1715991, and National Natural Science Foundation of
China under Grant No.11728103. Liming Ling is supported by National Natural Science Foundation of
China(Contact No. 11771151), Guangdong Natural Science Foundation (Contact No. 2017A030313008), Guangzhou Science and Technology Program(No. 201707010040). The author Liming Ling thanks the Professor Peter D. Miller for his encourage, help, guidance and discussion during the scholar visiting of one year in the University of Michigan. Meanwhile, he thanks Dr. Deniz Bilman for his valuable discussions.
The work of D.A.T. is supported by the Ministry of Education, Culture,
Sports, Science (MEXT)-Supported Program for the Strategic
Research Foundation at Private Universities ``Topological
Science'' (Grant No. S1511006).


\begin{thebibliography}{99}
\bibitem{AbramowitzStegun}
M. Abramowitz and I. A. Stegun, \emph{Handbook of Mathematical Functions with Formulas, Graphs, and Mathematical Tables} 9th edition (New York:Dover) 1965.

\bibitem{AshourCNB}O. A. Ashour, S. A. Chin, S. N. Nikoli\`c, and M. R. Beli\`c \emph{Higher-order Breathers as Quasi-rogue Waves on a Periodic Background.} arXiv:1810.02887v1

\bibitem{BelokolosBEIM94} E. D. Belokolos, A. I. Bobenko, V. Z. Enol'skii, A. R. Its, and V. B. Matveev, \emph{Algebro-geometric approach to nonlinear integrable equations. Springer series studies in nonlinear dynamics.} Berlin, Germany: Spinger. 1994

\bibitem{BertolaET16} M Bertola, G A El and A. Tovbis \emph{Rogue waves in multiphase solutions of the focusing nonlinear Schr\"odinger equation.} Proc. R. Soc. A, 2016, 472(2194): 20160340.

\bibitem{BertolaT13} M. Bertola and A. Tovbis, \emph{Universality for the focusing nonlinear Schr\"odinger equation at the gradient catastrophe point: rational breathers and poles of the tritronqu\'ee solution to Painlev\'e I.} Communications on Pure and Applied Mathematics, 2013, 66(5): 678-752.

\bibitem{BertolaT17} M. Bertola and A. Tovbis \emph{Maximal amplitudes of finite-gap solutions for the focusing nonlinear Schr\"odinger equation.} Communications in Mathematical Physics, 2017, 354(2): 525-547.

\bibitem{BilmanB18}D. Bilman and R. Buckingham, \emph{Large-order asymptotics for multiple-pole solitons of the focusing nonlinear Schr\"odinger equation.} arXiv preprint arXiv:1807.09058, 2018.
\bibitem{BilmanM17} D. Bilman and P. D. Miller,  \emph{A robust inverse scattering transform for the focusing nonlinear Schr\"odinger equation,} \texttt{arXiv:1710.06568}, 2017.

\bibitem{BilmanLM18} D. Bilman, L. Ling and P. D. Miller, \emph{Extreme Superposition: Rogue Waves of Infinite Order and the Painlev\'e-III Hierarchy.} arXiv preprint arXiv:1806.00545, 2018.

\bibitem{BiodiniM17} G. Biondini and D. Mantzavinos, \emph{Long-Time Asymptotics for the Focusing Nonlinear Schr\"odinger Equation with Nonzero Boundary Conditions at Infinity and Asymptotic Stage of Modulational Instability.} Communications on Pure and Applied Mathematics, 2017, 70(12): 2300-2365; G. Biondini, S. Li and D. Mantzavinos, \emph{Soliton trapping, transmission and wake in modulationally unstable media.} Phys. Rev. E 2018, 98, 042211

\bibitem{ByrdFriedman}
P. F. Byrd and M. D. Friedman, \emph{Handbook of Elliptic Integrals for Engineers and Scientists} 2nd edition (Berling:Springer) 1971.

\bibitem{ChenP18}Chen J, Pelinovsky D E. \emph{Rogue periodic waves of the focusing nonlinear Schr\"odinger equation.} Proc. R. Soc. A, 2018, 474(2210): 20170814.
\bibitem{ChengLC14} X. P. Cheng, S.Y. Lou,  C. Chen and X.Y. Tang, \emph{Interactions between solitons and other nonlinear Schr\"odinger waves.} Physical Review E, 2014, 89(4): 043202.

\bibitem{Carr2000}
L. D. Carr,  C. W. Clark, and W. P Reinhardt, \emph{Stationary solutions of the one-dimensional nonlinear
Schr\"odinger equation. II. Case of attractive nonlinearity.} Physical Review A, 2000, 62:063611

\bibitem{Cieslieski09}Jan L. Cieslieski, \emph{Algebraic construction of the Darboux matrix revisited.} Journal of Physics A: Mathematical and Theoretical 42. 40 (2009): 404003.

\bibitem{DeconinckS17}B. Deconinck B and  B. L. Segal \emph{The stability spectrum for elliptic solutions to the focusing NLS equation.} Physica D: Nonlinear Phenomena, 2017, 346: 1-19.

\bibitem{ElKT16} G. A. El, E. G. Khamis and A. Tovbis \emph{Dam break problem for the focusing nonlinear Schr\"odinger equation and the generation of rogue waves.} Nonlinearity, 2016, 29(9): 2798.

\bibitem{FaddeevTakhtajan}L.D. Faddeev and L.A. Takhtajan, \emph{Hamiltonian Methods in the Theory of Solitons}, Berlin, Germany: Spinger. 1987.

\bibitem{GrinevichS18} P. G. Grinevich, P. M. Santini,  \emph{The finite gap method and the periodic NLS Cauchy problem of the anomalous waves, for a finite number of unstable modes.} arXiv:1810.09247

\bibitem{GuHZ05}G.C. Gu, H. S. Hu and Z.X. Zhou, \emph{Darboux Transformations in Integrable Systems.} Theory and their Applications to Geometry (Dordrecht: Springer) 2005
\bibitem{GuoLL12} B. Guo, L Ling, Q. P. Liu, \emph{Nonlinear Schr\"odinger equation: generalized Darboux transformation and rogue wave solutions,} Physical Review E 85 (2), 02660, 2012

\bibitem{Hamner2011} C. Hamner, J. J. Chang, P. Engels, and M. A. Hoefer, \emph{Generation of Dark-Bright Soliton Trains in Superfluid-Superfluid Counterflow}, Physical Review Letters, 2011, 106:065302

\bibitem{Kamchatnov97} A M. Kamchatnov, \emph{New approach to periodic solutions of integrable equations and nonlinear theory of modulational instability.} Physics Reports, 1997, 286: 199-270.
\bibitem{KedzioraAA14} D. J. Kedziora, A. Ankiewicz and N. Akhmediev, \emph{Rogue waves and solitons on a cnoidal background.} The European Physical Journal Special Topics, 2014, 223: 43-62.

\bibitem{Khawaja2002} U. Al Khawaja, H. T. C. Stoof, R. G. Hulet, K. E. Strecker, and G. B. Partridge, \emph{Bright Soliton Trains of Trapped Bose-Einstein Condensates}, Physical Review Letters, 2002, 89:200404

\bibitem{KharchevZ15}Kharchev S, Zabrodin A. \emph{Theta vocabulary I.} Journal of Geometry and Physics, 2015, 94: 19-31.

\bibitem{Kibler2010} B. Kibler, J. Fatome, C. Finot, G. Millot, F. Dias, G. Genty, N. Akhmediev and J. M. Dudley, \emph{The Peregrine soliton in nonlinear fibre optics.} Nature Physics, 2010, 6:790-795.

\bibitem{KotlyarovS17}Kotlyarov V, Shepelsky D. \emph{Planar unimodular Baker-Akhiezer function for the nonlinear schr\"odinger equation.} Annals of Mathematical Sciences and Applications, 2017, 2(2): 343-384.

\bibitem{Kurosaki2007} T. Kurosaki, and M. Wadati, \emph{Matter-Wave Bright Solitons with a Finite Background in Spinor Bose-Einstein Condensates}, Journal of the Physical Society of Japan, 2007, 76:084002

\bibitem{MatveevS91}V.B. Matveev and M. A. Salle, \emph{Darboux Transformation sand Solitons.} (Berlin:Springer) 1991

\bibitem{Li2018} S. Li, B. Prinari, and G. Biondini, \emph{Solitons and rogue waves in spinor Bose-Einstein condensates.} Physical Review E, 2018, 97:022221

\bibitem{LouCT14}S.Y. Lou, X. P. Cheng and X. Y. Tang, \emph{Dressed dark solitons of the defocusing nonlinear Schr\"odinger equation.} Chinese Physics Letters, 2014, 31(7): 070201.
\bibitem{LyngM07}G. D. Lyng and P. D. Miller, \emph{The N-soliton of the focusing nonlinear Schr\"odinger equation for $N$ large.} Communications on Pure and Applied Mathematics, 2007, 60(7): 951-1026.

\bibitem{Shin12} H J. Shin \emph{Soliton dynamics in phase-modulated lattices.} Journal of Physics A: Mathematical and Theoretical, 2012, 45(25): 255206.


\bibitem{Strecker2002}K. E. Strecker, G. B. Partridge, A. G. Truscott and R. G. Hulet, \emph{Formation and propagation of matter-wave soliton trains.} Nature, 2002, 417:150-153

\bibitem{Takahashi16} D A. Takahashi \emph{Integrable model for density-modulated quantum condensates: Solitons passing through a soliton lattice.} Physical Review E, 2016, 93(6): 062224.

\bibitem{Takahashi12} D. A. Takahashi, S. Tsuchiya, R. Yoshii and M. Nitta, \emph{Fermionic solutions of chiral Gross-Neveu and Bogoliubov-de Gennes systems in nonlinear Schr\"odinger hierarchy.} Physics Letters B, 2012, 718: 632-637.

\bibitem{Wright17}III O C. Wright \emph{Bounded ultra-elliptic solutions of the defocusing nonlinear Schr\"odinger equation.} Physica D: Nonlinear Phenomena, 2017, 360: 1-16.


\end{thebibliography}
\end{document}